\documentclass{wspc-procs975x65}

\usepackage{latexsym}
\usepackage{graphicx}
\usepackage{axodraw}

\newcommand{\be}{\begin{equation}}
\newcommand{\ee}{\end{equation}}
\newcommand{\ba}{\begin{eqnarray}}
\newcommand{\ea}{\end{eqnarray}}
\def\nn{\nonumber}
\def\half{\frac{1}{2}}
\newcommand{\Dslash}{{D \hskip -7pt /}}

\def\olmass{\begin{picture}(10,10)(10,10)
\PhotonArc(15,10)(10,0,180)2 5
\Line(0,10)(30,10)
\end{picture}}

\def\tlscbffig{\begin{picture}(1,20)(1,-5)
\Photon(-5,15)(35,15)2 5
\CArc(15,15)(20,0,180)
\CArc(15,15)(20,180,360)
\CArc(15,15)(22,0,180)
\CArc(15,15)(22,180,360)
\end{picture}}

\def\tlscbf{\begin{picture}(10,10)(10,10)
\Photon(4,15)(26,15)2 4
\CArc(15,15)(10,0,180)
\CArc(15,15)(10,180,360)
\CArc(15,15)(12,0,180)
\CArc(15,15)(12,180,360)
\end{picture}}

\def\tlspbf{\begin{picture}(10,10)(10,10)
\Photon(4,15)(26,15)2 4
\CArc(15,15)(10,0,180)
\CArc(15,15)(10,180,360)
\CArc(15,15)(11,0,180)
\CArc(15,15)(11,180,360)
\CArc(15,15)(12,0,180)
\CArc(15,15)(12,180,360)
\end{picture}}

\def\tlscfrph{\begin{picture}(10,10)(10,10)
\DashLine(5,15)(25,15){1}
\CArc(15,15)(10,0,180)
\CArc(15,15)(10,180,360)
\end{picture}}

\def\tlscfr{\begin{picture}(10,10)(10,10)
\Photon(5,15)(25,15)2 4
\CArc(15,15)(10,0,180)
\CArc(15,15)(10,180,360)
\end{picture}}

\def\olscbf{\begin{picture}(10,10)(10,10)
\CArc(15,15)(10,0,180)
\CArc(15,15)(10,180,360)
\CArc(15,15)(12,0,180)
\CArc(15,15)(12,180,360)
\end{picture}}

\def\olscfr{\begin{picture}(10,10)(10,10)
\CArc(15,15)(10,0,180)
\CArc(15,15)(10,180,360)
\end{picture}}

\def\olscbfv{\begin{picture}(10,10)(10,10)
\CArc(15,15)(10,0,180)
\CArc(15,15)(10,180,360)
\CArc(15,15)(12,0,180)
\CArc(15,15)(12,180,360)
\Vertex(15,4)2
\end{picture}}

\def\olscfrv{\begin{picture}(10,10)(10,10)
\CArc(15,15)(10,0,180)
\CArc(15,15)(10,180,360)
\Vertex(15,5)2
\end{picture}}

\begin{document}

\title{Heisenberg-Euler Effective Lagrangians :\\ Basics and Extensions
\footnote{\uppercase{T}o appear in \uppercase{I}an \uppercase{K}ogan \uppercase{M}emorial \uppercase{V}olume, {\it \uppercase{F}rom \uppercase{F}ields to \uppercase{S}trings : \uppercase{C}ircumnavigating \uppercase{T}heoretical \uppercase{P}hysics}, \uppercase{M}. \uppercase{S}hifman, \uppercase{A}. \uppercase{V}ainshtein and \uppercase{J}. \uppercase{W}heater, \uppercase{E}ds., (\uppercase{W}orld \uppercase{S}cientific).}}

\author{Gerald V. Dunne}

\address{Department of Physics\\
University of Connecticut\\
Storrs, CT 06269-3046, USA\\ 
E-mail: dunne@phys.uconn.edu}



\maketitle

\vspace{1cm}

\abstracts{
I present a pedagogical review of Heisenberg-Euler effective Lagrangians, beginning with the original work of Heisenberg and Euler, and Weisskopf, for the one loop effective action of quantum electrodynamics in a constant electromagnetic background field, and then summarizing some of the important applications and generalizations to inhomogeneous background fields, nonabelian backgrounds, and higher loop effective Lagrangians.}


\vspace{1cm}

{\bf {\sl Dedicated to the memory of Ian Kogan, a great physicist and friend, whose enthusiasm for life and science is sorely missed.}}

\newpage
~
\medskip

\tableofcontents
\newpage

\section{Introduction: the Heisenberg-Euler Effective Lagrangian}
\label{intro}
\renewcommand{\theequation}{1.\arabic{equation}}
\setcounter{equation}{0}

\subsection{The Spinor and Scalar QED one loop results}
\label{1lspsc}

In classical field theory the Lagrangian encapsulates the relevant classical equations of motion and the symmetries of the system. In quantum field theory the effective Lagrangian encodes quantum corrections to the classical Lagrangian, corrections that are induced by quantum effects such as vacuum polarization. This can be used as a semi-phenomenological device, as in effective field theory, or as a fundamental approach in which one uses an external classical field as a direct probe of the vacuum structure of the quantum theory. The seminal work of Heisenberg and Euler\cite{he}, and Weisskopf \cite{viki1} produced the paradigm for the entire field of effective Lagrangians by computing the nonperturbative, renormalized, one-loop effective action for quantum electrodynamics (QED) in a classical electromagnetic background of constant field strength. This special soluble case of a constant field strength leads immediately to several important insights and applications.

\begin{figure}
\centerline{\includegraphics[scale=0.7]{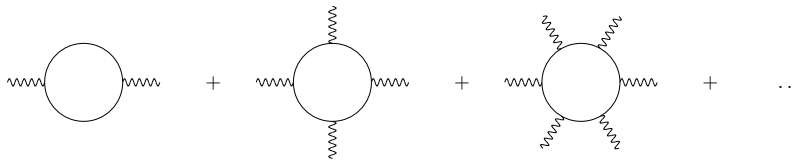}}
\caption{The diagrammatic perturbative expansion of the one loop effective action (\protect{\ref{action}}).}
\label{digexp}
\end{figure}

In spinor QED, the one-loop effective action for electrons in the presence of a background electromagnetic field is
\begin{equation}
S^{(1)}=-i \ln \det (i\Dslash-m)=-\frac{i}{2}\ln \det(\Dslash^2+m^2)
\label{action}
\end{equation}
where the Dirac operator is $\Dslash =\gamma^\nu\left(\partial_\nu+ie A_\nu\right)$,  $A_\nu$ is a
fixed classical gauge potential with field strength tensor $F_{\mu\nu}=\partial_\mu A_\nu- \partial_\nu A_\mu$, and $m$ is the electron mass. This one-loop effective action has a natural perturbative expansion in powers of the external photon field $A_\mu$, as illustrated diagrammatically in Figure \ref{digexp}. By Furry's theorem (charge conjugation symmetry of QED), the expansion is in terms of even numbers of external photon lines.
Heisenberg and Euler, and Weisskopf, showed that in the low energy limit for the external photon lines, in which case the background field strength $F_{\mu\nu}$ could be taken to be constant, it is possible to compute a relatively simple closed-form expression for the effective action, which generates {\sl all} the perturbative diagrams in Figure \ref{digexp}. Heisenberg and Euler  \cite{he} expressed their final answer for spinor QED in several equivalent ways:
\ba
{\mathcal L}_{\rm sp}^{(1)} &=&\frac{1}{h c}
\int_0^{\infty}\hskip -5pt \frac{d\eta}{\eta^3}
e^{-\eta e {\mathcal E}_c}
\left\{i e^2 \eta^2 (\vec{E}.\vec{B})\frac{\left[\cos\left(\eta e \sqrt{\vec{E}^2-\vec{B}^2+2 i (\vec{E}.\vec{B})}\right)+{\rm c.c.}\right]}{\left[\cos\left(\eta e \sqrt{\vec{E}^2-\vec{B}^2+2 i (\vec{E}.\vec{B})}\right)-{\rm c.c.}\right]} \right.\nn\\
&&\left.\hskip 5cm +1+ \frac{e^2 \eta^2}{3} (\vec{B}^2-\vec{E}^2)
\right\}\nn\\
&=& \frac{1}{hc}
\int_0^{\infty}\frac{d\eta}{\eta^3}
\,e^{-\eta\,e {\mathcal E}_c}
\left\{-i\,e^2 a\, b\, \eta^2 \left[\frac{\cosh [(b+i a)e \eta] +\cosh [(b-ia )e \eta]}{\cosh [(b+i a)e \eta] -\cosh [(b-ia )e \eta]}\right]\right.\nn\\
&&\left.\hskip 5cm 
+1+\frac{e^2\eta^2}{3} (b^2-a^2)
\right\}
\nn\\
&=& -\frac{1}{hc}
\int_0^{\infty}\frac{d\eta}{\eta^3}
\,e^{-\eta\,e {\mathcal E}_c}
\left\{
\frac{e^2a\,b\,\eta^2}{\tanh(e b \eta)\tan(e a \eta)} -1
-\frac{e^2\eta^2}{3} (b^2-a^2)
\right\}.\nn\\ \nn\\
\label{hesp}
\ea
Here ${\mathcal E}_c$ is the critical field strength 
\ba
{\mathcal E}_c=\frac{m^2 c^3}{e\hbar}\ ,
\label{critical}
\ea
and $a$ and $b$ are related to the Lorentz invariants\,\footnote{\,There 
has been a notational reversal \cite{csreview,dittrichgies,jentschura} of 
$a\leftrightarrow b$ since the original Heisenberg-Euler paper \cite{he}. 
I stick here with Heisenberg's original notation since in a frame in which 
$\vec{B}$ and $\vec{E}$ are parallel, we associate $b\leftrightarrow B$ and 
$a\leftrightarrow E$, which seems more natural. Also, modern formulations 
have adopted Schwinger's choice \cite{schwinger} of units in which the fine 
structure constant $\alpha=\frac{e^2}{\hbar c}\to\frac{e^2}{4\pi}$, in which 
case the prefactor in (\ref{hesp}) is $\frac{1}{hc}\to \frac{1}{8\pi^2}$.}  
characterizing the background electromagnetic field strength \cite{he}:
\ba
a^2-b^2&=&\vec{E}^2-\vec{B}^2= -\frac{1}{2}\,F_{\mu\nu}F^{\mu\nu}
\equiv -2 {\mathcal F}\ ,\\
a\, b&=& \vec{E}\cdot\vec{B}= -\frac{1}{4}\,F_{\mu\nu}\tilde{F}^{\mu\nu}\equiv -{\mathcal G}\ .
\label{ab}
\ea 
Thus
\ba
a= \sqrt{\sqrt{{\mathcal F}^2+{\mathcal G}^2}-{\mathcal F}} \ , \quad 
b=\sqrt{\sqrt{{\mathcal F}^2+{\mathcal G}^2}+{\mathcal F}}\ .
\label{abfg}
\ea
If ${\mathcal G}\neq 0$, it is possible to transform to a Lorentz frame in which the electric and magnetic fields are parallel or antiparallel, depending on the sign of ${\mathcal G}$. A suitable sign choice is implicit in (\ref{abfg}). The $a$ and $b$ parameters are significant  because $\pm b$ and $\pm i a$ are the eigenvalues of the $4\times 4$ antisymmetric constant matrix $F_{\mu\nu}$. This explains why this constant field strength case is exactly soluble; a constant $F_{\mu\nu}$ can be represented by a gauge field $A_\mu=-\frac{1}{2}F_{\mu\nu}x^\nu$, which is linear in $x$. Thus, in an appropriate basis, the Dirac operator factorizes into two independent Landau level problems, of ``cyclotron'' frequencies $b$ and $i\,a$ respectively. Hence the traces in (\ref{action}) can be done in closed form in terms of trigonometric functions, leading directly to (\ref{hesp}). For details see \cite{he,viki1,jentschura,greiner,dr-qed,blau,soldati,dittrichgies}.

Weisskopf \cite{viki1} computed the analogous quantity for scalar QED
\begin{equation}
S^{(1)}_{\rm scalar}=\frac{i}{2}\,\ln \det\,(D_\mu^2+m^2)
\label{sqedaction}
\end{equation}
which involves the Klein-Gordon operator rather than the Dirac operator. Weisskopf obtained (I have converted from his different units and different definition of the critical field, which introduces various factors of $2\pi$):
\ba
{\mathcal L}_{\rm scalar}^{(1)} &=&-\frac{1}{2 h c}
\int_0^{\infty}\frac{d\eta}{ \eta^3}
\,e^{-\eta e {\mathcal E}_c}
\left\{\frac{2 i e^2 \eta^2 (\vec{E}.\vec{B})}{\left[\cos\left(e \eta \sqrt{\vec{E}^2-\vec{B}^2+2 i (\vec{E}.\vec{B})}\right)-{\rm c.c.}\right]}\right.\nn\\[1mm]
&&\left.\hskip 5cm +1- \frac{e^2\eta^2}{6} (\vec{B}^2-\vec{E}^2)
\right\}\nn
\\[1mm]
&=&-\frac{1}{2 h c}
\int_0^{\infty}\frac{d\eta}{\eta^3}
\,e^{-\eta e {\mathcal E}_c}
\left\{\left[\frac{2 i e^2\, a\, b\, \eta^2}{\cosh[(b-i a)e \eta]-\cosh[(b+ia )e\eta]}\right]\right.\nn\\[1mm]
&&\left.\hskip 5cm 
+1-\frac{e^2 \eta^2}{6} (b^2-a^2)
\right\}
\nn
\\[1mm]
&=& \frac{1}{2 h c}
\int_0^{\infty}\frac{d\eta}{\eta^3}
\,e^{-\eta e {\mathcal E}_c}
\left\{
\frac{e^2a\,b\,\eta^2}{ \sinh(e b\eta)\sin(e a\eta)} -1
+\frac{e^2\eta^2}{6} (b^2-a^2)
\right\}.\nn\\
\label{hesc}
\ea
Note that the prefactor becomes $\frac{1}{16\pi^2}$ in Schwinger's units \cite{schwinger}.

\subsection{Physical Applications}
\label{apps}

The Heisenberg and Euler result (\ref{hesp}) leads immediately to a number of important physical insights and applications.

\subsubsection{Nonlinear QED Processes : Light-Light scattering} 
\label{photonscattering}

The Euler-Heisenberg effective Lagrangian (\ref{hesp}) is nonlinear in the electromagnetic fields,  the quartic and higher terms representing new nonlinear interactions, which do not occur in the tree level Maxwell action. The first of these new interactions is light-light scattering, represented diagrammatically by the second Feynman diagram in the expansion in Figure \ref{digexp}. Expanding the Euler-Heisenberg answer to quartic order we find
\ba
S^{(1)}=\frac{e^4}{360 \pi^2 m^4}\int d^4x\,
\left[(\vec{E}^2-\vec{B}^2)^2+7(\vec{E}\cdot\vec{B})^2\right]+\dots
\label{lightlight}
\ea
which gives the low energy limit (since the field strength was constant) of the amplitude for light-light scattering. As first discussed by Euler and K\"ockel \cite{ek}, these nonlinearities can be viewed as dielectric effects, with the quantum vacuum behaving as a polarizable medium. In Weisskopf's words \cite{viki1}:

\begin{quote}
``{\sl When passing through electromagnetic fields, light will behave as if the vacuum, under the action of the fields, were to acquire a dielectric constant different from unity.}"
\end{quote}

\noindent The full light-light scattering process in QED was not solved until 1951 by Karplus and Neuman \cite{karplus}.

\subsubsection{Pair-production from vacuum in an electric field.} 
\label{pairprod}

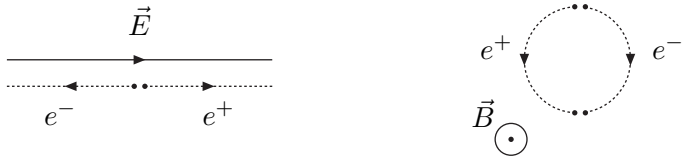
\begin{figure}[ht]
\vskip1cm
\hskip 4cm{\begin{picture}(20,20)(20,20)
\ArrowLine(-50,50)(50,50)
\DashArrowLine(-2,40)(-50,40)1
\DashArrowLine(2,40)(50,40)1
\Text(0,65)[c]{$\vec{E}$}
\Text(-30,30)[c]{$e^-$}
\Text(30,30)[c]{$e^+$}
\Vertex(-2,40)1
\Vertex(2,40)1
\DashArrowArc(165,50)(20,98,262)1
\DashArrowArcn(165,50)(20,82,278)1
\Vertex(168,70)1
\Vertex(168,30)1
\Vertex(164,70)1
\Vertex(164,30)1
\Text(135,55)[c]{$e^+$}
\Text(200,55)[c]{$e^-$}
\CArc(140,20)(6,0,360)
\Vertex(140,20)1
\Text(130,30)[c]{$\vec{B}$}
\end{picture}}
\vskip .5cm
\caption{A static electric field can tear apart a virtual $e^+ e^-$ pair from the vacuum, producing an asymptotic electron and positron, as shown on the left. On the other hand, a static magnetic field does not break this virtual dipole apart, as shown on the right for a magnetic field directed out of the page.}
\label{evsb}
\end{figure}

The presence of a background electric field accelerates and splits virtual vacuum dipole pairs, leading to $e^+e^-$ particle production: see Figure \ref{evsb}. This instability of the vacuum was realized already by Heisenberg and Euler \cite{he}, motivated in part by earlier work of Sauter on the Klein paradox \cite{sauter}. This pair production process was later formalized in the language of QED by Schwinger \cite{schwinger,schwinger2}.  Heisenberg and Euler \cite{he} deduced the leading pair production rate in a weak electric field to be
\begin{eqnarray}
\Gamma \sim\, \frac{e^2 E^2}{4\pi^3}
\exp\left[-\frac{m^2\pi }{eE}\right].
\label{imag}
\end{eqnarray}
This rate is deduced from the imaginary part of the effective Lagrangian (\ref{hesp}) when the background is purely electric
\ba
\Gamma=2\, Im {\mathcal L}\,.
\label{rate}
\ea
In modern language, this imaginary part gives the rate of vacuum non-persistence due to pair production \cite{schwinger,schwinger2}. The rate is extremely small for typical electric field strengths, becoming more appreciable when the $E$ field approaches a critical value $E_c\sim \frac{m^2 c^3}{e\hbar}\sim 10^{16}\,{\rm Vcm}^{-1}$, where the work done accelerating a virtual pair apart by a Compton wavelength is of the order of the rest mass energy for the pair. Such electric field strengths are well beyond current technological capabilities, even in the most intense lasers. For an excellent  recent review of the search for this remarkable phenomenon of vacuum pair production, see \cite{ringwald}. Even though the condition of a constant electric field is rather unrealistic, Heisenberg and Euler's result (\ref{imag}) provides the starting point for more detailed analyses which incorporate time-dependent electric fields, as is discussed below in Section \ref{inhomog}.

\subsubsection{Charge renormalization, $\beta$-functions and the strong-field limit.}
\label{charge}

Another remarkable thing about Heisenberg and Euler's result (\ref{hesp}) is that they correctly anticipated charge renormalization. The first term (on each line) on the the RHS of (\ref{hesp}) is the bare result, the second term is the subtraction of a field-free infinite term, and the third term is the subtraction of a logarithmically divergent term which has the same form as the classical Maxwell Lagrangian. This last subtraction corresponds precisely to what we now call charge renormalization, as was later formalized by Schwinger \cite{schwinger,schwinger2}. Indeed, the study of such logarithmically divergent terms was a major focus of the early quantum field theory work of both Heisenberg and Weisskopf. Weisskopf \cite{viki1} noted the characteristic strong-field limit behavior of the Heisenberg-Euler result (\ref{hesp}), for example for spinor QED in a strong magnetic background:
\ba
\frac{{\mathcal L}^{(1)}_{\rm spinor}}{{\mathcal L}_{\rm Maxwell}}\sim -\frac{e^2}{12\pi^2}\, \log\left(\frac{e  B}{m^2}\right)\, , \qquad B\to\infty\ .
\label{strong}
\ea
In modern language, the coefficient of the logarithmic dependence of this ratio is known as the one-loop QED $\beta$-function, and Weisskopf anticipated the importance of such logarithmic behavior. In later work \cite{viki2} he showed that for $n\geq 2$ loop order in perturbation theory, the divergence was at most ${\log}^{n-1}$, a fact that was important for the work of Gell-Mann and Low \cite{gellmannlow} and the development of the renormalization group. This connection between the strong field limit of effective Lagrangians and $\beta$-functions will be discussed in more detail below in Section \ref{betas}.

\subsubsection{Low-energy effective field theory} 
\label{eft}

The Heisenberg-Euler result (\ref{hesp}) is the paradigm of what is now called "low energy effective field theory" \cite{manohar,donoghue}. In this approach one describes the physics of some light degrees of freedom (here the photon field) at energies much lower than some energy scale above which one has integrated out the heavy degrees of freedom (here the electron field). Generically, the effective Lagrangian is expanded in terms of gauge and Lorentz invariant operators $O^{(n)}$ for the light fields, respecting the relevant remaining symmetries:
\begin{equation}
{\mathcal L}_{\rm eff}=m^4\, \sum_n a_n\, \frac{O^{(n)}}{ m^n}\ .
\label{effectivelag}
\end{equation}
By power counting, the operators $O^{(n)}$ are balanced by appropriate powers of the heavy mass scale $m$. We can see this structure directly in the  Heisenberg-Euler result (\ref{hesp}), for example in the first nontrivial term, the light-light scattering term (\ref{lightlight}). The full expansion is given below in (\ref{1lspweak}) and (\ref{1lspweakeg}). The light photon field is described in terms of gauge and Lorentz invariant  operators constructed from the field strength $F_{\mu\nu}$, and having mass dimension $n$. At mass dimension $8$, we can have
$(F_{\mu\nu}F^{\mu\nu})^2$ or $(F_{\mu\nu}\tilde{F}^{\mu\nu})^2$, as in (\ref{lightlight}). At
mass dimension $10$, we could have terms involving derivatives of the field strength, such as  $(\partial_\mu F_{\nu\rho}
\partial^\mu F^{\nu\rho})(F_{\alpha\beta}F^{\alpha\beta})$, which do not show up in the Heisenberg-Euler result, but which can be found in a derivative expansion about the constant field result, as discussed below in section \ref{inhomog}. The effective field theory interpretation of the Heisenberg-Euler Lagrangian also makes connection with the operator-product-expansion (OPE), where the polarization tensor is expanded as
\begin{eqnarray}
\Pi_{\mu\nu}=(q_\mu q_\nu -q^2
g_{\mu\nu})\sum_n c_n(Q^2) \,\langle O^{(n)}\rangle
\label{ope}
\end{eqnarray}
as is discussed in detail in \cite{novikov,itep,arkadyope}.

\subsection{Weak-field expansions of Heisenberg-Euler}
\label{weak}
In this section I present some results for various weak-field expansions of the Heisenberg-Euler spinor QED effective Lagrangian (\ref{hesp}), as well as for the corresponding scalar QED effective Lagrangian (\ref{hesc})  derived by Weisskopf.

\subsubsection{Spinor QED case}
\label{weakspinor}

The weak field expansion of  (\ref{hesp}), expressed in terms of the Lorentz invariants $a$ and $b$ defined in (\ref{ab}), is:
\ba
{\mathcal L}_{\rm spinor}^{(1)} \sim -\frac{m^4}{8\pi^2}\sum_{n=2}^\infty (2n-3)! \sum_{k=0}^n \frac{{\mathcal B}_{2k} {\mathcal B}_{2n-2k}}{(2k)!(2n-2k)!} \left(\frac{2e b }{m^2}\right)^{2n-2k}\left(\frac{2 i e a}{m^2}\right)^{2k} .\nn\\
\label{1lspweak}
\ea
Here the ${\mathcal B}_{2n}$ are the Bernoulli numbers \cite{bateman,abramowitz,gradshteyn,ww}, which arise from the Taylor expansions of the trigonometric functions appearing in (\ref{hesp}):
\ba
z\, \coth z=\sum_{k=0}^\infty \frac{2^{2k} 
{\mathcal B}_{2k}}{(2k)!}\, z^{2k} \, \quad z\, \cot z=\sum_{k=0}^\infty \frac{(-1)^k 2^{2k} {\mathcal B}_{2k}}{(2k)!}\, z^{2k}\ .
\label{coth}
\ea
At any given order in $m^{-2n}$, the combinations of $a$ and $b$ can be rewritten in terms of the more familiar Lorentz invariants $(\vec{E}^2-\vec{B}^2)$ and  $\vec{E}\cdot\vec{B}$ in (\ref{ab}).
For example, the first two orders yield:
\ba
{\mathcal L}_{\rm spinor}^{(1)}&\sim& \frac{e^4\left(\frac{a^4}{360} + \frac{a^2\,b^2}{72} + \frac{b^4}{360}\right)}{m^4 \pi^2} + 
  \frac{e^6 \left(\frac{a^6}{630} + \frac{a^4\,b^2}{180} - \frac{a^2\,b^4}{180} - \frac{b^6}{630}\right)}{m^8 \pi^2}+\dots \nn\\[2mm]
 &  \hskip -1cm = &~\hskip -.5cm\frac{e^4\left[(a^2-b^2)^2 + 7(a\,b)^2\right]}{360 m^4\pi^2 }
  +\frac{e^6 (a^2-b^2)\left[2(a^2-b^2)^2+13 (a b)^2\right]}{1260  m^8 \pi^2}
 +\dots\, .\nn\\
  \label{1lspweakeg}
\ea
The first term in (\ref{1lspweakeg}) corresponds to the light-light scattering result (\ref{lightlight}), while the second gives the low energy limit of the diagram in Figure \ref{digexp} with six external photon legs, and agrees with Eqn. (43) in \cite{he}.

Several special cases of the general expansion (\ref{1lspweak}) are of interest. 
\begin{enumerate}

\item\underline{Purely magnetic background} : If the background is purely magnetic, of strength $B>0$, then the integral representation (\ref{hesp}) reduces to
\begin{equation}
{\mathcal L}_{\rm spinor}^{(1)}=-\frac{e^2 B^2}{8\pi ^{2}} \int_{0}^{\infty} \frac{ds}{s^{2}}\; 
\left(\coth s-\frac{1}{s}-\frac{s}{3}\right)\,e^{-m^2s/(eB)}
\label{1lspmag}
\end{equation}
which has the asymptotic expansion
\begin{eqnarray}
{\mathcal L}_{\rm spinor}^{(1)} \sim -\frac{m^4}{8\pi^2} \sum_{n=0}^\infty 
\frac{{\mathcal B}_{2n+4}} {(2n+4)(2n+3)(2n+2)}\left(\frac{2eB}{m^2}\right)^{2n+4}\hspace{-3mm}.
\label{1lspmagexp}
\end{eqnarray}
This example illustrates clearly that the Heisenberg-Euler perturbative weak field expansion (\ref{1lspweak}) is a divergent asymptotic series, since the Bernoulli numbers ${\mathcal B}_{2n}$ alternate in sign and diverge factorially fast in magnitude \cite{bateman,abramowitz}:
\begin{eqnarray}
{\mathcal B}_{2n}=(-1)^{n+1}\,2 \,\frac{(2n)!}{(2\pi)^{2n}}\, \zeta(2n)\ .
\label{bernoulli}
\end{eqnarray}
Here $\zeta(n)$ denotes the Riemann zeta function \cite{abramowitz}
\begin{eqnarray}
\zeta(n)=\sum_{k=1}^\infty \frac{1}{k^n}
\label{rzeta}
\end{eqnarray}
which is exponentially close to 1 for large integers  $n$. In this magnetic field case, the perturbative series is divergent but alternating, and is Borel summable \cite{chadha}. In fact, the Borel sum of (\ref{1lspmagexp}) is simply the original Heisenberg-Euler integral representation (\ref{1lspmag}), as can be checked using the relation \cite{bateman,abramowitz}
\begin{eqnarray}
{\rm coth} s-\frac{1}{s}-\frac{s}{3}=\sum_{k=1}^\infty\, \frac{-2s^3}{k^2\pi^2(s^2+k^2\pi^2)}\ .
\label{cothsum}
\end{eqnarray}

\vskip.5cm

\item \underline{Purely electric background} : If the background is purely electric, of strength $E>0$, the integral representation (\ref{hesp}) reduces formally to
\begin{equation}
{\mathcal L}_{\rm spinor}^{(1)}=-\frac{e^2 E^2}{8\pi ^{2}} \int_{0}^{\infty} \frac{ds}{s^{2}}\; 
\left(\cot s-\frac{1}{s}+\frac{s}{3}\right)\,e^{-m^2s/(eE)}\ .
\label{1lspelec}
\end{equation}
This integral representation has the asymptotic expansion
\begin{eqnarray}
{\mathcal L}_{\rm spinor}^{(1)} \sim -\frac{m^4}{8\pi^2} \sum_{n=0}^\infty 
\frac{(-1)^n {\mathcal B}_{2n+4}} {(2n+4)(2n+3)(2n+2)}\left(\frac{2eE}{m^2}\right)^{2n+4}\hspace{-3mm}.
\label{1lspelecexp}
\end{eqnarray}
If only one of $E$ or $B$ is nonzero, the only Lorentz invariant is $\vec{E}^2-\vec{B}^2$; therefore, changing from the perturbative series (\ref{1lspmagexp}) to the perturbative series (\ref{1lspelecexp}) simply involves the replacement $B^2\to - E^2$.  Therefore, in the electric case the perturbative series (\ref{1lspelecexp}) is divergent and nonalternating, and thus is not Borel summable \cite{chadha}. This has a dramatic physical consequence. It means that the effective Lagrangian has an exponentially small nonperturbative contribution which is imaginary, even though it is real to all orders in perturbation theory. This is a classic signal of instability, and corresponds precisely to the vacuum instability of vacuum pair production first noted by Heisenberg and Euler. This imaginary part can also be seen to arise from the poles on the real $s$ axis encountered in the integral representation (\ref{1lspelec}), which shows that this integral representation must be defined more carefully, with an approriate $i\epsilon$ prescription \cite{jentschura,schwinger}. From this integral representation, or from a Borel dispersion relation, it is straightforward to derive the full nonperturbative imaginary part of the Heisenberg-Euler effective Lagrangian to be:
\begin{eqnarray}
Im{\mathcal L}_{\rm spinor}^{(1)}\sim \frac{e^2 E^2}{8\pi^3}\sum_{n=1}^\infty \frac{1}{n^2}\, 
\exp\left[-\frac{m^2\pi n}{eE}\right]\,.
\label{electricinstanton}
\end{eqnarray}
In modern language this result, first derived by Schwinger \cite{schwinger}, gives the higher order instanton contributions to the leading imaginary part (\ref{imag}) found by Heisenberg and Euler.
\vskip .5cm

\item \underline{Perpendicular magnetic and  electric background fields} : If $\vec{E}\cdot\vec{B}=0$, then  ${\mathcal G}=0$, and from (\ref{abfg}) we see that either $a=0$ or $b=0$, depending on the sign of ${\mathcal F}$. If $E^2>B^2$, then the situation is as for the purely electric field, with a vacuum instability. On the other hand, if $B^2>E^2$, then the system behaves like the case of a purely magnetic field. If $E^2=B^2$, then ${\mathcal L}_{\rm spinor}^{(1)}$ vanishes and there are no quantum corrections at all.
\vskip .5cm

\item \underline{Parallel magnetic and  electric background fields} : As mentioned previously, if $\vec{E}\cdot\vec{B}\neq 0$, then it is possible to make a Lorentz transformation to a frame in which $\vec{B}$ and $\vec{E}$ are parallel (or antiparallel). In this case we can take $a=E$, and $b=B$ [recall that we are using Heisenberg's original notation for $a$ and $b$, as in (\ref{ab})]. Then we can rewrite the effective Lagrangian (\ref{hesp}) as
\ba
{\mathcal L}_{\rm spinor}^{(1)}& =& \frac{e^2E B}{4\pi^3}\sum_{n=1}^\infty \int_0^\infty \frac{dt}{t}\, \frac{e^{-m^2 n\pi t}}{n}\left\{ \frac{(e B t)^2}{1+(e B t)^2} \coth\left(\frac{E}{B}\,n\pi\right) \right.\nn\\
&&\left. \hskip 3cm +\frac{(e E t)^2}{1-(e E t)^2} \coth\left(\frac{B}{E}\,n\pi\right)\right\}
\label{heram}
\ea
using a remarkable trigonometric identity due to Ramanujan and Sitaramachandrarao \cite{berndt,valluri}
\ba
(\pi x\cot \pi x)(\pi y \coth \pi y)&=&1+\frac{\pi^2}{3}(y^2-x^2)-2\pi x y^3\sum_{n=1}^\infty \frac{\coth\left(n\pi \frac{x}{y}\right)}{n(n^2+y^2)}\nn\\
&&\hskip 1.5cm - 2\pi x^3 y\sum_{n=1}^\infty \frac{\coth\left(n\pi \frac{y}{x}\right)}{n(n^2-x^2)}\ .
\label{ramcoth}
\ea
Thus, we see that in this case of parallel electric and magnetic fields, the imaginary part of the effective Lagrangian is
\ba
Im{\mathcal L}_{\rm spinor}^{(1)}\sim  \frac{e^2 E B}{8\pi^2}\sum_{n=1}^\infty \frac{1}{n}\,\coth\left(\frac{B}{E}\, n \pi\right) \exp\left[-\frac{m^2\pi n}{eE}\right]\,,
\label{ebimag}
\ea
a result stated by Bunkin and Tugov \cite{bunkin}, and obtained in a very different manner by Nikishov \cite{nikishov} using a virial representation of the imaginary part of ${\mathcal L}$. Note that as $B\to 0$ we recover the Schwinger instanton result (\ref{electricinstanton}), and that for $B>0$ we get a slight, but not strong, enhancement to the prefactor, with the exponent being unchanged.  
\end{enumerate}

\subsubsection{Scalar QED case}
\label{weakscalar}

The analysis of the previous section for the weak field expansions of the spinor QED effective Lagrangian (\ref{hesp}) can also be applied to Weisskopf's effective Lagrangian (\ref{hesc}) for scalar QED. The weak field expansion of  (\ref{hesc}), expressed in terms of the Lorentz invariants $a$ and $b$ defined in (\ref{ab}), is:
\ba
{\mathcal L}_{\rm scalar}^{(1)} \sim \frac{m^4}{16\pi^2}\sum_{n=2}^\infty (2n-3)! \sum_{k=0}^n \frac{\bar{{\mathcal B}}_{2k} \bar{{\mathcal B}}_{2n-2k}}{(2k)!(2n-2k)!} \left(\frac{2 e b}{m^2}\right)^{2n-2k}\left(\frac{2 i e a}{m^2}\right)^{2k}.\nn\\
\label{1lscweak}
\ea
Where we have defined the convenient shorthand
\ba
\bar{{\mathcal B}}_{2k}\equiv \frac{(1- 2^{2k-1})}{2^{2k-1}}{\mathcal B}_{2k}
\ ,
\label{bbar}
\ea
in terms of which the relevant expansions are [compare with (\ref{coth})]:
\ba
\frac{z}{\sinh z}=\sum_{k=0}^\infty \frac{ 2^{2k}\bar{{\mathcal B}}_{2k}}{(2k)!}\, z^{2k}\ , \quad \frac{z}{\sin z} =\sum_{k=0}^\infty \frac{(-1)^k 2^{2k}\bar{{\mathcal B}}_{2k}}{(2k)!}\, z^{2k}\ .
\label{1/sinh}
\ea
As in the spinor case, at any given order of (\ref{1lscweak}), the combinations of $a$ and $b$ can be rewritten in terms of the more familiar Lorentz invariants $(\vec{E}^2-\vec{B}^2)$ and  $\vec{E}\cdot\vec{B}$.
For example, the first two orders yield:
\ba
{\mathcal L}_{\rm scalar}^{(1)}&\sim& \frac{e^4\left(\frac{7 b^4}{360} - \frac{a^2\,b^2}{36} + \frac{7 a^4}{360}\right)}{16\,m^4 \pi^2} + 
  \frac{e^6\left(\frac{-31 b^6}{2520} + \frac{7 b^4\,a^2}{360} - \frac{7 b^2\,a^4}{360} + \frac{31 a^6}{2520}\right)}{16\,m^8 \pi^2}+\dots \nn\\[2mm]
  &\hskip -1.3cm =&~\hskip -0.8cm \frac{e^4\left[7(a^2-b^2)^2 + 4(a\,b)^2\right]}{5760 m^4\pi^2 }
  +\frac{e^6(a^2-b^2)\left[31(a^2-b^2)^2+77 (a b)^2\right]}{40320  m^8 \pi^2}
 +\dots\,.\nn\\
  \label{1lscweakeg}
\ea
The first term in (\ref{1lscweakeg}) agrees with the first term listed in \cite{viki1}.
\vskip .5cm

Several special cases of the general expansion (\ref{1lscweak}) are of interest. 
\begin{enumerate}

\item \underline{Purely magnetic background} : If the background is purely magnetic, of strength $B>0$, then the integral representation (\ref{hesc}) reduces to
\begin{equation}
{\mathcal L}_{\rm scalar}^{(1)}=\frac{e^2 B^2}{16\pi ^{2}} \int_{0}^{\infty} \frac{ds}{s^{2}}\; 
\left(\frac{1}{\sinh s}-\frac{1}{s}+\frac{s}{6}\right)\,e^{-m^2s/(eB)}
\label{1lscmag}
\end{equation}
which has the asymptotic expansion
\begin{eqnarray}
{\mathcal L}_{\rm scalar}^{(1)} \sim \frac{m^4}{16\pi^2} \sum_{n=0}^\infty 
\frac{\bar{{\mathcal B}}_{2n+4}} {(2n+4)(2n+3)(2n+2)}\left(\frac{2eB}{m^2}\right)^{2n+4}\hskip -0.3cm .
\label{1lscmagexp}
\end{eqnarray}
Once again, this is a divergent but Borel summable perturbative series, and the Borel sum gives precisely the integral representation (\ref{1lscmag}).
\vskip.5 cm

\item \underline{Purely electric background} : If the background is purely electric, of strength $E>0$, then the integral representation (\ref{hesc}) reduces formally to
\begin{equation}
{\mathcal L}_{\rm scalar}^{(1)}=\frac{e^2 E^2}{16\pi ^{2}} \int_{0}^{\infty} \frac{ds}{s^{2}}\; 
\left(\frac{1}{\sin s}-\frac{1}{s}-\frac{s}{6}\right)\,e^{-m^2s/(eE)}
\hskip 0.3cm .
\label{1lscelec}
\end{equation}
This integral representation has the asymptotic expansion
\begin{eqnarray}
{\mathcal L}_{\rm scalar}^{(1)} \sim \frac{m^4}{16\pi^2} \sum_{n=0}^\infty 
\frac{(-1)^n \bar{{\mathcal B}}_{2n+4}} {(2n+4)(2n+3)(2n+2)}\left(\frac{2eE}{m^2}\right)^{2n+4}\hskip -0.3cm .
\label{1lscelecexp}
\end{eqnarray}
As in the spinor case, this is a non Borel summable perturbative series, and correspondingly the effective Lagrangian has a nonperturbative imaginary part given by
\begin{eqnarray}
Im{\mathcal L}_{\rm scalar}^{(1)}\sim  \frac{e^2 E^2}{16\pi^3}\sum_{n=1}^\infty \frac{(-1)^{n-1}}{n^2}\, \exp\left[-\frac{m^2\pi n}{eE}\right]\,.
\label{scelectricinstanton}
\end{eqnarray}
\vskip.5cm 

\item \underline{Perpendicular magnetic and  electric background fields} : If $\vec{E}\cdot\vec{B}=0$, then  ${\mathcal G}=0$, and from (\ref{abfg}) we see that either $a=0$ or $b=0$, depending on the sign of ${\mathcal F}$. If $E^2>B^2$, then the situation is as for the purely electric field, with a vacuum instability. On the other hand, if $B^2>E^2$, then the system behaves like the case of a purely magnetic field. If $E^2=B^2$, then ${\mathcal L}_{\rm scalar}^{(1)}$ vanishes and there are no quantum corrections at all.
\vskip .5cm

\item \underline{Parallel magnetic and  electric background fields} : If $\vec{E}$ and $\vec{B}$ are parallel, we can rewrite the effective Lagrangian (\ref{hesc}) as
\ba
{\mathcal L}_{\rm scalar}^{(1)}& =& -\frac{e^2 E B}{8\pi^3}\sum_{n=1}^\infty \int_0^\infty \frac{dt}{t}\, \frac{e^{-m^2 n\pi t}}{n}\left\{ \frac{(e B t)^2}{1+(e B t)^2} \frac{(-1)^n}{\sinh\left(\frac{E}{B}\,n\pi\right)} \right.\nn\\
&&\left. \hskip 3cm +\frac{(e E t)^2}{1-(e E t)^2} \frac{(-1)^n}{\sinh\left(\frac{B}{E}\,n\pi\right)}\right\}.
\label{scheram}
\ea
using another remarkable trigonometric identity due to Ramanujan and Sitaramachandrarao \cite{berndt}
\ba
&&\left(\frac{\pi x}{\sin \pi x}\right)\left(\frac{\pi y}{\sinh \pi y}\right) =1 +\frac{\pi^2}{6}(x^2-y^2)\nn\\[1mm]
&&\hskip-10pt -2\pi x^3 y\sum_{n=1}^\infty \frac{(-1)^n}{n(n^2-x^2)}\,\frac{1}{\sinh\left(n\pi \frac{y}{x}\right)}-2\pi x y^3\sum_{n=1}^\infty \frac{(-1)^n}{n(n^2+y^2)}\,\frac{1}{\sinh\left(n\pi \frac{x}{y}\right)}\,.\nn\\
\label{ramcsch}
\ea
Thus, we see that in this case of parallel electric and magnetic fields, the imaginary part of the effective Lagrangian is
\ba
Im{\mathcal L}_{\rm scalar}^{(1)}\sim \frac{e^2 E B}{16\pi^2}\sum_{n=1}^\infty \frac{(-1)^{n-1}}{n}\,\frac{1}{\sinh\left(\frac{B}{E}\,n\pi\right)} \exp\left[-\frac{m^2\pi n}{eE}\right]\,,\nn\\
\label{scebimag}
\ea
a result first obtained in a different manner by Popov \cite{popov}. Once again, note that as $B\to 0$ we recover the pure electric field result (\ref{scelectricinstanton}), and that for $B>0$ we get a slight, but not strong, enhancement to the prefactor, with the exponent being unchanged.  

\end{enumerate}

\subsection{Strong-field expansions of Heisenberg-Euler}
\label{strongfield}

In order to discuss the strong-field expansions of the Heisenberg-Euler effective Lagrangians (\ref{hesp}) and (\ref{hesc}), it is useful to convert the proper-time integral representations into zeta function expressions and alternative integral representations which are more suited to the strong field limit. We concentrate here on the case of a magnetic background field.
In zeta function regularization \cite{dr-qed,blau,elizaldebook,klaus} we define the determinant of an operator ${\mathcal O}$ in terms of its $\zeta$-function
\ba
\zeta(s)=\sum_{\lambda} \lambda^{-s}
\label{genzeta}
\ea
where the sum is over the eigenvalues $\lambda$ of ${\mathcal O}$. Then the determinant of ${\mathcal O}$ can be defined as
\ba
\det {\mathcal O} =\exp\left[-\zeta^\prime(0)\right]\,.
\label{zetadet}
\ea
For the Heisenberg-Euler effective Lagrangians in a magnetic background, these $\zeta$-functions can be expressed in terms of the standard Hurwitz $\zeta$-function of number theory \cite{bateman,abramowitz,ww}:
\ba
\zeta_H(s,z)=\sum_{n=0}^\infty (n+z)^{-s} ,\quad Re(s)>1\ , \quad v\neq 0, -1, -2, \dots\,.
\label{hurwitz}
\ea
This Hurwitz $\zeta$-function can be analytically continued in the $s$ plane to define an analytic function with a single simple pole at $s=1$. 

\subsubsection{Spinor QED case}
\label{strongspinor}

For spinor QED in a magnetic background, the eigenvalues of the Dirac operator are 
\ba
\lambda_n^\pm =m^2+k_\perp^2+e B(2n+1\pm 1)\qquad , \quad n=0, 1, \dots 
\label{spevs}
\ea
where the $\pm$ refers to the two spin components, and $\vec{k}_\perp$ is the transverse  
2-momentum. Then the $\zeta$-function for this Dirac operator is \cite{dr-qed}
\ba
 \zeta_{\rm spinor}(s)&=& \frac{eB}{2\pi}\sum_{n=0}^\infty \sum_\pm \int\frac{d^2 k_\perp}{(2\pi)^2}\left(\frac{m^2+k_\perp^2+e B(2n+1\pm 1)}{\mu^2}\right)^{-s}\nn\\
\hskip-.5cm&=& \frac{m^4}{4\pi^2}\left(\frac{eB}{m^2}\right)^2 \frac{\left(\frac{\mu^2}{2 eB}\right)^{s}}{(s-1)}\left[ 2\zeta_H\left(s-1,\frac{m^2}{2eB}\right)-\left(\frac{m^2}{2eB}\right)^{1-s}\right].\nn\\
\label{spzeta}
\ea
Here the overall factor of $\frac{eB}{2\pi}$ is the Landau degeneracy factor, and the renormalization scale $\mu$ has been introduced to make the eigenvalues dimensionless. 
Given this form in terms of the Hurwitz zeta function, it is straightforward to obtain
\ba
\zeta_{\rm spinor}^\prime(0)&=&\frac{(e B)^2}{4\pi^2}\left\{ -2 \zeta^\prime_H\left(-1, \frac{m^2}{2eB}\right) -\frac{m^2}{2eB}\ln\left(\frac{m^2}{2eB}\right) \right.\nn\\
&&\left.+\left[\frac{1}{6}+\left(\frac{m^2}{2eB}\right)^2\right]
\left[1+\ln\left(\frac{\mu^2}{2eB}\right)\right]\right\}.
\ea
With on-shell renormalization ($\mu=m$), and subtracting the zero field contribution,  $-\frac{3 m^4}{32\pi^2}$, which ensures that ${\mathcal L}_{\rm spinor}^{(1)}$ vanishes when $B=0$, we obtain
\ba
{\mathcal L}_{\rm spinor}^{(1)}&=&\frac{(e B)^2}{2\pi^2}\left\{  \zeta^\prime_H\left(-1, \frac{m^2}{2eB}\right) + \zeta_H\left(-1, \frac{m^2}{2eB}\right)\ln\left(\frac{m^2}{2eB}\right) \right.\nn\\
&&\left.\hskip 2cm -\frac{1}{12}+\frac{1}{4}\left( \frac{m^2}{2eB}\right)^2\right\}.
\label{spzetalag}
\ea
Here we have used the fact \cite{bateman,ww} that $\zeta_H(-1,v)=-\frac{1}{12}+\frac{v}{2}-\frac{v^2}{2}$. 

We can exhibit directly the equivalence of this $\zeta$-function representation (\ref{spzetalag}) with the proper time integral representation (\ref{1lspmag}) by noting the following integral representations of the Hurwitz $\zeta$-function \cite{bateman}:
\ba
\zeta_H(s,z)&=& \frac{1}{\Gamma(s)}\int_0^\infty \frac{e^{-z\, t}\, t^{s-1}}{1-e^{-t}}\, dt \quad , \quad Re(s)>1\quad,\quad Re(z)>0\nn\\
&=&\frac{z^{1-s}}{s-1}+\frac{z^{-s}}{2}+\frac{s z^{-1-s}}{12} +\frac{2^{s-1}}{\Gamma(s)}\int_0^\infty \frac{dt}{t^{1-s}} e^{-2 z\, t}\left( \coth t-\frac{1}{t}-\frac{t}{3}\right)\nn\\
\label{hurwitzint}
\ea
where the second expression is valid for $Re(s)>-2$, as a result of the subtractions of the integrand \cite{adamchik}. Thus, we can evaluate the derivative at $s=-1$, as required:
\ba
\zeta_H^\prime(-1,z)= \frac{1}{12}-\frac{z^2}{4}-\zeta_H(-1,z)\ln z-\frac{1}{4} \int_0^\infty \frac{dt}{t^2}\, e^{-2 z\, t}\left( \coth t-\frac{1}{t}-\frac{t}{3}\right)\,.\nn\\
\label{hurwitzderint}
\ea
Comparing with (\ref{spzetalag}), we recover exactly the familiar proper-time integral representation (\ref{1lspmag}) for the one loop Heisenberg-Euler effective Lagrangian ${\mathcal L}_{\rm spinor}^{(1)}$.

Now to make a strong-field expansion of ${\mathcal L}_{\rm spinor}^{(1)}$ we use the following relation between the Hurwitz $\zeta$-function and the log of the $\Gamma$ function \cite{bateman,ww}:
\ba
\zeta_H^\prime(-1,z)= \zeta^\prime(-1)-\frac{z}{2}\ln (2\pi) -\frac{z}{2}(1-z)+\int_0^z \ln \Gamma(x) dx\ .
\label{zetalog}
\ea
This identity follows from an integration of Binet's integral representation \cite{ww} of $\ln\Gamma(z)$. Thus we can write
\ba
{\mathcal L}_{\rm spinor}^{(1)}&=&\frac{(eB)^2}{2\pi^2}\left\{ -\frac{1}{12}+\zeta^\prime(-1)-\frac{m^2}{4eB}+\frac{3}{4}\left(\frac{m^2}{2eB}\right)^2 -\frac{m^2}{4 eB}\, \ln (2\pi)\right.\nn\\
&&\left. +\left[-\frac{1}{12}+\frac{m^2}{4eB}-\frac{1}{2}\left(\frac{m^2}{2eB}\right)^2\right] \ln \left(\frac{m^2}{2eB}\right) +\int_0^{\frac{m^2}{2eB}} \ln\Gamma(x) dx\right\}.\nn\\
\ea
In the strong-field limit, the range of integration in the last term vanishes, so we can use the Taylor expansion \cite{bateman,abramowitz} of $\ln\Gamma(x)$:
\ba
\ln\Gamma(x)=-\ln x-\gamma x +\sum_{n=2}^\infty \frac{(-1)^n}{n}\zeta(n)\, x^n
\label{loggamma}
\ea
where $\zeta(n)$ is the usual Riemann $\zeta$-function. This leads to :
\ba
{\mathcal L}_{\rm spinor}^{(1)}&=&\frac{(eB)^2}{2\pi^2}\left\{ -\frac{1}{12}+\zeta^\prime(-1)-\frac{m^2}{4eB}+\frac{3}{4}\left(\frac{m^2}{2eB}\right)^2 -\frac{m^2}{4 eB}\, \ln (2\pi)\right.\nn\\
&&\left. +\left[-\frac{1}{12}+\frac{m^2}{4eB}-\frac{1}{2}\left(\frac{m^2}{2eB}\right)^2\right] \ln \left(\frac{m^2}{2eB}\right) -\frac{\gamma}{2}\left(\frac{m^2}{2eB}\right)^2 \right.\nn\\
&&\left.+\frac{m^2}{2eB}\left(1-\ln\left(\frac{m^2}{2eB}\right)\right) +\sum_{n=2}^\infty \frac{(-1)^n \zeta(n)}{n(n+1)} \left(\frac{m^2}{2eB}\right)^{n+1}\right\}.\nn\\
\label{spmagstrong}
\ea
The leading behavior in the strong field limit is
\ba
{\mathcal L}_{\rm spinor}^{(1)}\sim\frac{(eB)^2}{24\pi^2} \ln \left(\frac{2eB}{m^2}\right)+\dots
\label{1lspmagleading}
\ea
which agrees with Weisskopf's original observation  (\ref{strong}).

\subsubsection{Scalar QED case}
\label{strongscalar}

For scalar QED the analysis is very similar. There is no spin projection term in the eigenvalues of the Klein-Gordon operator, so the $\zeta$-function is
\ba
\zeta_{\rm scalar}(s)&=& \frac{eB}{2\pi}\sum_{n=0}^\infty  \int\frac{d^2 k_\perp}{(2\pi)^2}\left(\frac{m^2+k_\perp^2+e B(2n+1)}{\mu^2}\right)^{-s}\nn\\
&=& \frac{(e B)^2}{4\pi^2} \left(\frac{\mu^2}{2 eB}\right)^{s}\frac{1}{(s-1)}\, \zeta_H\left(s-1,\frac{1}{2}+\frac{m^2}{2eB}\right)\,.
\label{sczeta}
\ea
With on-shell renormalization ($\mu=m$), and subtracting the zero field contribution $\frac{3 m^4}{64\pi^2}$, we obtain
\ba
{\mathcal L}_{\rm scalar}^{(1)}&=&-\frac{(e B)^2}{4\pi^2}\left\{  \zeta^\prime_H\left(-1, \frac{1}{2}+\frac{m^2}{2eB}\right)+\frac{3}{4}\left(\frac{m^2}{2 eB}\right)^2\right.\nn\\
&&\left.  + \left[1+\ln\left(\frac{m^2}{2eB}\right) \right] \zeta_H\left(-1, \frac{1}{2}+\frac{m^2}{2eB}\right)\right\}.
\label{sczetalag}
\ea
We can exhibit directly the equivalence of this $\zeta$-function representation to the proper time integral representation (\ref{1lscmag}) by noting the following integral representations of the Hurwitz $\zeta$-function:
\ba
\zeta_H(s,\frac{1}{2}+z)&=& \frac{1}{\Gamma(s)}\int_0^\infty e^{-z\, t}\, t^{s-1}\, \frac{e^{-t/2}}{1-e^{-t}}\, dt \quad , \quad Re(s)>1, Re(z)>-\frac{1}{2}\nn\\
&=&\frac{z^{1-s}}{s-1}-\frac{s z^{-1-s}}{24} +\frac{2^{s-1}}{\Gamma(s)}\int_0^\infty \frac{dt}{t^{1-s}}\, e^{-2 z\, t}\left( \frac{1}{\sinh t}-\frac{1}{t}+\frac{t}{6}\right)\nn\\
\label{schurwitzint}
\ea
where the second expression is valid for $Re(s)>-2$ as a result of the subtractions of the integrand. Thus, we can evaluate the derivative at $s=-1$, as required:
\ba
\zeta_H^\prime(-1,\frac{1}{2}+z)&=& -\frac{1}{24}\left[1+\ln z\right] -\frac{z^2}{4}+\frac{z^2}{2}\ln z \nn\\
&& 
-\frac{1}{4}\int_0^\infty \frac{dt}{t^{2}}\, e^{-2 z\, t} \left( \frac{1}{\sinh t}-\frac{1}{t}+\frac{t}{6}\right)\,.
\label{schurwitzderint}
\ea
Comparing with (\ref{sczetalag}), we recover exactly the familiar proper-time integral representation (\ref{1lscmag}) for the one loop Heisenberg-Euler effective Lagrangian ${\mathcal L}_{\rm scalar}^{(1)}$.

Now to make a strong-field expansion of ${\mathcal L}_{\rm scalar}^{(1)}$ we use the following relation between the Hurwitz $\zeta$-function and the log of the $\Gamma$ function \cite{bateman,ww}:
\ba
\zeta_H^\prime(-1,\frac{1}{2}+z)=-\frac{1}{2}\, \zeta^\prime(-1)-\frac{z}{2}\,\ln (2\pi) -\frac{\ln 2}{24}+\frac{z^2}{2}+\int_0^z \ln \Gamma\Big(x+\frac{1}{2}\,\Big) \,dx\nn\\
\label{sczetalog}
\ea
where we have used the fact that \cite{barnes,adamchik}
\ba
\int_0^{\frac{1}{2}}\ln \Gamma(x) \, dx=\frac{5}{24}\,\ln 2+\frac{1}{4}\,\ln \pi +\frac{1}{8}-\frac{3}{2}\,\zeta^\prime(-1)\ .
\label{hint}
\ea
The strong-field expansion is obtained using the Taylor expansion:
\ba
\ln\Gamma\Big(x+\frac{1}{2}\Big)=\frac{1}{2}\,\ln \pi -(\gamma+2\ln 2)\, x+\sum_{n=2}^\infty \frac{(-1)^{n-1}(1-2^n)}{n}\,\zeta(n)\, x^n\ . \nn\\
\label{scloggamma}
\ea
Thus, we obtain the strong-field expansion
\ba
{\mathcal L}_{\rm scalar}^{(1)}&=&-\frac{(eB)^2}{4\pi^2}\left\{\frac{5}{4}\left(\frac{m^2}{2eB}\right)^2 
+\left[\frac{1}{24}-\frac{1}{2}\left(\frac{m^2}{2eB}\right)^2\right]\left[1+ \ln \left(\frac{m^2}{2eB}\right)\right]\right.\nn\\[1mm]
&&\left. -\frac{1}{2}\zeta^\prime(-1) -\frac{\ln 2}{24} -\frac{m^2}{4 eB}\, \ln (2)  -\frac{1}{2} \left(\frac{m^2}{2eB}\right)^2\left(\gamma+2\ln 2\right)\right.\nn\\[1mm]
&& \left.+\sum_{n=2}^\infty \frac{(-1)^{n-1} (1-2^n) \zeta(n)}{n(n+1)} \left(\frac{m^2}{2eB}\right)^{n+1}\right\}.\nn\\
\label{scmagstrong}
\ea
The leading behavior in the strong field limit is
\ba
{\mathcal L}_{\rm scalar}^{(1)}\sim\frac{(eB)^2}{96\pi^2} \ln \left(\frac{2eB}{m^2}\right)+\dots\,.
\label{1lscmagleading}
\ea

\section{Inhomogeneous backgrounds : beyond the constant field strength limit}
\label{inhomog}
\renewcommand{\theequation}{2.\arabic{equation}}
\setcounter{equation}{0}

The Heisenberg-Euler and Weisskopf results, (\ref{hesp}) and (\ref{hesc}) respectively, are computed for the special case where the background field strength $F_{\mu\nu}$ is constant. Clearly, it is of interest to be able to compute the effective Lagrangian in more general cases, where $F_{\mu\nu}$ is inhomogeneous. The general formalism  for computing such determinants was developed by Schwinger \cite{schwinger,schwinger2}, Salam and Matthews \cite{salam}, and Salam and Strathdee \cite{salamstrathdee}, among others. It is, unfortunately, cumbersome to do such calculations, and so several physically relevant approximate techniques have been developed. I briefly review these in this section.

\subsection{Solvable inhomogeneous backgrounds}
\label{solvable}

There exist a limited number of solvable cases in which one can compute the effective Lagrangian in a more-or-less closed-form, analogous to the Heisenberg-Euler result (\ref{hesp}). These solvable cases reveal some interesting new physics, beyond the constant field approximation, as well as providing useful checks on the more general approximate techniques. In order for a particular background to lead to such an explicit form for the effective Lagrangian, we need to be able to solve the spectral problem for the corresponding Dirac or Klein-Gordon operator appearing in (\ref{action}) or (\ref{sqedaction}), respectively. Solvable Dirac and Klein-Gordon equations have been studied exhaustively, so one can simply look through the list of solvable cases and decide if the corresponding electromagnetic background is physically relevant.

\subsubsection{Plane-wave background}
\label{plane}

The simplest case of an inhomogeneous background is that of a single monochromatic plane wave field. This was shown to lead to a solvable Dirac equation by Volkov in 1935 \cite{volkov}, and is described in \cite{landauqed}.
Such a gauge field can be written $A_\mu=A_\mu (k\cdot x)$, where $k_\mu$ is light-like  ($k\cdot k=0$), and $k\cdot A=0$. Then the solutions to the Dirac equation can be written explicitly as
\begin{eqnarray}
\psi_p =e^{i\Omega}\,\left[1+\frac{e(\gamma\cdot k)(\gamma\cdot A)}{2 
(k\cdot p)}\right]\frac{u(p)}{\sqrt{2p_0}}
\label{volkovsol}
\end{eqnarray}
where $p$ is the momentum, $u(p)$ is a fundamental free spinor, and the phase factor $\Omega$ is
\begin{eqnarray}
\Omega=-p\cdot x-\int_0^{k\cdot x}\left(\frac{e(p\cdot A)}{(k\cdot p)}-
\frac{e^2 A^2}{2(k\cdot p)}\right) d(k\cdot x)\ .
\label{volkovphase}
\end{eqnarray}
Having the exact solutions means one can compute the exact Green's function, which in turn means one can compute the one-loop effective action. Schwinger \cite{schwinger} showed that the effective action in fact vanishes for such a plane wave background. (This is consistent with the fact that for a plane wave both Lorentz invariants ${\mathcal F}$ and ${\mathcal G}$ vanish, so that the low frequency limit Heisenberg-Euler result vanishes.) In Schwinger's words \cite{schwinger}: ``there are no nonlinear vacuum phenomena for a single plane wave, of arbitrary strength and spectral composition''. In particular, Schwinger's result means that there is no pair production in a single monochromatic plane wave background, which can be understood from simple energy and momentum conservation -- at least two plane waves are necessary to produce a pair \cite{schwinger,brownkibble}. The problem of pair production in crossed laser beams has been addressed recently in \cite{friedavan}.

\subsubsection{Time dependent electric field $E(t)=E\, {\rm sech}^2(\omega t)$}
\label{timedepe}

With such an electric field, pointing in a fixed direction, spatially uniform, but with a special prescribed ${\rm sech}^2(\omega t)$ time dependence, the Dirac and Klein-Gordon equations reduce to hypergeometric equations. 
Narozhnyi and Nikishov \cite{nn} (see also \cite{popov}) used this fact to express the effective action as
\begin{eqnarray}
Im\,  S= \pm \frac{(2s+1)}{2(2\pi)^3}\int 
d^2p_\perp \int dp_z \log\left(1\pm w_{p}\right)
\label{nnaction}
\end{eqnarray}
where the upper/lower signs $\pm$ refers to spinor/scalar QED, respectively, and $(2s+1)$ is an overall spin factor of $2$ for spinor QED and $1$ for scalar QED. The momentum has been divided into $p_z$ along the direction of the electric field, with $\vec{p}_\perp$ being the transverse momentum.  
The $w_p$ can be computed as a scattering amplitude factor for the corresponding Dirac or Klein-Gordon equation. For spinor QED one finds \cite{nn}:
\begin{eqnarray}
w_p^{\rm spinor}=\frac{{\rm sinh} \,\pi(\lambda-\mu+\nu)\,{\rm sinh}\,
\pi(\lambda+\mu-\nu)} {{\rm sinh}\, \pi(-\lambda+\mu+\nu)\,{\rm sinh}\,
\pi(\lambda+\mu+\nu)}
\label{nnspw}
\end{eqnarray}
where $\lambda\equiv -\frac{eE}{\omega^2}$, and 
\begin{eqnarray}
\mu&\equiv&
\frac{1}{2\omega}\sqrt{m^2+p_\perp^2+(p_z+eE/\omega)^2}\ ,\nonumber\\[1mm]
\nu&\equiv&
\frac{1}{2\omega}\sqrt{m^2+p_\perp^2+(p_z-eE/\omega)^2}\ .
\label{nnsppars}
\end{eqnarray}
For  scalar QED one finds  \cite{nn}
\begin{eqnarray}
w_p^{\rm scalar}=\frac{{\rm cosh}\, \pi(\lambda^\prime-\mu+\nu)\,{\rm cosh}\,
\pi(\lambda^\prime+\mu-\nu)} {{\rm cosh} \,\pi(\lambda^\prime -\mu-\nu)\,{\rm cosh}\,
\pi(\lambda^\prime+\mu+\nu)}
\label{nnscw}
\end{eqnarray}
where $\lambda^\prime=\sqrt{\lambda^2-1/4}$, and $\mu$ and $\nu$ are as defined in (\ref{nnsppars}).

When $\omega\to 0$, the time scale of the electric field tends to infinity and we recover the constant field results (\ref{electricinstanton}) and (\ref{scelectricinstanton}). To see this, note that 
\begin{eqnarray}
w_{p}\to    \frac{\eta_p}{1\mp \eta_p}\ , \quad {\rm where}\quad \eta_p \equiv  \exp\left[-\frac{\pi}{eE}(m^2+p_\perp^2)\right]\,.
\label{nneta}
\end{eqnarray}
We get a factor of $eE$ from $ \int dp_z$,  a factor of $\frac{(\pm 1)^{n-1}}{n}\,\eta_p^n$ from the expansion of the log in (\ref{nnaction}), and finally a Landau degeneracy factor of $\frac{eE}{n\pi}$ from $\int d^2 p_\perp$. Thus, the virial representation (\ref{nnaction}) yields:
\begin{eqnarray}
Im \, {\mathcal L}= (2s+1)\, \frac{e^2 E^2}{16\pi^3}\sum_{n=1}^\infty \frac{(\pm 1)^{n-1}}{n^2}\, \exp\left[-\frac{m^2\pi n}{eE}\right]
\label{nnresult}
\end{eqnarray}
in agreement with (\ref{electricinstanton}) and (\ref{scelectricinstanton}).

For nonzero frequency $\omega$, the Narozhni-Nikishov answer (\ref{nnaction}) is expressed as the integral representation (\ref{nnaction}) with $w_p$ given by (\ref{nnspw}) or (\ref{nnscw}) for spinor or scalar QED respectively. In \cite{dhelectric}, for spinor QED, a more explicit form of this exact answer was found, involving just a single integral representation, analogous to (\ref{1lspelec}) for the constant field case. 
  \begin{eqnarray}
S&=& \frac{2m^4 L^3}{3\pi^2\omega }\int_0^\infty
\frac{dt}{e^{2\pi t}-1} \, \cdot \nonumber\\\nonumber\\
&&
\left[
\left(\frac{eE}{m^2}+\frac{t\,\omega^2}{m^2}\right) \left(1-v^2\right)\hskip -5pt 
~_2 F_1\left(1,\frac{1}{2};\frac{3}{2};\frac{-v^2}{1-v^2}\right) +
(E\to -E)\right]
\label{exacte}
\end{eqnarray}
where $v^2\equiv \frac{t^2\omega^2}{m^2}+2 \frac{e E}{m^2}\, t$. This can be compared with the proper-time representation of the constant field result (\ref{1lspelec}), which can be rewritten as
\begin{eqnarray}
{\mathcal L} &=&-\frac{(eE)^2}{8\pi^2}\int_0^\infty \frac{ds}{s^2}\, e^{-\frac{m^2}{e E}s }\left(\cot\, s-\frac{1}{s}+\frac{s}{3}\right)
\nonumber\\
&=& -\frac{m^4}{8\pi^2}\left(\frac{2 e E}{m^2}\right)  \int_0^\infty
\frac{dt}{e^{2\pi t}-1} \left\{ \frac{2 e E}{m^2}\, t\, \ln\left[1-\left(\frac{2e E t}{m^2}\right)^2 \right]    -\frac{4 e E}{m^2}\, t\right.\nn\\
&&\left. \hskip 3cm + 2\, {\rm arctanh}\left(\frac{2 e E t}{m^2}\right)\right\}.
\label{constante}
\end{eqnarray}

\subsubsection{Space dependent magnetic field $B(x)=B\, {\rm sech}^2(\frac{x}{\lambda})$}
\label{spacedepb}

 As in the previous case, this case is soluble because the Dirac and Klein-Gordon equations reduce to hypergeometric equations. In \cite{cangemi2+1} an exact closed-form expression was found for the spinor effective Lagrangian in $2+1$ dimensional QED in such a background, and in \cite{dh3+1} this was extended to $3+1$ dimensions. In the $2+1$ dimensional case \cite{cangemi2+1}, 
\ba
S_\pm^{(2+1)} = \frac{L}{4\pi\lambda^2} \int_0^\infty \frac{dt}{e^{2 \pi t}\pm 1} \left ( (b_\pm-it)
    \frac{(\lambda ^2 m^2+v_\pm^2)}{v_\pm} \ln \frac{\lambda m
      -iv_\pm}{\lambda m +iv_\pm} + c.c. \right )\nn\\
\label{2+1exact}
\ea
where $+$ denotes scalar QED, $-$ denotes spinor QED, $c.c.$ denotes the complex conjugate, and
\ba
  b_\pm&=&\left\{\begin{array}{ll} \sqrt{(eB\lambda^2)^2+1/4} & \;\;\;
      (+) \text{\ \ bosons} \\ eB\lambda^2 & \;\;\; (-) \text{\ \
        fermions} \end{array} \right.\\
  v_\pm^2&=&\left\{\begin{array}{ll} t^2 + 2i\,t\,b_+-1/4 & \;\;\; (+)
      \text{\ \ bosons}\\ t^2 + 2i\,t\,b_- & \;\;\; (-) \text{\ \
        fermions} \end{array} \right.
        \label{bpm}
\ea
For spinor QED this was extended to $3+1$ dimensions \cite{dh3+1}, to give
 \begin{eqnarray}
S^{(3+1)}_{\rm spinor}&=& -\frac{2m^4 L^3\lambda}{3\pi^2} \int_0^\infty \hskip -7pt 
\frac{dt}{ e^{2\pi t}-1} \left[
\left(\frac{e B}{m^2}-\frac{i t}{m^2\lambda^2}\right)\frac{(1+v^2)^{\frac{3}{2}}}{v}{\rm arcsin}(i v) +
c.c.\right]\nn\\
\label{exactb}
\end{eqnarray}
where $v^2\equiv \frac{t^2}{m^2 \lambda^2}+2i \frac{eB}{m^2}\, t$. This can be compared directly with the constant magnetic field Heisenberg-Euler result (\ref{1lspmag}), which can be written in a modified form: 
\begin{eqnarray}
{\mathcal L} &=&-\frac{(e B)^2}{8\pi^2}\int_0^\infty \frac{ds}{s^2}\, e^{-\frac{m^2}{e B}s}\left(\coth\, s-\frac{1}{s}-\frac{s}{3}\right)
\nonumber\\[1mm]
&=&  \frac{m^4}{8\pi^2}\left(\frac{2 e B}{m^2}\right)  \int_0^\infty
\frac{dt}{e^{2\pi t}-1} \left\{ \left(\frac{2 e B}{m^2}-i\right) \ln\left[1+i\left(\frac{2e B t}{m^2}\right) \right]    +c.c.\right\}.\nn\\
\label{constantb}
\end{eqnarray}

\subsubsection{Other solvable backgrounds}
\label{other}

Another solvable case, which is very similar to the cases described in Sections \ref{timedepe} and \ref{spacedepb}, is that of an electric field pointing in a fixed direction, static in time, but having spatial dependence
\ba
E(x)=E\, {\rm sech}^2(\frac{x}{\lambda})
\label{kpelectric}
\ea
along the direction of the electric field. Once again, this case is solvable because with the gauge field $A_0(x)= -E \lambda \, \tanh(\frac{x}{\lambda})$, the Dirac and Klein-Gordon equations reduce to hypergeometric equations. For details see \cite{kimpage} and \cite{nikishovpreprint}. Such a background in Euclidean $2$-dimensional space was studied in \cite{devega}.

An interesting class of solvable electric fields was recently found by Tomaras {\it et al} \cite{tomaras,fried}, who found that when the electric field depends on a light-cone coordinate $x^0\pm x^3$, then the one loop effective Lagrangian can be found exactly, and has essentially the same form as the constant field Heisenberg-Euler result. Specifically, define the light-cone coordinate
\ba
x^+=\frac{1}{\sqrt{2}}(x^0-x^3)
\label{xplus}
\ea
and choose the vector potential
\ba
A_-(x^+)=\int_0^{x^+} du\, E(u)
\label{lca}
\ea
with $A_\perp=0$, and use the gauge $A_+=0$. Then the spinor one loop effective action is \cite{tomaras,fried}
\ba
S_{\rm spinor}^{(1)}=\frac{1}{8\pi^2} \int d^4x \, \int_0^\infty \frac{ds}{s^3}\, e^{-i s m^2}\,\left[e E(x^+)\,s \, \coth\left(e E(x^+) s\right) -1\right]\,.\nn\\
\label{tomarasaction}
\ea
This leads to a pair production rate
\ba
2 Im\, S_{\rm sp}^{(1)} =\int d^4 x\left\{ -\frac{e^2 E^2(x^+)}{24\pi}+\frac{1}{4\pi}\sum_{n=1}^\infty \left(\frac{eE(x^+)}{n\pi}\right)^2 \exp\left[-\frac{n\, \pi \, m^2}{|e E(x^+)|}\right]\right\}.
\nn\\
\label{tomarasimag}
\ea

\subsection{Semiclassical approximation}
\label{wkb}

The examples listed in the previous section exhaust the analytically solvable cases with physically realistic inhomogeneous electromagnetic backgrounds. For more general forms of inhomogeneity in $F_{\mu\nu}$ a powerful approach is the semiclassical approach. The semiclassical method works best if the background field depends on just one coordinate. Here I concentrate on the example provided by an electric background field which points in a fixed direction in space (say the $z$ direction), but has a time dependent profile. This approach was pioneered by Brezin and Itzykson \cite{brezin}, and Popov and Marinov \cite{popov,popovmarinov}, building on earlier ionization work by Keldysh \cite{keldysh} and Perelomov {\it et al} \cite{perelomov} which will be discussed below. 

For a time dependent electric field we can choose the gauge
\ba
A_z(t)=-\frac{E}{\omega}\, f(\omega t) \quad \Rightarrow \quad E_z(t)=E \dot{f}(\omega t)\ .
\label{gent}
\ea
For example, the solvable case in Section \ref{timedepe} corresponds to the choice: $A_z(t)=-\frac{E}{\omega}\, \tanh(\omega t)$. I also concentrate in this section on scalar QED; spinor QED adds some technical details concerning the spin components, but the basic techniques and results are very similar. Then the Klein-Gordon operator for such a background field is
\ba
D_\mu^2+m^2= m^2+p_\perp^2
+\partial_0^2+\left(p_z-eA_z\right)^2\ .
\label{kgt}
\ea
We expand the field operator in solutions of \cite{brezin}
\begin{eqnarray}
-\ddot{\phi}-(p_z-eA_z)^2\phi=(m^2+p_\perp^2)\phi
\label{kgeq}
\end{eqnarray}
with scattering boundary conditions
\begin{eqnarray}
\phi&\sim & e^{-it\sqrt{m^2+p^2}}+b_{\vec{p}}\,
e^{it\sqrt{m^2+p^2}}\quad , \quad t\to-\infty
\nonumber\\[1mm]
&\sim & a_{\vec{p}}\, e^{-it\sqrt{m^2+p^2}}\quad , \hskip 2.5cm
t\to+\infty
\label{scattbc}
\end{eqnarray}
so that particles are viewed as propagating forwards in time and antiparticles as propagating backwards in time. Then the pair-creation probability is given in terms of the reflection coefficient $b_{\vec{p}}$
\begin{eqnarray}
P\approx \int \frac{d^3 p}{(2\pi)^3}\, |b_{\vec{p}}|^2\ .
\label{scattprob}
\end{eqnarray}
This is an example of  ``over-the-barrier'' scattering \cite{landauqm}, and the semiclassical WKB expression for this amplitude gives
\begin{eqnarray}
|b_{\vec{p}}|^2\approx \exp\left[-2\,{ \mathcal
I}m\oint\sqrt{m^2+p_\perp^2+[p_z-e A_z(t)]^2}\,\,dt\right]
\label{bpwkb}
\end{eqnarray}
Changing variables to $u=i[p_z-eA_z(t)]/m$, we can write
\begin{eqnarray}
|b_{\vec{p}}|^2\approx \exp\left[-2\,\frac{m^2}{e}
\int\frac{\sqrt{1-u^2+p_\perp^2/m^2}}{|E(u)|}\,\,du\right]
\label{bpapprox}
\end{eqnarray}
where the limits of the  $u$ integral are set by the turning points. The exponential term for the pair production rate comes from neglecting the momenta, which instead affect the prefactor (in fact, different prescriptions for treating the momenta lead to slightly different forms of the WKB prefactor \cite{brezin,popov,popovmarinov}.) The leading WKB approximation for the pair-creation probability (which is directly related to the imaginary part of the effective Lagrangian) is then
\begin{eqnarray}
P\sim \exp\left[-\pi \,E_{\rm c}\, g(\gamma)\right]
\label{wkbprob}
\end{eqnarray}
where $E_c=\frac{m^2c^3}{e\hbar}$ is the critical electric field (\ref{critical}) defined by Heisenberg and Euler, and 
\ba
g(\gamma)=\frac{2}{\pi} \int_{-1}^1\frac{\sqrt{1-u^2}}{|E(u)|}\,\,du\ .
\label{gg}
\ea
Here $\gamma$ is the  "adiabaticity" parameter : 
\ba 
\gamma\equiv \frac{m \omega}{eE} \ .
\label{reladiab}
\ea
For example, in the sinusoidal case,
\ba
 A_z=-\frac{E}{\omega}\,\sin(\omega t) \quad \Rightarrow \quad E_z(t)=E\cos(\omega t)
\label{cosine}
\ea
we can evaluate $g(\gamma)$ exactly \cite{popov,brezin,popovmarinov},
\ba
E\, g(\gamma)&=&\frac{2}{\pi} \int_{-1}^1\sqrt{\frac{1-u^2}{1+\gamma^2 u^2}}\,du\nn\\[1mm]
&=&\frac{4}{\pi}\frac{\sqrt{1+\gamma^2}}{\gamma^2}
\left[{\bf K}\left(\frac{\gamma^2}{1+\gamma^2}\right)-
{\bf E}\left(\frac{\gamma^2}{1+\gamma^2}\right)\right]\nn\\[1mm]
&\sim& \begin{cases}1-\frac{1}{8}\gamma^2
\quad  , \quad\gamma\ll 1\cr
\frac{4}{\pi\gamma}\ln \gamma \quad , \quad \gamma\gg 1
\end{cases}
\label{coscase}
\ea
where ${\bf K}(x)$ and ${\bf E}(x)$ are the complete elliptic integral functions \cite{ww}.

This is a truly remarkable result as, when inserted into the pair production rate (\ref{wkbprob}), it leads to an interpolation of the pair production probability between a nonperturbative limit when the adiabaticity parameter $\gamma\ll 1$, and a perturbative limit when $\gamma \gg 1$:
\begin{eqnarray}
P\sim
\begin{cases}\exp\left[-\pi\,\frac{m^2c^3}{eE\hbar}\right]
\ , \hskip .3cm
\gamma\ll 1\quad {\rm (nonperturbative)}\cr\cr \left(\frac{eE}{\omega
mc}\right)^{4mc^2/\hbar
\omega}~\ ,\hskip .3cm \gamma \gg 1 \quad {\rm (perturbative)}
\end{cases}
\label{pnp}
\end{eqnarray}
where we have re-inserted the $\hbar$ and $c$ factors. 
When $\gamma\ll 1$ the time dependent field is slowly varying and we recover the leading nonperturbative exponential behavior of the Heisenberg-Euler result (\ref{imag}). On the other hand, when $\gamma\gg 1$ the result in (\ref{pnp}) has the perturbative form of the square of the gauge field strength $(\frac{eA}{mc})^2$ raised to a power equal to the number of factors of the photon energies $\hbar \omega$ needed to make up the pair creation threshold energy $2mc^2$.

It is worth pausing here to recall the corresponding nonrelativistic ionization problem, where this type of interpolation behavior was first found in a classic paper by Keldysh \cite{keldysh}. Oppenheimer \cite{opp} first estimated the ionization probability for a hydrogen atom in a static electric field $E$ to have the exponential form:
\ba
 P\sim \exp\left[-\frac{2}{3}\frac{m^2 e^5}{E\hbar^4}\right]\,.
 \label{oppion}
 \ea
More generally, for tunneling through a barrier with binding energy ${\mathcal E}_b$,  the WKB expression is 
\begin{eqnarray}
P&\sim& \exp\left[-\frac{2}{\hbar}\int_0^{{ \mathcal E}_b/eE}dz\sqrt{2m({ \mathcal
E}_b-eE z)}\right]\nonumber\\
&=& \exp\left[-\frac{4}{3}\frac{\sqrt{2m}\,{ \mathcal
E}_b^{3/2}}{eE\hbar}\right]\,.
 \label{fowl}
\end{eqnarray}
 For a hydrogen atom,  ${ \mathcal
E}_b=\frac{me^4}{2\hbar^2}$, leading to  Oppenheimer's estimate (\ref{oppion}).
 
In a seminal paper \cite{keldysh}, Keldysh generalized this ionization analysis to a time dependent electric field, and in particular a monochromatic sinusoidal field $E(t)=E\cos(\omega t)$. This approach was further developed by Perelomov {\it et al} in \cite{perelomov}. With the weak field condition, $E\ll \frac{\sqrt{2m}{ \mathcal E}_b^{3/2}}{e\hbar}$, and the condition for a classical electromagnetic  field, $\hbar \omega \ll { \mathcal E}_b$, Keldysh defined the ratio of these two small ratios to be the "adiabaticity parameter" :
\ba
 \gamma=\frac{\hbar\omega/{ \mathcal
E}_b}{e\hbar E/(\sqrt{2m}{ \mathcal
E}_b^{3/2})}=\frac{\omega
\sqrt{2m{ \mathcal
E}_b}}{eE}\ .
 \label{keldad}
 \ea
Using WKB he computed the leading behavior of the ionization probability to be
\begin{eqnarray}
P\sim \exp\left[-\frac{2{ \mathcal
E}_b}{\hbar \omega}\, g(\gamma)\right]
\label{kprob}
\end{eqnarray}
where for the sinusoidal electric field
\begin{eqnarray}
g(\gamma)&=&\left(1+\frac{1}{2\gamma^2}\right){\rm arcsinh}\gamma
-\frac{\sqrt{1+\gamma^2}}{2\gamma}\nn\\
 &\sim&
 \begin{cases}
 \frac{2}{3}\gamma \ ,\quad\quad ~~\gamma\ll 1\,.\cr
\ln(2\gamma)\ ,\quad \,\gamma\gg 1\,.
 \end{cases}
\label{kgg}
\end{eqnarray}
Keldysh noted that the ionization probability (\ref{kprob}) interpolates, as $\gamma$ ranges from  $\gamma\ll 1$ to $\gamma\gg 1$, between the nonperturbative tunneling form (\ref{fowl}) and a perturbative multiphoton ionization form:
 \begin{eqnarray}
P\sim
\begin{cases}
 \exp\left[-\frac{4}{3}\frac{\sqrt{2m}\,{ \mathcal
E}_b^{3/2}}{eE\hbar}\right]
\,,\quad
\gamma\ll 1 \quad ({\rm nonperturbative})\,.\cr\cr \left(\frac{eE}{2\omega \sqrt{2m
{ \mathcal
E}_b}}\right)^{2{ \mathcal
E}_b/\hbar
\omega},\quad ~~\gamma \gg 1 \quad ({\rm perturbative})\,.
 \end{cases}
\label{kpnp}
\end{eqnarray}
 For $\gamma\ll 1$ we recover the tunneling form (\ref{fowl}), and for $\gamma\gg 1$ we see a multiphoton form of the ionization rate.
The analogy of Keldysh's result (\ref{kpnp}) to the QED case for the WKB vacuum pair-production result (\ref{pnp}) becomes clear taking the "binding energy" for the pair production process to be the pair rest energy : 
${ \mathcal E}_b=2mc^2$.

\subsection{Derivative expansion}
\label{deriv}

Another approach to the problem of an inhomogeneous background field strength $F_{\mu\nu}$ is to expand about the solvable constant field case, in terms of derivatives of $F_{\mu\nu}$. Thus, generically one obtains an expression for the one-loop effective action of the form
\begin{eqnarray}
S=S_{[0]}[F]+S_{[2]}[F,(\partial F)^2]+S_{[4]}[F,(\partial F)^2,(\partial F)^4]+\dots
\label{derivexp}
\end{eqnarray}
where $S_{[2n]}$ involves up to $2n$ derivatives of the field strength.
This formal expansion should of course be understood as a multi-dimensional expansion, because even the constant field term $S_{[0]}[F]$, which is the Heisenberg-Euler result, has itself a doubly-infinite weak-field expansion, as shown in (\ref{1lspweak}). At higher orders in the derivative expansion, things rapidly become more complicated since there are many more Lorentz and gauge invariant terms that can (and do) appear once we include derivatives of $F$. In the next two subsections \ref{despb} and \ref{destde} we consider the question of the convergence of the derivative expansion and show that it should be understood as an asymptotic expansion.

Nevertheless, the derivative expansion approach is a very powerful one. The leading order approximation consists of taking the Heisenberg-Euler constant field result for the effective Lagrangian, substituting the inhomogeneous $F_{\mu\nu}$ for the constant one, and then performing the spacetime integrations. In many cases, this gives an excellent approximation to the full effective action. We expect the derivative expansion to be ``good'' (in the sense of an asymptotic expansion \cite{carlbook}) when the scale of variation of the background field is large compared to the electron Compton wavelength, or when the scale of variation of the background field is large compared to the length scale set by the average magnitude of the background field. For example, for a spatially varying magnetic background, with variation scale $\lambda$, and peak field $B$ (which defines a "magnetic length" scale $l_B=\sqrt{\frac{\hbar c}{eB}}$), the derivative expansion is expected to be good when 
\ba
\lambda\gg l_B \qquad {\rm or} \qquad \lambda\gg \lambda_{\rm Compton}=\frac{\hbar}{mc}\ .
\label{dergood}
\ea

A very useful tool for systematizing the derivative expansion is to use the Fock-Schwinger gauge choice
\ba
x_\mu A_\mu=0\ .
\label{fsgauge}
\ea
In this gauge it is possible \cite{novikov,itep} to express the gauge field in terms of the field strength and its derivatives (evaluated at some reference point $x=0$):
\ba
A_\mu(x)&=&\frac{1}{2}x^\alpha F_{\alpha\mu}+\frac{1}{3}x^\alpha x^\beta \partial_\beta F_{\alpha\mu}+\dots\nn\\
&=&\sum_{n=0}^\infty \frac{1}{(n+2)n!}\, x^\alpha x^{\alpha_1} x^{\alpha_2} \dots x^{\alpha_n} \, \partial_{\alpha_1} \partial_{\alpha_2} \dots \partial_{\alpha_n} F_{\alpha \mu}\nn\\
&=& x^\alpha \int_0^1d\eta\, \eta\, e^{\eta\, x\cdot\partial} F_{\alpha \mu}\ .
\label{fsg}
\ea
In this gauge one can expand the effective action (\ref{action}) in powers of derivatives of the field strength. A very convenient formalism for doing such an expansion is the world-line formalism \cite{strassler,ss1,ss2,gerry,csreview} in which the effective action is written as a quantum mechanical path integral. 
For scalar QED, 
\ba
S_{\rm sc}[A]=\int_0^\infty \frac{dT}{T}e^{-m^2 T}\int {\mathcal D}x\, \exp\left\{ -\int_0^T d\tau \left[\frac{1}{4}\dot{x}^2+i e A_\mu(x(\tau))\dot{x}^\mu(\tau)\right]\right\}\nn\\
\label{wlscaction}
\ea
where the path integral is over closed spacetime loops $x_\mu(\tau)$. Keeping just the first term in the derivative expansion  (\ref{fsg}) of $A_\mu$, this quantum mechanical path integral is Gaussian and exactly solvable. Higher order terms in the derivative expansion are then generated by expanding the non-Gaussian interaction terms and computing the resulting expansion terms using the known  worldline Green's functions for the constant background field about which one is perturbing. For spinor QED one must also include Grassmann valued fields $\psi_\mu(\tau)$ which are antiperiodic in the propertime parameter $\tau$: $\psi_\mu(0)=-\psi_\mu(T)$. Then for spinor QED
\ba
S_{\rm sp}[A]&=&-\frac{1}{2}\int_0^\infty \frac{dT}{T}e^{-m^2 T}\int {\mathcal D}x\, \int {\mathcal D}\psi\, \nn\\
&&\hskip -1cm \exp\left\{ -\int_0^T d\tau \left[\frac{1}{4}\dot{x}^2+\frac{1}{2}\psi\cdot\dot{\psi}+i e A_\mu\dot{x}^\mu-ie\psi^\mu F_{\mu\nu}\psi^\nu \right]\right\}
\label{wlspaction}
\ea
which can once again be expanded systematically about the constant field case. The resulting next-to-leading order (NLO) derivative expansion terms for $2+1$ dimensions can be found in \cite{cangemider}, and for $3+1$ dimensions in \cite{leepacshin,igor}. 

The general expressions, even at the next-to-leading order level,  are too long to be usefully included here. But for a given inhomogeneous background  it is straightforward (albeit tedious) to evaluate the first few orders of the derivative expansion. Perhaps a more important question is to ask : is the derivative expansion convergent? As it stands, this is not a well-posed question because of the proliferation of terms with completely new Lorentz structures at higher orders. However, this question can be made well-posed if we restrict the background field to depend on only one spacetime coordinate, with the inhomogeneity characterized by a single scale. In this case the general  derivative expansion expression (\ref{derivexp}) reduces to a double sum, which we can hope to analyze in more detail. To be even more specific, consider some of  the solvable inhomogeneous cases from Section \ref{solvable} -- by comparing the closed-form answers with the derivative expansion we learn something about how the derivative expansion behaves.

\subsubsection{Spatially inhomogeneous magnetic field $B(x)=B\,{\rm sech}^2(\frac{x}{\lambda})$}
\label{despb}

For this spatially inhomogeneous magnetic background field the exact closed form answer for the spinor QED effective action was given in (\ref{exactb}). It is a simple exercise to make a double asymptotic expansion of this answer:
\begin{eqnarray}
S\sim -\frac{L^{2}\lambda m^{4}}{8\pi ^{3/2}}
{\sum_{j=0}^{\infty}}\frac{1}{j!} \frac{1}{(m\lambda)^{2j} }\sum_{k=1}^{\infty }\frac{\Gamma
(2k+j)\Gamma (2k+j-2){\mathcal B}_{2k+2j}}{\Gamma (2k+1)\Gamma (2k+j+\frac{1}{2})}
\left( \frac{2eB}{m^{2}}\right) ^{2k}\nn\\
\label{full3+1}
\end{eqnarray}
(where the $j=0$ and $k=1$ term is excluded from the sum as it corresponds to a charge renormalization term.)
The sum over $j$ corresponds to the derivative expansion, with expansion parameter  $\frac{1}{(m\lambda)^2}=(\frac{\lambda_{\rm Compton}}{\lambda})^2$, while the sum over $k$ corresponds to a perturbative expansion with expansion parameter $\frac{eB}{m^2}$, as in the constant-field Heisenberg-Euler case (\ref{1lspmagexp}). The $j=0$ term in (\ref{full3+1}) is the leading order derivative expansion term, and the $j=1$ term is the NLO term. The LO effective Lagrangian is just the Heisenberg-Euler expression (\ref{1lspmag}) for a constant magnetic field, 
\ba
{\mathcal L}_{[0]}&=&-
\frac{e^2 B^2}{8\pi ^{2}}\int_{0}^{\infty} \frac{ds}{s^{2}}\;e^{-m^2s/(eB)}
(\coth s-\frac{1}{s}-\frac{s}{3})\nn\\[1mm]
&\sim&-\frac{m^{4}}{8\pi^2} \sum_{k=2}^{\infty }\frac{{\mathcal B}_{2k}}
{2k(2k-1)(2k-2)}\left(\frac{2eB}{m^{2}}\right)^{2k}
\label{zero}
\ea
while the NLO effective Lagrangian for a static but spatially inhomogeneous magnetic field is \cite{cangemider,igor}
\ba
{\mathcal L}_{[2]}&=&-e\, \frac{\partial _{i}B\partial _{i}B} {64\pi^{2}B}
\int_{0}^{\infty}
\frac{ds}{s}e^{-m^2 s/(eB)}(s\coth s)^{\prime \prime \prime }\nn\\[1mm]
&\sim& -e^2\,\frac{\partial _{i}B\partial _{i}B}{4\pi^2
m^2}\sum_{k=1}^{\infty}
\frac{{\mathcal B}_{2k+2}}{(2k-1)}\left( \frac{2eB}{m^{2}}\right)^{2k-2}
\label{nlo}
\ea
which involves two derivatives of the magnetic field.
To compare with the expansion (\ref{full3+1}) of the exact result, we substitute the inhomogeneous magnetic field, $B(x)=B{\rm sech}^2(\frac{x}{\lambda})$, for $B$ in the derivative expansion expressions (\ref{zero}) and (\ref{nlo}), and then integrate over the inhomogeneity direction $x$. This yields the LO and NLO derivative expansion results for the effective action as
\ba
S_{[0]}\sim 
-\frac{L^2\lambda m^4}{8\pi^{3/2}}\sum_{k=2}^{\infty}\frac{1}{2k}
\frac{{\mathcal B}_{2k}\Gamma (2k-2)}{\Gamma (2k+\frac{1}{2})}
\left(\frac{2eB}{m^{2}}\right)^{2k},
\label{full3+1j=0}
\ea
\ba
S_{[2]}\sim -\frac{L^2 m^2}{8\lambda \pi ^{3/2}}\sum_{k=1}^{\infty }\frac{{\mathcal
B}_{2k+2} \Gamma (2k-1)}{\Gamma (2k+\frac{3}{2})}\left( \frac{2eB}{m^{2}}
\right)^{2k}.
\label{full3+1j=1}
\ea
These results agree exactly with the $j=0$ and $j=1$ terms from the asymptotic expansion (\ref{full3+1}) of the exact result. This example clearly illustrates the approach of substituting the inhomogeneous fields for the fields appearing in the derivative expansion expressions for the effective Lagrangian.

\subsubsection{Time dependent  electric field $E(t)= E \, {\rm sech}^2(\omega \, t)$}
\label{destde}

Another interesting comparison of the derivative expansion can be made for this time dependent solvable case, whose exact solution was given in Section \ref{timedepe}. One can do the same thing described above for the inhomogeneous magnetic case, and compare the asymptotic expansion of the exact result (\ref{exacte}) with the LO and NLO field theoretic derivative expansion results from \cite{igor}. Instead, here I show \cite{dhborel} how the derivative expansion agrees with the semiclassical WKB results of Section \ref{wkb}.

An asymptotic expansion of the exact result (\ref{exacte}) yields the derivative expansion
\begin{eqnarray}
  S\sim -\frac{m^4}{8\pi^{3/2}\omega }
\sum_{j}^\infty \sum_{k}^\infty
\frac{(-1)^{j+k}}{(m/\omega)^{2j}}
\left(\frac{2eE}{m^2}\right)^{2k}\frac{\Gamma(2k +
j)\Gamma(2k+ j -2) {\mathcal B}_{2k+2j}}{ j!(2k)!\Gamma(2k+j+\frac{1}{2})}\ .\nn\\
\label{derivel}
\end{eqnarray}
Fixing the order $j$ of the derivative expansion, the remaining sum over $k$ is divergent and nonalternating, and hence has a nonperturbative imaginary part. 
For fixed $j$, the expansion coefficients behave for large $k$ as
\begin{eqnarray}  
c_k^{(j)}&=&\frac{(-1)^{j+k}\Gamma(2k+j)
\Gamma(2k+j+2){\mathcal B}_{2k+2j+2}} {\Gamma (2k+3)\Gamma
(2k+j+\frac{5}{2})}\nn\\
&\sim&  2 \,
\frac{\Gamma(2k+3j-\frac{1}{2})}{(2\pi)^{2j+2k+2}}\ , \qquad k\to\infty\ .
\label{dc}
\end{eqnarray}
Using standard Borel dispersion relations \cite{zinnborel,thooftborel}, one finds \cite{dhborel}
\ba
{\mathcal I}m S^{(j)} \sim  \frac{1}{j!}\,
\left(\frac{\pi m^4 \omega^2}{4e^3 E^3}\right)^{j}
\exp\left[-\frac{m^2\pi}{eE}\right]
\label{jsum}
\ea
Remarkably, the sum over $j$ exponentiates to give the leading term:
\ba
{ \mathcal I}m S\sim \exp\left [-\frac{m^2\pi}{eE}
\left(1-\frac{1}{4}\left(\frac{m \omega}{eE}\right)^2\right)\right]\,.
\label{kthenj}
\ea
On the other hand, if we first considered a fixed order of the $k$ summation, then the remaining $j$ sum is divergent and nonalternating. For large $j$, the coefficients behave as 
\begin{eqnarray}  
c_j^{(k)}&=&(-1)^{j+k}\frac{\Gamma(j+2k)
\Gamma(j+2k-2){\mathcal B}_{2k+2j}} {\Gamma (j+1)\Gamma
(j+2k+\frac{1}{2})}\nn\\
&\sim& 2^{\frac{9}{2}-2k} 
\frac{\Gamma(2j+4k-\frac{5}{2})}{ (2\pi)^{2j+2k}}\ , \qquad j\to\infty\ .
\label{dce}
\end{eqnarray}
This leads to the nonperturbative imaginary part:
\ba
{ \mathcal I}m S^{(k)} \sim  \frac{1}{(2k)!}\left(\frac{2\pi e E}{\omega^2}\right)^{2k} \,
\exp[-2\pi m/\omega]\ .
\label{ksum}
\ea
The remaining $k$ sum also exponentiates, yielding the leading behavior
\ba
{ \mathcal I}m S\sim \exp\left [-\frac{2\pi m}{\omega}
\left(1-\frac{eE}{m\omega}\right)\right]\,.
\label{jthenk}
\ea

How do we reconcile these two different expressions (\ref{kthenj}) and (\ref{jthenk}) for the imaginary part of the effective action? The answer can be found in the WKB analysis of Section \ref{wkb} and the Keldysh adiabaticity parameter, which in this case is
\ba
\gamma=\frac{m\,\omega}{eE}=\frac{\omega/m}{eE/m^2}\ ,
\label{ggder}
\ea
the ratio of the two expansion parameters in (\ref{derivel}).

For this particular time dependent electric field $E(t)=E\, {\rm sech}^2(\omega t)$, we can return to the WKB analysis of Section \ref{wkb} and evaluate \cite{popov}  the exponent $g(\gamma)$ in (\ref{gg}) which appears in the leading WKB pair production rate (\ref{wkbprob}):
\ba
E\, g(\gamma)&=& \frac{2}{\pi}\int_{-1}^1 \frac{\sqrt{1-u^2}}{1+\gamma^2 u^2}\, du \nn\\
&=& \frac{2}{1+\sqrt{1+\gamma^2}} \sim \begin{cases} 1-\frac{\gamma^2}{4}+\dots \qquad ,\quad \gamma\ll 1\ .\cr
\frac{2}{\gamma}\left(1-\frac{1}{\gamma}+\dots\right) \,, \quad \gamma\gg 1\ .\end{cases}
\label{ggcompare}
\ea
Then the WKB approximation (\ref{wkbprob}) for the pair-production probability, $P\sim \exp[-\pi m^2 g(\gamma)/(eE)]$, gives precisely (\ref{kthenj}) in the $\gamma\ll 1$ limit, and (\ref{jthenk}) in the $\gamma\gg 1$ limit. In terms of the Borel resummation, it is a matter of competing exponential factors, with the dominant exponential being determined by the size of the adiabaticity parameter $\gamma$. Thus, this derivative expansion example is completely consistent with the WKB analysis of the imaginary part of the effective action. It also demonstrates clearly that the derivative expansion is an {\it asymptotic} expansion, rather than a {\it convergent} expansion, because if it were convergent there would be no imaginary parts generated from the derivative expansion.

\subsection{Large mass expansion}
\label{largemass}

The large mass expansion is very closely related to the derivative expansion, but is a different way of organizing the expansion. For example, consider the all-orders derivative expansion double series (\ref{full3+1}). This could clearly be written as an expansion in inverse powers of the mass $m$, with another series that involves different powers of the magnetic field strength and its derivatives:
\ba
S=-\frac{L^2\lambda m^4}{8\pi^{3/2}} \sum_{l=3}^\infty \frac{\Gamma(l)\Gamma(l-2)}{\Gamma(l+\frac{1}{2})}\frac{1}{m^{2l}}\sum_{k=1}^{[l/2]} \frac{(2eB)^{2k}}{\lambda^{2l-4k}}\frac{{\mathcal B}_{2l-2k}}{(2k)!(l-2k)!}\ .
\label{magmassexp}
\ea
The general structure of the large mass expansion of the effective action is
\ba
\Gamma[F]=\int_0^\infty\frac{dT}{T}\frac{e^{-m^2 T}}{(4\pi T)^{d/2}}\,  \int d^dx \sum_{n=1}^\infty \frac{(-T)^n}{n!} {\mathcal O}_n[F]
\label{genmassexp}
\ea
where the functions ${\mathcal O}_n[F]$ have mass dimension $2n$ and are expressed in terms of the field strength $F_{\mu\nu}$ and its derivatives. There are many different ways to derive such an inverse mass expansion. Perhaps the most efficient \cite{fliegnermass} is the string-inspired worldline approach in which the effective action is expressed as a quantum mechanical path integral over a proper-time interval $\tau\in [0,T]$, as shown for scalar QED in (\ref{wlscaction}) and for spinor QED in (\ref{wlspaction}). These worldline forms of the effective action clearly separate the mass dependence from the  background field dependence, and with the Fock-Schwinger gauge (\ref{fsg}) lead to manifestly gauge invariant expressions for the functions ${\mathcal O}_n[F]$. A key step in this process is finding a minimal basis for the ${\mathcal O}_n[F]$, since these can clearly be rewritten in various ways using the Bianchi identity, the antisymmetry of $F_{\mu\nu}$, and integrations-by-parts. Such a minimal basis was constructed by U. M\"uller in \cite{muller}, and is described in detail in \cite{fliegnermass}. The expressions for the high order terms ${\mathcal O}_n[F]$ are very complicated and are typically generated using a symbolic manipulation program such as FORM \cite{form}. Here I list, from \cite{fliegnermass}, the first few terms ${\mathcal O}_n[F]$ for the case of charged scalar particles in an inhomogeneous nonabelian background field :
  \ba
 O_1 &=& 0\ ,\nn\\
 O_2 &=&  \;\frac{1}{6}\; {\rm tr}\,F_{\kappa\lambda}F_{\lambda\kappa}\ ,
\nn\\
 O_3 &=& 
 \;\frac{1}{20}\;{\rm tr}\,F_{\kappa\lambda\mu}F_{\kappa\mu\lambda} 
 - \;\frac{2}{15}\;\mbox{i}\;{\rm tr}\, F_{\kappa\lambda}F_{\lambda\mu}F_{\mu\kappa}\ , 
\nn\\
 O_4 &=& 
 -\;\frac{1}{21}\;{\rm tr}\,F_{\kappa\lambda}F_{\lambda\mu}F_{\mu\nu}F_{\nu\kappa}   
 + \;\frac{1}{70}\;{\rm tr}\,F_{\kappa\lambda\mu\nu}F_{\lambda\kappa\nu\mu} \nn\\
 &&+ \;\frac{2}{35}\;{\rm tr}\,F_{\kappa\lambda}F_{\lambda\kappa}F_{\mu\nu}F_{\nu\mu} 
 + \;\frac{4}{35}\;{\rm tr}\,F_{\kappa\lambda}F_{\lambda\mu}F_{\kappa\nu}F_{\nu\mu}  \nn\\ 
&& -\;\frac{6}{35}\;\mbox{i}\;{\rm tr}\,F_{\kappa\lambda}F_{\mu\lambda\nu}F_{\mu\nu\kappa}
 -\;\frac{8}{105}\;\mbox{i}\;{\rm tr}\,F_{\kappa\lambda}F_{\lambda\mu\nu}F_{\kappa\nu\mu}\nn\\
&& + \;\frac{11}{420}\;{\rm tr}\,F_{\kappa\lambda}F_{\mu\nu}F_{\lambda\kappa}F_{\nu\mu}\ . 
\label{Ons}
\ea 
 Here, $F_{\mu\nu\rho}\equiv D_\mu F_{\nu\rho}$, so it is clear that ${\mathcal O}_n[F]$ has mass dimension $2n$. The next two terms, ${\mathcal O}_5[F]$ and ${\mathcal O}_6[F]$, are also presented in \cite{fliegnermass}, and the dependence on a background scalar field is included also. The number of terms appearing in a given ${\mathcal O}_n[F]$ grows rapidly with $n$. For example, for a gauge background, using the minimal basis, the number of independent terms in ${\mathcal O}_n[F]$ is found \cite{fliegnermass} to be equal to $0$, $1$, $2$, $7$, $36$, $300$, ..., for $n=1, 2, \dots, 6$, which looks like factorial growth.

\subsection{Worldline loops}
\label{worldloops}

A recent interesting proposal for evaluating effective actions is to implement the worldline expressions (\ref{wlscaction}) and (\ref{wlspaction}) numerically, by doing the quantum mechanical path integral as a Monte-Carlo integration over a randomly generated set of worldline loops \cite{gieslangfeld,schmidtstamatescu}. In principle this is a very powerful and general technique, since it is not constrained by the "slowly varying" restriction for the applicability of the derivative expansion or the large mass condition for the inverse mass expansion. Moreover, it can be applied to both scalar and spinor, and to both abelian and nonabelian systems. So far, the method has been tested on some fairly simple abelian models (as well as to the Casimir effect for surfaces of nontrivial shape \cite{casimir}), and compares favorably with other methods when a comparison is possible, and produces robust answers in other inaccessible cases. Technical aspects of this method are still being developed, but this is an approach whose generality makes it very promising.

\subsection{Bounds}
\label{bounds}

Another approach to the evaluation of one-loop effective actions in nontrivial inhomogeneous backgrounds is to seek rigorous mathematical bounds using operator theory and functional analysis. This approach has been developed recently by M. Fry \cite{fry}, and also has great potential to yield important information about the behavior of fermion determinants in general backgrounds. For example, in Euclidean $2$-dimensions, for a unidirectional abelian field strength $F(\vec{x})$, the fermion determinant is bounded below by \cite{fry}
\ba
\ln \det (-i \Dslash +m)\geq \frac{1}{4\pi}\int d^2 x\left[ e F(\vec{x}) -\left(m^2+e F(\vec{x})\right) \ln\left(1+\frac{e F(\vec{x})}{m^2}\right)\right]\,.\nn\\
\label{frybound}
\ea
For higher dimensional systems, such general bounds are considerably more difficult to establish, but once again their appeal is their generality.

\section{Heisenberg-Euler beyond QED}
\label{beyondqed}
\renewcommand{\theequation}{3.\arabic{equation}}
\setcounter{equation}{0}

In 1965, Vanyashin and Terent'ev \cite{terentev} performed a Heisenberg-Euler computation beyond QED, in a nonrenormalizable theory of vector electrodynamics. They computed the Heisenberg-Euler effective Lagrangian  for a massive charged vector boson of gyromagnetic ratio $2$, in a constant electromagnetic background field. For an electric background field they found a scalar pair production rate of the form in (\ref{nnresult}) for spin $1$. More significantly, they noted the anti-screening effect of the vector bosons on the renormalized electric charge. In hindsight, this is a precursor to asymptotic freedom, but at the time the physical significance was not appreciated, in large part due to the fact that this was a nonrenormalizable theory. Later, after the discovery of asymptotic freedom \cite{grosswilczek,politzer}, Skalozub \cite{skalozub} explained the physical significance of Vanyashin and Terent'ev's results by doing  a related Heisenberg-Euler effective Lagrangian computation in a renormalizable $SU(2)$ theory with spontaneous symmetry breaking. Once the principles of consistently quantizing nonabelian gauge theories were established, a number of Heisenberg-Euler applications and calculations followed.

\subsection{Covariantly constant nonabelian backgrounds}
\label{nonabelian}

The natural nonabelian generalization of Heisenberg and Euler's  condition of constant abelian field strength is that of a {\sl covariantly constant} nonabelian field strength:
\ba
D_\mu F_{\nu\rho}=0\ .
\label{covconstant}
\ea
This condition implies that the corresponding nonabelian gauge field $A_\mu$ can be expressed as a quasi-abelian gauge field, up to a gauge transformation \cite{brownduff,bms,shore}:
\ba
A_\mu=-\frac{1}{2}\,F_{\mu\nu} x^\nu (n^a T^a) -\frac{1}{g}\, \partial_\mu U\, U^{-1}\ .
\label{quasi}
\ea
Thus we can take the field strength to point in a particular direction $n^a$ (with $n^a n^a=1$) in color space, and characterized by an "abelian" field strength $F_{\mu\nu}$:
\ba
F_{\mu\nu}^a=F_{\mu\nu}\, n^a \ , \qquad A_\mu^a=-\frac{1}{2}\, F_{\mu\nu}x^\nu\, n^a\ .
\label{quasif}
\ea
Brown and Duff \cite{brownduff} presented the general formalism for computing one-loop effective actions in nonabelian theories, generalizing the approaches of Heisenberg and Euler, and Schwinger. Then Duff and Ram\'on-Medrano \cite{drm} applied this to the one-loop effective action in Yang-Mills theory, finding an ultraviolet finite expression
\ba
{\mathcal L}_{\rm YM}&=& \frac{1}{2(4\pi)^2} {\rm Tr} \int_0^\infty \frac{ds}{s^3}\left\{\left[e^{-X s} e^{-F(Y;s)}-1-\frac{s^2}{2}X^2-\frac{s^2}{12}Y_{\mu\nu}Y_{\mu\nu}\right]\right.\nn\\
&&\left.\hskip 3cm -2\left[e^{-F({\mathcal Y};s)}-1-\frac{s^2}{12} {\mathcal Y}_{\mu\nu}{\mathcal Y}_{\mu\nu}\right]\right\}
\label{drm}
\ea
where
\ba
F(Y;s)\equiv \frac{1}{2}\,{\rm tr}\ln \left(\frac{\sin(\gamma s)}{\gamma s}\right) \ , \qquad \gamma^2_{\mu\nu}\equiv Y_{\mu\alpha}Y_{\nu\alpha}
\label{fandy}
\ea
and $X$, $Y$ and ${\mathcal Y}$ are expressed in terms of the covariantly constant background field strength $F_{\mu\nu}^a$ as \cite{drm}: 
\ba
X_{\mu\nu}^{ab}&=&2 g f^{abc} F^c_{\mu\nu}\ ,\nn\\[1mm]
Y_{\mu\nu\alpha\beta}^{ab}&=& g f^{acb} F_{\mu\nu}^c \, \delta_{\alpha\beta}\ ,\nn\\[1mm]
{\mathcal Y}_{\mu\nu}^{ab}&=& g\, f^{acb}\, F_{\mu\nu}^c\ .
\label{xyy}
\ea
Duff and Ram\'on-Medrano noted that the one-loop effective action (\ref{drm}) is UV finite but infrared divergent, due to the massless nature of Yang-Mills theory.

Savvidy et al \cite{bms,ms,savvidy} considered the case of $SU(2)$ Yang-Mills theory and 
wrote a formal unrenormalized expression for the one-loop effective Lagrangian, motivated by the original Heisenberg-Euler form (\ref{hesp}):
\ba
{\mathcal L}_{SU(2)}&=& \frac{1}{8\pi^2}\int_0^\infty \frac{ds}{s^3} \frac{g f_1s}{\sinh(g f_1 s)} \frac{g f_2 s}{\sin(g f_2 s)} \nn\\[1mm]
&& +\frac{1}{4\pi^2}\int_0^\infty \frac{ds}{s^3} (g f_1 s)(g f_2 s)\left(\frac{\sin(g f_1s)}{\sinh(g f_2 s)} - \frac{\sin(g f_2 s)}{\sinh(g f_1 s)}\right)
\label{su2}
\ea
where $f_1$ and $f_2$ are essentially Heisenberg and Euler's $a$ and $b$ parameters (\ref{abfg}), with ${\mathcal F}=\frac{1}{4}F_{\mu\nu}^a F_{\mu\nu}^a$, and ${\mathcal G}=\frac{1}{4}F_{\mu\nu}^a \tilde{F}_{\mu\nu}^a$. In the chromomagnetic case where $f_1=H$ and $f_2=0$, the effective Lagrangian has the strong field limit
\ba
{\mathcal L}\sim -\frac{11}{48\pi^2} (g H)^2 \left[\ln\left(\frac{gH}{\mu^2}\right)-\frac{1}{2}\right]\,.
\label{11}
\ea
Identifying the coefficient of the log term with the $\beta$-function, in the spirit of Weisskopf's original observation \cite{viki1,viki2} and the work of Coleman and Weinberg \cite{cw} and Ritus \cite{ritusspin,ritusscal,ginzburg}, Savvidy et al noted that this Heisenberg-Euler effective action computation gives the correct one-loop Yang-Mills $\beta$-function \cite{grosswilczek,politzer}. The negative sign of the $\beta$-function, which signals asymptotic freedom in Yang-Mills theories, arises physically because of the magnetic moment coupling of the charged vector nonabelian field. The significance of this magnetic moment coupling had been emphasized previously by Tsai et al \cite{tsai}, as noted by Salam and Strathdee \cite{salamstrathdee} in their paper which also helped to introduce zeta function techniques to the computation of Heisenberg-Euler effective Lagrangians. Savvidy et al also noted the interesting fact that for a covariantly constant chromoelectric field there is a vacuum instability, very much like that found by Heisenberg and Euler for QED in a constant electric field. 

N. K. Nielsen and P. Olesen \cite{nielsen} found that the situation is even more interesting than this, as they showed that in the covariantly constant nonabelian case there is in fact also an instability in a chromomagnetic background, not just in a  chromoelectric background. This is in distinct contrast to the QED case where a constant magnetic background has no vacuum instability. Nielsen and Olesen noted that the difference is directly due to the anomalous magnetic moment coupling in the Yang-Mills equations, which contributes to the effective action:
\ba
S=-\int dk \sum_{n=0}^\infty \sum_{S_3=\pm 1}\sqrt{2g H(n+\frac{1}{2}-S_3)+k^2}\ .
\label{maginst}
\ea
The mode with $n=0$ and spin aligned along the chromomagnetic background ($S_3=+1$) gives an imaginary contribution to the effective action, indicating an instability. They further showed that this result can be established by a precise treatment of the physical integration contours required in the formal bare expression (\ref{su2}) of Savvidy et al, thereby making direct connection with the proper-time results of Heisenberg and Euler, and Schwinger. Subsequently, Olesen and collaborators \cite{olesen} developed a color magnetic condensate picture of the QCD vacuum based on this chromomagnetic instability. This chromomagnetic instability was also pointed out by Yildiz {\it et al} \cite{yildiz1}, who identified an imaginary contribution in the associated Heisenberg-Euler effective Lagrangian, and who also computed the massive spinor and scalar contributions to the effective Lagrangian in a covariantly constant background. Then, in a development that will be of interest later in this review, Leutwyler \cite{leutwyler} showed that for a covariantly constant nonabelian background there is an instability unless the background field is self-dual (or anti self-dual): $F_{\mu\nu}=\pm \tilde{F}_{\mu\nu}$. At about the same time, the significance of self-dual backgrounds was becoming appreciated in gauge theory \cite{duffisham1,duffisham2}, having already been realized to be important in gravity and supergravity \cite{grisaru,kallosh}. Dittrich and Reuter \cite{drzeta} and Elizalde \cite{elizalde} further developed the zeta function approach, and applied it to the computation of the QCD effective Lagrangian in covariantly constant nonabelian backgrounds. Other important developments have been the derivative expansion and large mass expansion for nonabelian theories \cite{novikov,itep,fliegnermass,mcarthur}, and also for nonabelian supersymmetric theories \cite{mcarthur2}.

\subsection{Supersymmetric Heisenberg-Euler effective Lagrangians}
\label{susylags}

The Heisenberg-Euler approach is of great use in studies of supersymmetry and supersymmetry breaking. It is possible to apply the Heisenberg-Euler techniques, as generalized by Schwinger, Tsai and others \cite{rss} to higher spins, to study effective Lagrangians of supersymmetric theories in constant background fields. This is now a vast branch of quantum field theory \cite{buchbook}, so it is only possible to give  a small taste of the approach here. An important paper in this development was by Fradkin and Tseytlin \cite{ftsusy}, who used dimensional reduction from $d=10$ to $d=4$ to compute the one loop effective Lagrangian for $4$ dimensional ${\mathcal N}=4$ SYM (supersymmetric Yang-Mills). Fradkin and Tseytlin considered a quasi-abelian covariantly constant $SU(2)$ background as in (\ref{quasif}) in the $4$ spacetime dimensions, with $A_i=0$ in the remaining $6$ dimensions. Then the one loop effective action has the form
\ba
\Gamma^{(1)}=-\frac{1}{2}\frac{V_d}{(4\pi)^{d/2}}\int_0^\infty \frac{ds}{s^{1+d/2}}\, \Phi(s)
\label{ftaction}
\ea
where the important observation is that the form of the integrand $\Phi(s)$ is independent of the dimension $d$, with the dimension dependence entering in the measure of the propertime $s$ integration. Thus, one can compute the $d=4$ SYM case beginning from $d=10$. Analogous to the Heisenberg-Euler definitions (\ref{abfg}), Fradkin and Tseytlin defined the (Euclidean) invariants
\ba
F_{1,2}^2=J_1\pm\sqrt{J_1^2-J_2^2}
\label{ftinvs}
\ea
where $J_1=\frac{1}{4}F_{\mu\nu}F_{\mu\nu}$, and $J_2=\frac{1}{4}F_{\mu\nu}\tilde{F}_{\mu\nu}$. Then for spin $0$, $\frac{1}{2}$, and $1$, one finds that for gauge group $SU(2)$:
\ba
\Phi(\Delta_0)&=&2\,\frac{s\, F_1}{\sinh (s F_1)}\,\frac{s\, F_2}{\sinh (s F_2)}\ ,\nn\\[1mm]
\Phi(\Delta_{1/2})&=&2^{[d/2]}\, \cosh (s F_1)\, \cosh (s F_2) \, \Phi(\Delta_0)\ ,\nn\\[1mm]
\Phi(\Delta_1)&=& \left[d+4\left(\sinh^2 (s F_1) +\sinh^2 (s F_2)\right)\right]\Phi(\Delta_0)\ .
\label{ftphis}
\ea
where $\Delta_{0,1/2,1}$ are the corresponding operators for fields of spin $0$, $\frac{1}{2}$, and $1$.
Thus for $d=10$ we obtain
\ba
\Phi(s)&=&-2\Phi(\Delta_0)-\frac{1}{4}\Phi(\Delta_{1/2})+\Phi(\Delta_1)\nn\\
&=& 4\left(\cosh s F_1 -\cosh s F_2\right)^2 \, \Phi(\Delta_0)\ .
\label{ftphi}
\ea
Therefore, in the dimensionally reduced $d=4$ theory one has \cite{ftsusy}
\ba
\hskip -1cm {\mathcal L}^{(1)}_{\rm SYM}=-\frac{1}{4\pi^2} \int_0^\infty \frac{ds}{s^3}\, \frac{s\, F_1}{\sinh s F_1}\,\frac{s\, F_2}{\sinh s F_2}\, \left(\cosh s F_1 -\cosh s F_2\right)^2\,.
\label{ftlag}
\ea
This expression (\ref{ftlag}) is UV finite, owing to the supersymmetry. It is, however, IR divergent unless $F=\tilde{F}$, in which case $F_1=F_2$. This IR divergence is due to the anomalous magnetic moment term in the spin $1$ operator $\Delta_1$, as has been noted already in our discussion of QCD effective Lagrangians in Section \ref{covconstant}.

Another illustrative supersymmetric application of Heisenberg-Euler techniques is to the computation of the one loop effective Lagrangian for $4$ dimensional ${\mathcal N}=4$ SYM in a constant ${\mathcal N}=2$ SUSY background \cite{bkt}. This ${\mathcal N}=2$ background is characterized by a vector multiplet ${\mathcal W}=\left\{W_\alpha , \Phi \right\}$, and if ${\mathcal W}$ satisfies the constancy conditions
\ba
D_\alpha W_\beta = D_{(\alpha}W_{\beta)}={\rm constant}\ , \qquad \Phi={\rm constant}\ ,
\label{susyconstant}
\ea
then the superfield analogue of Schwinger's proper-time formalism \cite{buchbook} can be applied to compute the one loop effective action \cite{bkt}
\ba
\Gamma^{(1)}&=& \frac{1}{16\pi^2}\int d^8 z \frac{W^2\,\bar{W}^2}{\Phi^2\, \bar{\Phi}^2} 
+ \frac{1}{8\pi^2}\int d^8 z\int_0^\infty dt\, t\, e^{-t} \frac{W^2\,\bar{W}^2}{\Phi^2\, \bar{\Phi}^2} \, \omega(t\,\Psi, t \,\bar{\Psi})\ .\nn\\
\label{1lbkt}
\ea
Here $\omega(x,y)$ is the trigonometric function
\ba
\omega(x,y)=\left(\frac{\cosh x -1}{x^2}\right)\left(\frac{\cosh y -1}{y^2}\right)\left(\frac{x^2-y^2}{\cosh x -\cosh y}\right)-\frac{1}{2}
\label{bktomega}
\ea
and $\Psi$ and $\bar{\Psi}$ are the fields
\ba
\bar{\Psi}^2\equiv \frac{1}{4\bar{\Phi}^2}\, D^2\left(\frac{W^2}{\Phi^2}\right)\ , \qquad {\Psi}^2\equiv \frac{1}{4\Phi^2}\, \bar{D}^2\left(\frac{\bar{W}^2}{\bar{\Phi}^2}\right)\,.
\label{bktpsi}
\ea
Notice the clear similarity to the basic Heisenberg-Euler structures in the integrand. Supersymmetry combines contributions from different spin fields with particular weights. Buchbinder {\it et al} show \cite{bkt} that this result can be made manifestly ${\mathcal N}=2$ superconformal invariant to yield
\ba
\Gamma^{(1)}&=& \frac{1}{16\pi^2}\int d^{12} z \ln \frac{{\mathcal W}}{\mu}\, \ln \frac{\bar{\mathcal W}}{\mu}
+ \frac{1}{8\pi^2}\int d^{12} z\int_0^\infty dt\, t\, e^{-t} \, \Omega(t\,{\bf \Psi}, t \,{\bf \bar{\Psi}})\ .\nn\\
\label{finalbkt}
\ea
Here ${\bf \Psi}$ is the full superconformal field, and the function $\Omega(x,y)$ is related to the function $\omega(x,y)$ as follows \cite{bkt}:
\ba 
{\rm If}\quad \omega(x,y)&=&\sum_{m,n=1}^\infty c_{m,n}x^{2m} y^{2n} \nn\\
 {\rm then} \quad \Omega(x,y)&=&\sum_{m,n=1}^\infty \frac{c_{m,n}}{(2m)(2m+1)(2n)(2n+1)}\, x^{2m} y^{2n}\ .
\label{bigo}
\ea
The interesting thing about this computation is that the fact that the theory is superconformal allows one to use the classification of superconformal invariants in order to go beyond the Heisenberg-Euler constant field approximation. This is an example where added symmetry, combined with the basic Heisenberg-Euler structure is a very powerful computational tool \cite{buchbook}.

\subsection{Instanton background}
\label{instcal}

An important example of nonabelian effective actions concerns the spinor and scalar effective action in an instanton background. This was first computed in the small mass limit by 't Hooft \cite{thooft}.
While not strictly-speaking of Heisenberg-Euler form, there are many parallels, since the spectra of the massless Dirac and Klein-Gordon operators are known in closed-form for an instanton background, as are the associated Green's functions. Nevertheless, unlike the Heisenberg-Euler cases, there is no known exact solution for the one loop effective action in an instanton background for arbitrary quark (or scalar) mass. Consider the case of an $SU(2)$ single instanton background:
\ba
A_\mu(x)=A^a_\mu\,\frac{\tau^a}{2} =\frac{1}{g}\, \frac{\eta_{\mu\nu a}\,\tau^a \, x_\nu}{x^2+\rho^2}
\label{insta}
\ea
where $\rho$ denotes the scale of the instanton, $\tau^a$ are the Pauli matrices, and $\eta_{\mu\nu a}$ are the 't Hooft symbols, which mix the spacetime and group indices \cite{thooft,itep}. The corresponding field strength is
\ba
F_{\mu\nu}(x)=-\frac{2}{g}\,  \frac{\rho^2\, \eta_{\mu\nu a}\,\tau^a }{(x^2+\rho^2)^2}\ .
\label{instf}
\ea
Due to the self-duality of the background field, the Dirac operator has a quantum mechanical supersymmetry \cite{thooft,jackiwrebbi,schwarz,dadda}, which means that the spectrum of the Dirac operator coincides with that of the corresponding Klein-Gordon operator, except for a multiplicity factor of $4$, and the presence of zero modes for the Dirac operator. The net effect is that the the one-loop spinor and scalar effective Lagrangians in such a background are related by:
\ba
{\mathcal L}_{\rm spinor}^{(1)}=-2\, {\mathcal L}_{\rm scalar}^{(1)} +\frac{1}{2} \left(\frac{g^2}{32\pi^2}\,  F_{\mu\nu}^a F_{\mu\nu}^a \right)\, \ln\left(\frac{m^2}{\mu^2}\right)\,.
\label{instrelation}
\ea
Here $\mu$ is a renormalization scale mass, and $\frac{g^2}{32\pi^2}\, F_{\mu\nu}^a F_{\mu\nu}^a$ is the number density of zero modes, with the total number of zero modes being the instanton number
\ba 
N_0=\frac{g^2}{32\pi^2}\int d^4x\, F_{\mu\nu}^a F_{\mu\nu}^a \quad ,
\label{nzero}
\ea
where $N_0=1$ for the single instanton field strength in (\ref{instf}). The implication of (\ref{instrelation}) is that one can compute the (somewhat simpler) scalar effective Lagrangian and deduce the corresponding spinor one. 

Using a combination of the solubility of the zero mass Green's functions and spectra, and spacetime-dependent mass regulators, 't Hooft computed the leading small mass behavior of the scalar effective action as \cite{thooft}
\ba
S_{\rm scalar}^{(1)}\sim \frac{1}{6}\ln (m\,\rho)+\alpha\Big(\frac{1}{2}\Big)+\dots
\label{thooftresult}
\ea
where the constant $\alpha\Big(\frac{1}{2}\Big)$ is given by
\ba
\alpha\Big(\frac{1}{2}\Big)=\frac{1}{6}\,(\gamma+\ln\,\pi)-\frac{\zeta^\prime(2)}{\pi^2}-\frac{17}{72}\approx  0.145873\ .
\label{alphahalf}
\ea
Carlitz and Creamer \cite{cc} computed the next order term in the small mass expansion, and Kwon {\it et al} \cite{kwon} the next-to-next order term, so that we now have the small mass expansion
\ba
S_{\rm scalar}^{(1)}\sim \frac{1}{6}\ln (m\,\rho)+\alpha\Big(\frac{1}{2}\Big)+\frac{1}{2}\,(m\,\rho)^2\left(\ln (m\,\rho)+\gamma-\ln\, 2\right) + O\left((m\,\rho)^4\right)\,.\nn\\
\label{smallmassinst}
\ea
Gross {\it et al} \cite{gpy} extended the 't Hooft style determinant calculation to the important case of finite temperature instantons, or {\it calorons}. Very recently \cite{diakonov}, the analogous computation has been performed for the more complicated situation of a background caloron field with nontrivial holonomy.

The large mass expansion of the one loop effective action in an instanton background can be computed using various techniques, ranging from the operator product expansion and the derivative expansion \cite{novikov,itep}, to the heat kernel \cite{kwon} and inverse mass expansion of \cite{fliegnermass}. The first two terms were computed in \cite{novikov}, and the next two in \cite{kwon}:
 \ba
S_{\rm scalar}^{(1)}\sim -\frac{1}{75}\,\frac{1}{(m\, \rho)^2}-\frac{17}{735}\,\frac{1}{(m\, \rho)^4}
+\frac{232}{2835}\,\frac{1}{(m\, \rho)^6}- \frac{7916}{148225}\,\frac{1}{(m\, \rho)^8}+\dots \ .\nn\\
\label{largemassinst}
\ea
A smooth interpolating approximation between the small and large mass limits (\ref{smallmassinst}) and (\ref{largemassinst}), based on a modified Pad\'e approximation, was proposed in \cite{kwon}. Such an interpolation is of physical interest for the extrapolation of lattice results, obtained at unphysically high quark masses, to lower physical quark masses. Details of this type of interpolation were studied for the soluble Heisenberg-Euler case in \cite{chiral}.

\section{Two-Loop Heisenberg-Euler effective Lagrangian}
\label{twoloop}
\renewcommand{\theequation}{4.\arabic{equation}}
\setcounter{equation}{0}

\begin{figure}
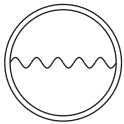

\centerline{\tlscbffig}
\vskip.5cm
\caption{The two loop diagram for the two loop effective Lagrangian. The double line refers to a propagator in the presence of the constant background field, while the wavy line represents the internal virtual photon.}
\label{twoloopfig}
\end{figure}

All the previous results have been for the one-loop effective Lagrangian, which neglects virtual lines inside the single scalar or spinor loop. In this Section I summarize what is known about the two-loop effective Lagrangian. In principle, the computation of the two-loop Heisenberg-Euler effective Lagrangian in QED is completely straightforward, as we only need to compute a single vacuum diagram (see Figure \ref{twoloopfig}) with an internal photon line and a single fermion (or scalar) loop, where these spinor or scalar propagators are in the presence of the background field. These background field propagators have been known in closed-form for a long time\cite{fock,schwinger,nambu}. 
\vskip 1cm

The problem is, however, not completely straightforward because at two-loop we need to perform mass renormalization in addition to charge renormalization. Nevertheless, in seminal work in the mid-1970's, Ritus found exact integral representations for the fully renormalized two loop Heisenberg-Euler effective Lagrangian in both spinor \cite{ritusspin} and scalar \cite{ritusscal} QED. These were impressive computations, but unfortunately the answers are rather complicated double-parameter integrals [see (\ref{2lspr}) and (\ref{2lscr}) below]. I first review the structure of Ritus's solutions and then discuss some recent simplifications that have been found.

\subsection{Two loop spinor QED Heisenberg-Euler effective Lagrangian}
\label{twoloopspinor}

Using Feynman gauge for the internal photon line in Figure \ref{twoloopfig}, the two-loop spinor QED effective Lagrangian is \cite{ritusspin,ginzburg}
\ba
{\mathcal L}^{(2)}_{\rm spinor}&=&i\,\frac{e^2}{2}\int d^4x^\prime \, {\rm tr}\left[\gamma_\mu G(x,x^\prime)\gamma^\mu G(x^\prime,x)\right]\, D(x-x^\prime)\nn\\
&=& i\,\frac{e^2}{2}\int \frac{d^4p}{(2\pi)^4} \int \frac{d^4q}{(2\pi)^4}  {\rm tr}\left[\gamma_\mu G(p)\gamma^\mu G(q)\right]\, D(p-q)
\label{2lsplag}
\ea
where the position and momentum space propagators are related by
\ba
G(x,x^\prime) =e^{i\Phi(x,x^\prime)} \, \int \frac{d^4p}{(2\pi)^4} \, e^{ip\cdot(x-x^\prime)}\, G(p)\ .
\label{ft}
\ea
Here $\Phi(x,x^\prime)$ is a gauge dependent phase factor coming from the line integral of the gauge  field from $x$ to $x^\prime$. For a constant background field strength $F_{\mu\nu}$, using the Fock-Schwinger gauge (\ref{fsg}), one finds
\cite{ritusspin,ginzburg}
\begin{eqnarray}
G_{\rm spinor}(x,x^\prime)&=& -i\,
\frac{e^{-i\frac{e}{2}x_\mu F^{\mu\nu}x^\prime_\nu}}{(4\pi)^2}\int_0^\infty
\frac{ds}{s^2}\,\left[ m-\frac{i}{2}\gamma_\mu (\beta^{\mu\nu}+eF^{\mu\nu})z_\nu\right] \times 
\nn\\
&&\hskip -1cm \exp\left\{-im^2
s-L(s)+\frac{i}{4}z_\mu\beta^{\mu\nu}(s) z_\nu+\frac{ie}{2} \sigma_{\mu\nu}F^{\mu\nu} s  \right\}
\label{focksp}
\end{eqnarray}
where we have defined $z_\mu\equiv x_\mu-x_\mu^\prime$, and $\beta_{\mu\nu}$ and $L$ are functions of the proper time $s$ involving trigonometric functions of the $4\times 4$ constant field strength matrix $F_{\mu\nu}$ as follows:
\ba
\beta_{\mu\nu}(s)\equiv  \left[eF\, \coth(eFs)\right]_{\mu\nu}\ , \qquad 
L(s)\equiv \frac{1}{2}{\rm tr}\,\ln\left(\frac{\sinh(eFs)}{eFs}\right)\,.
\label{bl}
\ea
Note that the contour of the proper time $s$ integration is defined to pass below the singularities on the real axis, in accordance with the $i\epsilon$ prescription $m^2\to m^2-i\epsilon$ \cite{ginzburg}.

It is now straightforward, but somewhat tedious, to do the Dirac traces and spacetime integrals in (\ref{2lsplag}) to obtain the following expression for the bare two-loop effective Lagrangian as an integral over the proper time parameters $s$ and $s^\prime$ of the two spinor propagators:
\ba
{\mathcal L}^{(2)}_{\rm spinor}&=&-\frac{ie^2}{128\pi^4} \int_0^\infty ds\int_0^\infty ds^\prime \frac{(e\eta)^2(e\epsilon)^2 \, e^{-im^2(s+s^\prime)}}{\sin(e\eta s)\sin(e\eta s^\prime)\sinh(e\epsilon s)\sinh(e\epsilon s^\prime)}\nn\\[1mm]
&& \hskip 4.1cm \times \Bigl\{ 4m^2 \left(S \, S^\prime +P\, P^\prime\right) I_0-i\, I\Bigr\}\,.
\label{2lspbare}
\ea
Here (following Ritus's conventions \cite{ritusspin}), the two eigenvalues of $F_{\mu\nu}$ are called $\eta$ and $\epsilon$, rather than $b$ and $a$ respectively, as used by Heisenberg and Euler (\ref{ab}). Thus, in a frame in which the magnetic and electric fields are parallel we can take $\eta=B$, and $\epsilon=E$.  The functions $S$, $S^\prime$, $P$, $P^\prime$, $I$ and $I_0$ appearing in the two-loop bare effective Lagrangian (\ref{2lspbare}) are defined as:
\ba
S(s)=\cos(e\eta s)\,\cosh(e\epsilon s) &,&\qquad
P(s)=\sin(e\eta s)\,\sinh(e\epsilon s)\ .\nn\\
S^\prime \equiv S(s^\prime) &,&\qquad
P^\prime \equiv P(s^\prime)\ ,\nn\\
I_0= \frac{1}{b-a}\ln\frac{b}{a} &,&\qquad
I= \frac{(q-p)}{(b-a)^2}\ln \frac{b}{a}-\left(\frac{\frac{q}{b}-\frac{p}{a}}{b-a}\right)\, ,\nn\\
a=e\eta \left(\cot(e\eta s)+\cot(e\eta s^\prime)\right) &,&\qquad
b=e\epsilon\left(\coth(e\epsilon s)+\coth(e\epsilon s^\prime)\right)\, ,\nn\\
p= \frac{2(e\eta)^2\cosh(e\epsilon(s-s^\prime))}{\sin(e\eta s)\, \sin(e\eta s^\prime)} &,&\qquad
q= \frac{2(e\epsilon)^2\cos(e\eta(s-s^\prime))}{\sinh(e\epsilon s)\, \sinh(e\epsilon s^\prime)}
\ .
\label{spii}
\ea
The bare effective Lagrangian (\ref{2lspbare}) has several different types of divergences. These must be regulated and then interpreted physically. Proper time regularization involves cutting off each proper time integral at a lower bound $s_0$, as was done by Ritus \cite{ritusspin}. The divergences can then be isolated systematically. First there is a subtraction of the field free case, so that ${\mathcal L}^{(2)}$ vanishes when the background fields vanish. Then there is charge and wavefunction renormalization, just as for the one-loop effective Lagrangian, which involves identifying a divergent term in ${\mathcal L}^{(2)}$ of the form of the zero-loop Maxwell Lagrangian. This is done simply by expanding the integrand to quadratic order in the fields $\eta$ and $\epsilon$. This divergence can be absorbed by redefining the electric charge and the fields as
\ba
e=e_0\, Z_3^{1/2}\quad , \quad \eta=\eta_0\, Z_3^{-1/2}\quad ,\quad \epsilon=\epsilon_0\, Z_3^{-1/2}
\label{2lcharge}
\ea
where $Z_3$ is the usual multiplicative renormalization factor \cite{itzyksonzuber}. 
With these redefinitions, we can express ${\mathcal L}^{(2)}$ in terms of the renormalized charges and fields. In fact, note that the combinations $e\eta$ and $e\epsilon$ are renormalization invariant. Third, and more complicated, is mass renormalization. Even after dealing with the charge renormalization divergence, there remains a logarithmic divergence associated with taking one or other of $s$ and $s^\prime$ to $0$, while keeping the other one fixed and nonzero. This divergence can be separated and one finds that it has precisely the correct form for the one-loop renormalization of the electron mass:
\ba
m_R^2&=&m_0^2+\delta m^2\ ,\nn\\
{\mathcal L}_R^{(1)}(m_R^2)&=&{\mathcal L}_R^{(1)}(m_0^2)+\delta m^2 \,\frac{\partial {\mathcal L}_R^{(1)}(m_0^2)}{\partial m_0^2}\ .
\label{2lmass}
\ea
Note that the second term in (\ref{2lmass}) is of order $\alpha^2$, because $\delta m^2$ and ${\mathcal L}^{(1)}_R$ are each of order $\alpha$. Thus it is natural to see this term arising in the computation of the two loop effective Lagrangian.
The identification of the mass renormalization is quite an involved manipulation, and we refer the reader to \cite{ritusspin,ginzburg} for details. In particular, finding the correct finite part of the mass shift $\delta m^2$ for the relevant renormalization scheme is not straightforward. It relies on an independent computation of  $\delta m^2$, or can be done {\it a posteriori} by studying the imaginary part of the two-loop effective Lagrangian in an electric background, as is discussed in more detail below in Section \ref{2lspinormag}. The final answer for the renormalized two loop effective Lagrangian is \cite{ritusspin}:
\ba 
{\mathcal L}_{\rm R,\,spinor}^{(2)}&=&-\frac{ie^2}{64\pi^4}\int_0^\infty ds \int_0^s ds^\prime \left\{ K(s,s^\prime)-\frac{K_0(s)}{s^\prime}\right\} \nn\\
&&-\frac{ie^2}{64\pi^4}\int_0^\infty ds K_0(s) \left\{ \ln (im^2 s)+\gamma-\frac{5}{6}\right\}
\label{2lspr}
\ea
where $\gamma\approx 0.577...$ is Euler's constant (note that Ritus calls Euler's constant $\ln \gamma$), and the functions $K(s,s^\prime)$ and $K_0(s)$ are 
\ba
K(s,s^\prime)&=& e^{-im^2(s+s^\prime)}\left\{ \frac{(e\eta)^2(e\epsilon)^2}{P\, P^\prime}\left[4m^2(S\, S^\prime+P\, P^\prime)I_0-iI\right] \right. \nn\\
&&\hskip -2cm \left.-\frac{1}{s s^\prime(s+s^\prime)}\left[4m^2-\frac{2i}{s+s^\prime}+\frac{e^2(\eta^2-\epsilon^2)}{3}\left(2m^2(s s^\prime-2s^2-2(s^\prime)^2)\right.\right.\right.\nn\\
&&\left.\left.\left. \hskip 5cm-\left(\frac{5is s^\prime}{s+s^\prime}\right)\right)\right]\right\}\nn\\
K_0(s)&=&e^{-im^2 s}\left(4m^2+i\frac{\partial}{\partial s}\right) \left[\frac{(e\eta)(e\epsilon)}{\tan(e\eta s)\, \tanh(e\epsilon s)}-\frac{1}{s^2}+\frac{e^2(\eta^2-\epsilon^2)}{3}\right]\,.\nn\\
\label{spkss}
\ea
Ritus's expression (\ref{2lspr}) for the renormalized two loop Heisenberg-Euler  effective Lagrangian is finite and expressed in terms of renormalized quantities. It is, nevertheless, a very complicated expression, and it is significantly more difficult to extract physical information from it, along the lines of the applications discussed in Section \ref{apps}. For example, one can extract the two-loop light-light scattering low energy effective Lagrangian by expanding Ritus's answer to quartic order in the fields. One finds \cite{ritusspin}
\ba
{\mathcal L}_{\rm spinor}^{(2)}\sim \frac{e^6}{64\pi^4 m^4}\left[\frac{16}{81}(\eta^2-\epsilon^2)^2+\frac{263}{162}(\eta\,\epsilon)^2\right]+\dots
\label{2lspweak}
\ea
which should be compared to the one loop result (\ref{lightlight}). Similarly, in the limit of a strong magnetic field, one finds \cite{ritusspin}
\ba
{\mathcal L}_{\rm spinor}^{(2)}\sim \frac{e^4 \eta^2}{128\pi^4}\left[\ln\left(\frac{e\eta}{\pi m^2}\right) +{\rm constant}\right]+\dots
\label{2lspstrong}
\ea
from which we can extract the two loop spinor QED $\beta$-function, as discussed below in Section \ref{betas}. This leading strong-field behavior (\ref{2lspstrong}) should be compared with (\ref{1lspmagleading}).  However, the general weak and strong field expansions of Ritus's result (\ref{2lspr}) are very cumbersome, due to the double integral structure and the complicated integrand. In Section \ref{twoloopselfdual} we shall see that for a self-dual background the corresponding expression simplifies dramatically and much more can be said about the weak and strong field behavior of ${\mathcal L}^{(2)}$.

\subsubsection{Two loop spinor case for magnetic background}
\label{2lspinormag}

If we specialize Ritus's result (\ref{2lspr}) to a constant magnetic background, by taking $\eta=B$ and $\epsilon=0$, then ${\mathcal L}^{(2)}$ simplifies a bit, but not much. Then one finds (after rotating the contours and changing integration variables)
\begin{eqnarray}
{\mathcal L}^{(2)}_{\rm R,\, spinor}&=& \frac{e^2m^4}{(4\pi)^4}
\left(\frac{eB}{m^2}\right)^2
\int_0^\infty \frac{ds}{s^3}\, e^{-m^2 s/(eB)}\, \int_0^1 du\,
\left[ L(s,u)-2s^2\right. \nonumber\\
&&\left. \hskip 4cm+\frac{6}{u(1-u)}\left(\frac{s^2}{{\rm
sinh}^2s} +s\, {\rm coth}s\right)\right] \nonumber\\
&& -12\, \frac{e^2 m^4}{(4\pi)^4} \frac{eB}{m^2} \int_0^\infty
\frac{ds}{s}\, e^{-m^2 s/(eB)}\, \left[ {\rm
coth}s-\frac{1}{s}-\frac{s}{3}\right]\nonumber\\
&&\hskip 3cm\times \left[
\frac{3}{2}-\gamma-\log\left(\frac{m^2 s}{eB}\right) +\frac{eB}{m^2 
s}\right]
\label{2lspmag}
\end{eqnarray}
where
\begin{eqnarray}
L(s,u)&=&s\,{\rm coth}s\left[ \frac{\log\left(\frac{u(1-u)}{G(u,s)}\right)}{
[u(1-u)-G(u,s)]^2}\, F_1+\frac{F_2}{G(u,s)[u(1-u)-G(u,s)]}\right.\nn\\
&&\left. \hskip 3cm +\frac{F_3}{u(1-u)[u(1-u)-G(u,s)]}\right]\,,\nonumber\\[1mm]
 G(u,s)&=&\frac{{\rm cosh}
s- {\rm cosh}((1-2u)s)}{2 s\, {\rm sinh}s}\,,\\[1mm]
F_1&=&4 s({\rm
coth}s-{\rm tanh}s) G(u,s) - 4u(1-u)\,,\nonumber\\
F_2&=&2(1-2u) \frac{{\rm
sinh}((1-2u)s)}{{\rm sinh}s} +s (8 {\rm tanh}s-4 {\rm coth}
s)G(u,s)-2\,,\nonumber\\
F_3&=&4u(1-u)-2(1-2u)\frac{{\rm
sinh}((1-2u)s)}{{\rm sinh}s} -4 s \, G(u,s) {\rm tanh} s+2\ .\nn
\label{lg}
\end{eqnarray}
While simpler than the general expression (\ref{2lspr}), this magnetic background two loop effective Lagrangian (\ref{2lspmag}) is still not particularly simple. By a direct expansion of the integrand one can compute a number of terms in the weak-field expansion \cite{csreview,dunsch}:
\ba
{\mathcal L}_{\rm spinor}^{(2)}&\sim&\frac{e^2 m^4}{(4\pi)^4}\frac{1}{81}\left[ 64 \left(\frac{eB}{m^2}\right)^4-\frac{1219}{25}\left(\frac{eB}{m^2}\right)^6+\frac{135308}{1225} \left(\frac{eB}{m^2}\right)^8\right.\nn\\[1mm]
&&\left. -\frac{791384}{1575}\left(\frac{eB}{m^2}\right)^{10}+\frac{8519287552}{2223375} \left(\frac{eB}{m^2}\right)^{12} - \dots\right]\,.
\label{2lspmabweak}
\ea
The coefficients of this expansion alternate in sign and grow factorially fast, and the Borel summation properties of the series have been studied in \cite{dunsch}. At present, no closed formula is known for these expansion coefficients.

For an electric background, ${\mathcal L}_{\rm spinor}^{(2)}$ is given by the replacement $B^2\to -E^2$. Lebedev and Ritus \cite{lebedev} analyzed the singularity structure of the integrand in (\ref{2lspmag}) to deduce that including the the two loop contribution led to:
\ba
{\rm Im}\left({\mathcal L}^{(1)}_{\rm sp}+ {\mathcal
L}^{(2)}_{\rm sp}\right)\hskip -4pt = \hskip -4pt
\frac{e^2 E^2}{8\pi^3}\sum_{k=1}^\infty \left[ 
\frac{1}{k^2}+\alpha\pi\left(-\frac{c_k}{\sqrt{\frac{eE}{m^2}}}+
\hskip -4pt 1\hskip -4pt +\hskip -4pt O\left(\sqrt{\frac{eE}{m^2}}\right) 
\right)\right] e^{-\frac{m^2 \pi k}{e E}}\nn\\
\label{2lsppair}
\ea
where $\alpha=\frac{e^2}{4\pi}$ is the fine structure constant, and the coefficients $c_k$ are \begin{eqnarray}
c_1=0 \ ; \qquad 
c_k=\frac{1}{2\sqrt{k}}\sum_{l=1}^{k-1}\frac{1}{\sqrt{l(k-l)}} 
\ ,\qquad  k\geq 2\ .
\label{rituslebedevck}
\end{eqnarray}
Note that the two loop contribution has the same exponential form as at one loop, but with an overall factor of $\alpha \pi$, and with a (perturbative) expansion in the electric field strength in the prefactor. Ritus and Lebedev were  able to extract information about the coefficient $c_k$ of the first term in this prefactor. 

For $k\geq 2$ the prefactor expansions in (\ref{2lsppair}) begin with terms that
are singular in the limit of vanishing field $E\to 0$,
which seems potentially unphysical.
In \cite{lebedev} a physically intuitive solution
was offered to this apparent dilemma. Their proposal is
that if one could take into account all
contributions from even higher loop orders to the prefactor
of the $k$-th exponential, then one would find them
to exponentiate in the following way,
\begin{eqnarray}
\hskip -25pt \Bigl[
\frac{1}{k^2}
+\alpha\pi \,K_k\left(\frac{eE}{m^2}\right)
+\ldots
\Bigr]
\exp\left[-\frac{k\pi m^2}{eE}\right]\!
=
\frac{1}{k^2}
\exp\left[-\frac{k\pi m^2_{\ast}(k,E)}{eE}\right].
\label{shiftm}
\end{eqnarray}
Thus, it should be possible to absorb their effect into a field-dependent shift of the electron mass.
Using just the lowest order coefficients (\ref{rituslebedevck}) in the
small $E$ expansion of $K_k(\frac{eE}{m^2})$, this
mass shift reads
\ba
m_{\ast}(k,E) &=&
m +\half\,\alpha\, k\, c_k\sqrt{eE}-\half\,\alpha k\,\frac{eE}{m}\ .
\label{mass}
\ea
As shown in \cite{ritusmass,lebedev} these
contributions to the mass shift have a simple meaning in
the coherent tunneling picture : the negative term can be interpreted as the total 
Coulomb energy of attraction between opposite charges in
a coherent group; the positive one, which is present only in the case $k\geq 2$,  represents the
energy of repulsion between like charges. The picture is consistent with the work of Affleck {\it et al} \cite{affleck} who showed that in the weak field limit, the dominant result of all higher loop corrections to the leading one-loop pair production rate (\ref{imag}) is a multiplicative factor of $e^{e^2/4}=e^{\alpha \pi}\approx 1+\alpha \pi+\dots\;$.

It is important to note that this interpretation of the mass shift requires the mass $m$ on the right hand side of (\ref{mass}) to be the {\sl physical} renormalized electron mass of the vacuum theory. Only in this case does the expansion of $K_k(\frac{eE}{m^2})$ have the form indicated in eqs. (\ref{2lsppair}) and (\ref{rituslebedevck}). It is an interesting corollary of the Lebedev-Ritus analysis that the physical electron mass can be recognized from an inspection of the two-loop effective Lagrangian alone, without ever considering the one-loop electron mass operator. Thus, one can fix even the finite part of the mass renormalization by demanding that the physical mass be that which appears in the leading exponential of the imaginary part of the effective Lagrangian in an electric field \cite{lebedev}.

Later, in a Borel analysis \cite{dunsch} of the large order perturbation theory behavior of the two loop result (\ref{2lspmag}),  a numerical value for the next coefficient was extracted, but only for the leading single-instanton term:
\begin{eqnarray}
 \hskip -20pt {\rm Im}\left({\mathcal L}^{(1)}_{\rm sp}+{\mathcal L}^{(2)}_{\rm sp}\right)\sim \frac{e^2 
E^2}{8\pi^3}
\left[ 1+\alpha\pi \left(1-0.44 \sqrt{\frac{eE}{m^2}}+\dots \right) 
\right]\, e^{-\frac{m^2 \pi }{e E}}\ .
\label{2lpairborel}
\end{eqnarray}
This numerical approach was not sensitive to higher order terms in the instanton expansion [the sum over $k$ in (\ref{2lsppair})], as such terms are exponentially suppressed.

Dittrich \cite{dittrichjpa}, and Dittrich and Reuter \cite{dr-qed} also computed this two-loop effective Lagrangian, for the case of a magnetic background, and their final answer was in a slightly different form from Ritus's result. The strong field limits agreed, but it was later found that the weak field expansions did not agree, and the cause was found to be a difference in the finite part of the mass renormalization \cite{fliegner3}. This latter approach \cite{fliegner3} used the worldline formalism and dimensional regularization, which was also applied to the case of a general constant electromagnetic background in both spinor and scalar QED in \cite{kors}. In Section \ref{massren} a much simpler, almost trivial, method of mass renormalization is presented, for the case of scalar QED.

\subsection{Two loop scalar QED Heisenberg-Euler effective Lagrangian}
\label{twoloopscalar}

Ritus also computed \cite{ritusscal} the two loop renormalized Heisenberg-Euler effective Lagrangian for scalar QED, generalizing Weisskopf's one loop result (\ref{hesc}). For scalar QED, the scalar propagator in a constant background field $F_{\mu\nu}$ is simpler in form than the corresponding spinor one (\ref{focksp}), but the vertices are more complicated as the momentum dependence of the vertex changes from $p_\mu$ to $(p_\mu-eA_\mu)$ in a background field. 
The scalar propagator in coordinate space is
\begin{eqnarray}
G_{\rm scalar}(x,x^\prime)= -i\,
\frac{e^{-i\frac{e}{2}x_\mu F^{\mu\nu}x^\prime_\nu}}{(4\pi)^2} \int_0^\infty
\frac{ds}{s^2}\,\exp\left\{-im^2
s-L(s)+\frac{i}{4}z\beta(s) z\right\}\nn\\
\label{focksc}
\end{eqnarray}
where $\beta_{\mu\nu}(s)$ and $L(s)$ are the same functions defined in (\ref{bl}) for the spinor propagator.
Using Feynman gauge for the internal photon propagator, the two loop effective action is \cite{ritusscal}
\ba
S^{(2)}_{\rm scalar}&=&ie^2\int d^4x\int d^4x^\prime D(x-x^\prime)\left\{ \langle x| \Pi_\mu G | x^\prime\rangle \langle x^\prime | \Pi_\mu G | x\rangle\right. \nn\\ 
&&\left. +\langle x| \Pi_\mu G \Pi_\mu | x^\prime\rangle \langle x| G | x^\prime\rangle +4 i \delta(x-x^\prime) \langle x| G |x^\prime\rangle  \right\}
\label{2lscaction}
\ea
where we have dropped a term $\langle x| \Pi_\mu G|x\rangle \langle x^\prime| \Pi_\mu G|x^\prime\rangle$, which vanishes. The delta function term in (\ref{2lscaction}) corresponds to the tadpole diagram, a new feature of scalar QED compared to spinor QED. Using the basic matrix elements \cite{ritusscal}
\ba
&&\langle x| \Pi_\mu G | x^\prime\rangle=\nn\\
&&  -i\,
\frac{e^{-i\frac{e}{2}x_\mu F^{\mu\nu}x^\prime_\nu}}{(4\pi)^2} \int_0^\infty
\frac{ds}{s^2}\frac{1}{2}\left[\beta_{\mu\nu}+eF_{\mu\nu}\right]z_\nu\exp\left\{-im^2
s-L(s)+\frac{i}{4}z\beta(s) z\right\}\,.\nn\\
&&\langle x| G\Pi_\mu | x^\prime\rangle=\nn\\
&&  -i\,
\frac{e^{-i\frac{e}{2}x_\mu F^{\mu\nu}x^\prime_\nu}}{(4\pi)^2} \int_0^\infty
\frac{ds}{s^2}\frac{1}{2}\left[\beta_{\mu\nu}-eF_{\mu\nu}\right]z_\nu\exp\left\{-im^2
s-L(s)+\frac{i}{4}z\beta(s) z\right\}\,,\nn\\
&&\langle x| \Pi_\mu G \Pi_\mu | x^\prime\rangle=-i\delta(x-x^\prime)-m^2 \langle x| G | x^\prime\rangle +\frac{e^2}{2}(z\, F\, F\,z)\langle x| G | x^\prime\rangle
\label{mes}
\ea
one finds that the bare, unrenormalized, two loop effective Lagrangian for scalar QED is
\ba
{\mathcal L}^{(2)}_{\rm scalar}&=&\frac{ie^2}{256\pi^4} \int_0^\infty ds\int_0^\infty ds^\prime \frac{(e\eta)^2(e\epsilon)^2 \, e^{-im^2(s+s^\prime)}\, \left[m^2 I_0+\frac{i}{2} I\right]}{\sin(e\eta s)\sin(e\eta s^\prime)\sinh(e\epsilon s)\sinh(e\epsilon s^\prime)}\nn\\
\label{2lscbare}
\ea
where the forms of the functions $I_0$, $I$, $a$ and $b$ are the same as in the spinor case (\ref{spii}), but the functions $p$ and $q$ entering in the definition of $I$ are changed from those in (\ref{spii}) to
\ba
p&=&2(e\eta)^2\left(\cot(e\eta s)\, \cot(e\eta s^\prime)+3\right)\,,\nn\\
q&=& 2(e\epsilon)^2\left(\coth(e\epsilon s)\, \coth(e\epsilon s^\prime)-3\right)\,.
\label{spiisc}
\ea
The bare effective Lagrangian (\ref{2lscbare}) also contains several divergences, which may be regulated using a proper time cut-off $s_0$. These divergences correspond, as in the spinor QED case, to charge and mass renormalization. Proceeding as in the spinor case, one finds \cite{ritusscal}
\ba 
{\mathcal L}_{\rm R,\,scalar}^{(2)}&=&-\frac{ie^2}{128\pi^4}\int_0^\infty ds \int_0^s ds^\prime \left\{ K(s,s^\prime)-\frac{K_0(s)}{s^\prime}\right\} \nn\\
&&-\frac{ie^2}{128\pi^4}\int_0^\infty ds K_0(s) \left\{ \ln (im^2 s)+\gamma-\frac{7}{6}\right\}
\label{2lscr}
\ea
where the functions $K(s,s^\prime)$ and $K_0(s)$ are 
\ba
K(s,s^\prime)&=& e^{-im^2(s+s^\prime)}\left\{ \frac{(e\eta)^2(e\epsilon)^2\left[-m^2I_0-\frac{i}{2}I\right] }{\sin(e\eta s)\sin(e\eta s^\prime)\sinh(e\epsilon s)\sinh(e\epsilon s^\prime)} +\right. \nn\\
&&\hskip -2cm \left.\frac{1}{s s^\prime(s+s^\prime)}\left[m^2+\frac{i}{s+s^\prime}+\frac{e^2(\eta^2-\epsilon^2)}{6}\left(m^2(s +s^\prime)^2-m^2 s s^\prime+\frac{11 i s s^\prime}{s+s^\prime}\right)\right]\right\}\nn\\
K_0(s)&=&e^{-im^2 s}\left(-m^2+\frac{i}{2}\frac{\partial}{\partial s}\right) \left[\frac{(e\eta)(e\epsilon)}{\sin(e\eta s)\, \sinh(e\epsilon s)}-\frac{1}{s^2}-\frac{e^2(\eta^2-\epsilon^2)}{6}\right]\,.\nn\\
\label{sckss}
\ea
Just as in the spinor case (\ref{2lspr}), the general weak and strong field expansions of Ritus's scalar QED result (\ref{2lscr}) are very cumbersome, due to the double integral structure and the complicated integrand. Also, note that it is very different from the spinor result, in the sense that the integrands are very different. We return to this comment later when we consider self-dual backgrounds in Section \ref{twoloopselfdual}.

Nevertheless, it is still possible to extract from (\ref{2lscr}) the leading weak-field and leading strong-field behavior \cite{ritusscal}. For weak fields, expanding the integrand to quartic order in the fields,
\ba
{\mathcal L}_{\rm scalar}^{(2)}\sim \frac{e^6}{64\pi^4 m^4}\left[\frac{275}{2592}(\eta^2-\epsilon^2)^2+\frac{4}{81}(\eta\,\epsilon)^2\right]+\dots\quad ,
\label{2lscweak}
\ea
which should be compared to the one loop result (\ref{1lscweakeg}). Similarly, in the limit of a strong magnetic field, 
\ba
{\mathcal L}_{\rm scalar}^{(2)}\sim \frac{e^4 \eta^2}{128\pi^4}\left[\ln\left(\frac{e\eta}{\pi m^2}\right) +{\rm constant}\right]+\dots
\label{2lscstrong}
\ea
from which we can extract the two loop spinor QED $\beta$-function, as discussed below in Section \ref{betas}. Indeed, the fact that (\ref{2lspstrong}) and (\ref{2lscstrong}) have the same leading coefficient tells us that the two loop $\beta$-function coefficients in spinor and scalar QED are equal [see (\ref{spbeta}) and (\ref{scbeta}) below].

\subsubsection{Two loop scalar case for magnetic background}
\label{2lscalarmag} 

The general expression (\ref{2lscr}) can be specialized to just a magnetic background by setting $\epsilon=0$ and $\eta=B$, and one obtains an expression similar in form to (\ref{2lspmag}) \cite{csreview}. This expression is, however, still not particularly simple. By a direct expansion of the integrand one can compute a number of terms in the weak-field expansion \cite{csreview}:
\ba
{\mathcal L}_{\rm scalar}^{(2)}&\sim&\frac{e^2 m^4}{(4\pi)^4}\frac{1}{81}\left[\frac{275}{8} \left(\frac{eB}{m^2}\right)^4-\frac{5159}{200}\left(\frac{eB}{m^2}\right)^6+\frac{2255019}{39200} \left(\frac{eB}{m^2}\right)^8\right.\nn\\
&&\left. -\frac{931061}{3600}\left(\frac{eB}{m^2}\right)^{10}+\frac{139252117469}{71148000} \left(\frac{eB}{m^2}\right)^{12} - \dots\right]\,.
\label{2lscmagweak}
\ea
The coefficients of this expansion alternate in sign and grow factorially fast, just as in the spinor case discussed in the previous section. At present, no closed formula is known for these expansion coefficients.

To conclude this section on the two loop effective Lagrangians, we briefly mention that the world-line  techniques have been applied to two loop effective Lagrangians in Yang-Mills theory \cite{sato}, where there are infrared divergences due to the masslessness of the gluon fields. Kuzenko and McArthur \cite{km} have computed the two-loop Heisenberg-Euler effective Lagrangian for ${\mathcal N}=2$ SUSY QED, finding a result expressed as a double parameter integral similar in structure to those in the spinor and scalar QED results of Ritus.

\section{\boldmath{$\beta$}-functions and the strong-field limit of effective Lagrangians}
\label{betas}
\renewcommand{\theequation}{5.\arabic{equation}}
\setcounter{equation}{0}

In quantum field theory there is a close connection between the
short-distance behavior of renormalized Green's functions and the
strong-field limit of associated quantities calculated using the
background field method. This correspondence
leads, for example, to a direct relation between the perturbative
$\beta$-function and the strong-field asymptotics of the effective
Lagrangian, as was first established by Coleman and Weinberg \cite{cw}. Ritus developed this relation for QED using
the renormalization group, with the assumption that the strong-field
limit of the renormalized effective Lagrangian is mass-independent
\cite{ritusspin,ritusscal,ginzburg}. Another, equivalent, derivation which
invokes the scale anomaly \cite{crewther,anomaly} in a massless
limit, has been given in various forms by many authors
\cite{ms,pagels,fujikawa,hansson,dittrichgies}. This derivation is reviewed below for QED in Section \ref{general}.

I stress that computing the $\beta$-function, which characterizes the scale dependence of the running coupling,  is a much simpler problem than finding the full renormalized effective Lagrangians,  (\ref{2lspr}) and (\ref{2lscr}), as one only needs the leading strong field behavior of ${\mathcal L}$.
In this approach, one uses the external field as a probing scale, instead of the external momentum of a self-energy diagram (which is the usual text-book approach). 
One adapts the Gell-Mann Low \cite{gellmannlow} renormalization group argument for $\Pi_{\mu\sigma}(q^2,\mu^2)$ at large $q^2$, to the effective Lagrangian ${\mathcal L}(eF,\mu^2)$ to relate the strong-field limit of ${\mathcal L}$ to the $\beta$-function coefficients. A similar idea was used at two loop by Shifman and Vainshtein to compute the QED $\beta$-function using the operator product expansion\cite{shifman}, and has recently been extended to the three-loop Yang-Mills $\beta$-function\cite{bornsen}. There is an immediate combinatorial advantage to this background field approach, as there are many fewer diagrams at a given order, since there are no external lines. Furthermore, each diagram has fewer vertices and propagators. The price, of course, is that the spinor (or scalar) propagators are not free ones, but are in the background field. 

Here I concentrate on QED, but these ideas generalize, and have been applied to, nonabelian theories. Begin by recalling some basic facts about the QED $\beta$-function \cite{itzyksonzuber,peskin}. The $\beta$-function encapsulates the scale dependence of running coupling, and is defined as
\begin{eqnarray}
\beta(a)=\frac{d\, a}{d\, \ln \mu^2}
\label{running}
\end{eqnarray}
where the natural expansion parameter turns out to be
$a\equiv \frac{\alpha}{4\pi}$, 
with $\alpha=\frac{e^2}{4\pi}$ being the fine structure constant. At present, the spinor QED $\beta$-function is known to four loop order. Beyond two loop order the $\beta$-function coefficients are scheme dependent; for the $MS$ scheme the four loop result is \cite{gorishnii}:
\begin{eqnarray}
\beta_{\rm MS}(a)= \frac{4}{3} a^2+4 a^3 -\frac{62}{9} a^4 -\left[\frac{5570}{243}+\frac{832}{9}\,\zeta(3)\right] a^5 +\dots\ .
\label{betams}
\end{eqnarray}
It is worth emphasizing how difficult these computations are. The two loop result, $4a^3$, was obtained in 1950 by Jost and Luttinger \cite{jost}, the three loop result by Rosner in 1966 \cite{rosner}, and the four loop result in 1991 by Gorishny et al \cite{gorishnii}. (One is tempted to extrapolate that the five loop result will require 36 years to compute!) The four loop result was computed using computer programs to organize and manipulate the many hundreds of Feynman diagrams involved. This complexity contributed to the earlier publication of a different result for the four loop term by the same group \cite{larin87}, this result being retracted four years later after errors were found in some of the computer subroutines.  This illustrates how difficult these calculations are, and motivates other ways to understand and perform the computations.  Nevertheless, the techniques \cite{gorishnii} used to produce this landmark result (\ref{betams}) for the four loop QED $\beta$-function coefficient have now been applied to QCD where we now know the $\beta$-function and the anomalous quark mass dimension to four loop order \cite{4loop}, a truly remarkable achievement.

Another $\beta$-function of interest is the so-called "quenched" $\beta$-function (or $F_1$ function), in which one only includes a single fermion (or scalar) loop, neglecting self energy corrections to internal photon lines. This $\beta$-function is scheme independent, essentially because the corresponding single-fermion-loop diagrams only diverge like a single logarithm. At four loop order, the quenched spinor QED $\beta$-function is
\begin{eqnarray}
\beta_Q(a)= \frac{4}{3} a^2+4 a^3 -2 a^4 -46 a^5 +\dots
\label{quenched}
\end{eqnarray}
which has the remarkable property that the coefficients are rational numbers. This is despite the fact that at intermediate stages of the computation individual diagrams, and groups of diagrams, contribute irrational terms such as $\zeta(3)$ and $\zeta(5)$, but these all cancel at the end when the quenched $\beta$-function is computed. This fascinating property was already noticed at the $3$-loop level by Rosner in 1966, and has been studied in detail over the years \cite{bender}. Despite impressive recent progress by Broadhurst {\it et al} \cite{broadhurst} who have identified the cancellation of such irrationals with a relation between Feynman diagrams and knots, a truly deep understanding of 
this behavior is still lacking. A promising approach is the worldline formalism \cite{csreview} which combines the relevant Feynman diagrams into a much more compact representation, making the large cancellations somewhat more `natural'. This has been studied at the two-loop level \cite{ss2}, but little is known beyond two loop at present. Even at a more basic level, we do not know how the perturbative series for the $\beta$-function behaves. It is presumably divergent, but we do not know if it is Borel summable. The first four signs of $-$, $-$, $+$, $+$, do not give us much basis for predicting the alternating or nonalternating behavior of the full series. These comments  motivate the exploration of various different ways to compute $\beta$-functions, and the strong field limit of a Heisenberg-Euler effective Lagrangian technique gives a particularly efficient way to do such a computation.

\subsection{General Argument}
\label{general}

In this section I review the general argument
\cite{ritusspin,ginzburg,ms,pagels,fujikawa,hansson,dittrichgies,zmb}
relating the strong-field asymptotic behavior of the effective
Lagrangian to the perturbative $\beta$ function. 
I present the argument for QED, but it is more  general.
Consider an abelian gauge field coupled to spinor or
scalar matter fields, which are either explicitly massless or which have
a well-defined massless limit. Then the trace anomaly for the
energy-momentum tensor states that \cite{crewther,anomaly}
\begin{eqnarray}
\langle \Theta^{\mu}_{\mu}\rangle = \frac{\beta(\bar{e})}{2\bar{e}}\,
\frac{e^2}{\bar{e}^2}\, (F_{\mu\nu})^2,
\label{anomaly}
\end{eqnarray}
where $\bar{e}$ is the running coupling, and $\beta(\bar{e})$ is the
$\beta$ function, defined below in (\ref{betadef}). The
expectation value of the energy-momentum tensor can also be related
to the effective Lagrangian for a constant background field strength
$F_{\mu\nu}$:
\begin{eqnarray}
\langle \Theta^{\mu\nu}\rangle =-\eta^{\mu\nu}\,{\mathcal L}_{\rm
eff}+2\frac{\partial {\mathcal L}_{\rm eff}}{\partial \eta_{\mu\nu}}\ .
\label{emom}
\end{eqnarray}
These two relations, (\ref{anomaly}) and (\ref{emom}), determine the
effective Lagrangian to be of the form
\begin{eqnarray}
{\mathcal L}_{\rm eff}=-\frac{1}{4}\, \frac{e^2}{\bar{e}^2(t)}\,
F_{\mu\nu}F^{\mu\nu}\, ,
\label{efflag}
\end{eqnarray}
where the "renormalization group time", $t$, is expressed in terms of the
scale set by the field strength, serving as the renormalization scale
parameter $\mu^2\sim e|F|$,
\begin{eqnarray}
t=\frac{1}{4}\,\ln\left(\frac{e^2 |F^2|}{\mu_0^4}\right)\,,
\label{tscale}
\end{eqnarray}
and $\mu_0$ denotes a fixed reference scale at which, for example, the
value of the coupling may be measured.  Note that in this
argument the field strength plays the role which is usually
associated with a momentum transfer $Q^2$. 
This already suggests at a very
basic level why the strong-field and short-distance limits are
related.

The $\beta$ function is defined in terms of the running of the
coupling as 
\begin{eqnarray}
\beta(\bar{e}(t))&=&\frac{d\bar{e}(t)}{dt}\ .
\label{betadef}
\end{eqnarray}
To see how this solution (\ref{efflag}) leads to an explicit
connection between the strong-field asymptotics of ${\mathcal L}_{\rm
eff}$ and the perturbative
$\beta$ function, note that  (\ref{betadef}) can also be expressed as
\begin{eqnarray}
t=\int^{\bar{e}(t)}_e \, \frac{de^\prime}{\beta(e^\prime)}\ ,
\label{invdef}
\end{eqnarray}
where $e\equiv \bar{e}(0)$. Making a perturbative expansion of the
$\beta$ function
\begin{eqnarray}
\beta(e)=\beta_1 e^3 +\beta_2 e^5 +\dots
\label{pertbeta}
\end{eqnarray}
the relation (\ref{invdef}) determines the running coupling,
$\bar{e}(t)$, in terms of $e$ as
\begin{eqnarray}
\frac{1}{\bar{e}^2(t)}=\frac{1}{e^2}-2\beta_1\, t-2 \beta_2 e^2\, t+
O(e^4 t^2)\ .
\label{ebar}
\end{eqnarray}
Inserting this into (\ref{efflag}), the strong-field asymptotics of the
effective Lagrangian is, to two-loop order, 
\begin{eqnarray}
{\mathcal L}_{\rm eff}\sim \frac{1}{16}\left(2\beta_1e^2+2 \beta_2
e^4+\dots\right) \, F_{\mu\nu}F^{\mu\nu}\,
\ln\left(\frac{e^2|F^2|}{\mu_0^4}\right),
\label{corr}
\end{eqnarray}
where, as is conventional, we have subtracted the classical Lagrangian,
$-\frac{1}{4}F_{\mu\nu}F^{\mu\nu}$, from ${\mathcal L}_{\rm eff}$. Note the appearance of the $\beta$-function coefficients $\beta_1$ and $\beta_2$ in the prefactor of the strong-field behavior of the ${\mathcal L}$. At higher loops the structure is slightly more complicated, but knowledge of the leading strong-field behavior up to a given loop order determines the $\beta$-function coefficients to that order, within the corresponding renormalization scheme \cite{ginzburg}.  

In order to illustrate this correspondence explicitly, we recall that
the QED
$\beta$ functions, for spinor and scalar QED, to two-loop order, are
\begin{eqnarray}
\beta_{\rm spinor}
  &=&\frac{e^3}{12\pi^2}+\frac{e^5}{64\pi^4}+\dots\ ,
  \label{spbeta} \\
\beta_{\rm scalar}
  &=&\frac{e^3}{48\pi^2}+\frac{e^5}{64\pi^4}+\dots\ .
  \label{scbeta}
\end{eqnarray}

\subsection{Explicit example: constant magnetic field background}
\label{twoloopmag}

Equation (\ref{corr}) gives a direct correspondence between the
perturbative $\beta$ function coefficients and the strong-field
behavior of the effective Lagrangian. We now compare this with some
explicit results where the effective Lagrangian is known. First, consider
the Euler-Heisenberg effective Lagrangian for a constant background
magnetic field, of strength $B$. 

At one loop, the leading strong-field asymptotics is [recall (\ref{1lspmagleading}) and (\ref{1lscmagleading})]:
\begin{eqnarray}
{\mathcal L}_{\rm spinor}^{(1)\,{\rm magnetic}}\sim 
\frac{e^2B^2}{24\pi^2}\ln\left(\frac{eB}{m^2}\right)+\dots\ ,
\label{1loopspinorasym}\\[2mm]
{\mathcal L}_{\rm scalar}^{(1)\,{\rm magnetic}}\sim 
\frac{e^2B^2}{96\pi^2}\ln\left(\frac{eB}{m^2}\right)+\dots\ .
\label{1loopscalarasym}
\end{eqnarray}
Noting that $-\frac{1}{4}F_{\mu\nu}F^{\mu\nu}=-\frac{1}{2}B^2$, and
comparing with the correspondence (\ref{corr}), we deduce that
$\beta_1^{\rm spinor}=\frac{1}{12\pi^2}$ and $\beta_1^{\rm
scalar}=\frac{1}{48\pi^2}$, in agreement with
the one-loop
$\beta$ function coefficients quoted in (\ref{spbeta}) and (\ref{scbeta}).

At two loop, the corresponding leading behaviors are deduced from Ritus's results. Recalling the leading strong field behaviors (\ref{2lspstrong}) and (\ref{2lscstrong}) we have
\begin{eqnarray}
{\mathcal L}_{\rm spinor}^{(2)\,{\rm magnetic}}\sim 
\frac{e^4 B^2}{128\pi^4}\ln\left(\frac{eB}{m^2}\right)+\dots\ ,
\label{2loopspinorasym}\\[2mm]
{\mathcal L}_{\rm scalar}^{(2)\,{\rm magnetic}}\sim 
\frac{e^4 B^2}{128\pi^4}\ln\left(\frac{eB}{m^2}\right)+\dots\ .
\label{2loopscalarasym}
\end{eqnarray}
Once again, comparing with the correspondence (\ref{corr}), 
we deduce that $\beta_2^{\rm spinor}=\beta_2^{\rm
scalar}=\frac{1}{64\pi^4}$, in agreement with 
the two-loop $\beta$ function coefficients quoted in (\ref{spbeta})
and (\ref{scbeta}). 

\section{Two loop Effective Lagrangian in a self-dual background}
\label{twoloopselfdual}
\renewcommand{\theequation}{6.\arabic{equation}}
\setcounter{equation}{0}

Ritus's results for the two loop Heisenberg-Euler effective Lagrangian are significantly more complicated in form than the one-loop results. This restricts somewhat the investigation of  two loop effects. For example, as discussed in Section \ref{2lspinormag},
we have only very limited information about the two loop generalization of the pair production formulae (\ref{electricinstanton}) and (\ref{scelectricinstanton}). In this section I review some recent developments in which the two loop Heisenberg-Euler effective Lagrangian has been shown to simplify dramatically when the background field is self-dual.

\subsection{Closed-form effective Lagrangians in a self-dual background}
\label{closedform}

Consider restricting the constant electromagnetic background to be self-dual
\ba
F_{\mu\nu}=\tilde{F}_{\mu\nu}
\label{selfdual}
\ea
where $\tilde{F}_{\mu\nu}\equiv \frac{1}{2}\epsilon_{\mu\nu\rho\sigma}F^{\rho\sigma}$ is the standard dual electromagnetic field strength. In QED, which is an abelian theory without instantons, such a classical background field is unphysical in Minkowski space as it means $E=i B$ (or $B=-i E$). However, Duff and Isham showed that such nonhermitian classical fields arise as vacuum to coherent state matrix elements of hermitian quantum field operators \cite{duffisham2}, which means that effective actions in such backgrounds are physically meaningful generating functions. We will discuss this further below in terms of helicity amplitudes. Also, Euclidean classical configurations can be used as probes of real QED properties, such as the $\beta$-function and the anomalous mass dimension.

When the constant background is restricted to be self-dual, the fully renormalized two-loop Heisenberg-Euler effective Lagrangian takes a remarkably simple closed-form, for both spinor and scalar QED  \cite{ds1,ds2,ds3}:
\begin{eqnarray}
{\mathcal L}_{\rm spinor}^{(2)}
&=&
-\alpha^2 \,\frac{f^2}{2\pi^2}\left[
3\,\xi^2 (\kappa)
-\xi'(\kappa)\right]\,,
\label{2lsp}\\[2mm]
{\mathcal L}_{\rm scalar}^{(2)}
&=&
\alpha^2 \,\frac{f^2}{4\pi^2}\left[
\frac{3}{2}\,\xi^2 (\kappa)
-\xi'(\kappa)\right]\,.
\label{2lsc}
\end{eqnarray}
Here $f$ is the field strength parameter, $\frac{1}{4}F_{\mu\nu}F^{\mu\nu}=f^2$, and $\kappa$ is the natural  dimensionless parameter
\begin{eqnarray}
\kappa\equiv \frac{m^2}{2e f}\ . 
\label{kappa}
\end{eqnarray}
The ubiquitous function $\xi(\kappa)$ is essentially the Euler digamma function $\psi(\kappa)=\frac{d}{d\kappa}\ln \Gamma(\kappa)$:
\begin{eqnarray}
\xi(\kappa)\equiv -\kappa\left(\psi(\kappa)-\ln(\kappa)+\frac{1}{2\kappa}\right)\,. 
\label{xi}
\end{eqnarray}
The subtraction of the first two terms of the asymptotic expansion \cite{bateman,abramowitz} of $\psi(\kappa)$ correspond to renormalization subtractions, as will be shown below in Section \ref{loops}.

It comes as something of a surprise that in this self-dual background both parameter integrals in each of the expressions (\ref{2lspr}) and (\ref{2lscr}) for the two loop effective Lagrangian can be done in closed form to produce the simple answers (\ref{2lsp}) and (\ref{2lsc}). If we simply take Ritus's expressions (\ref{2lspr}) and (\ref{2lscr}) and specialize to the self-dual case by taking $\eta=i\epsilon$, then it is not at all obvious that the resulting integrals can be reduced to such a simple closed form as in (\ref{2lsp}) and (\ref{2lsc}). 

It is also interesting to note that in such a self-dual background, the one-loop Heisenberg-Euler effective Lagrangians  (\ref{hesp}) and (\ref{hesc}) for spinor and scalar QED are also naturally expressed in terms of this same function $\xi(\kappa)$:
\begin{eqnarray}
{\mathcal L}_{\rm spinor}^{(1)}
&=&-\frac{e^2f^2}{8\pi^2}
\int_0^\infty \frac{dt}{t^3}\,e^{-2\kappa t} \left[t^2 \coth^2 t
-1-\frac{2t^2}{3}\right]\nn\\
&=& -\frac{e^2f^2}{8\pi^2} \left[-\frac{1}{12}\ln \kappa +\zeta'(-1)+\Xi(\kappa)\right]\,,
\label{1lspsd}\\[2mm]
{\mathcal L}_{\rm scalar}^{(1)}
&=&\frac{e^2f^2}{(4\pi)^2}
\int_0^\infty \frac{dt}{t^3}\,e^{-2\kappa t} \left[\frac{t^2}{\sinh^2 t }
-1+\frac{t^2}{3}\right]\nn\\
&=&\frac{e^2f^2}{(4\pi)^2} \left[-\frac{1}{12}\ln \kappa +\zeta'(-1)+\Xi(\kappa)\right]\,.
\label{1lscsd}
\end{eqnarray}
Here we have defined the function (which is closely related to the Barnes G function \cite{barnes,adamchik})
\ba
\Xi(x)\equiv 
-x\ln\Gamma(x)+\frac{x^2}{2}\ln x-\frac{x^2}{4}-\frac{x}{2}+\int_0^x
dy\,\ln\Gamma(y)
\label{defxi}
\ea
which has the property that $\Xi^\prime(\kappa)=\xi(\kappa)$.

The dramatic simplicity of the two loop results (\ref{2lsp}) and (\ref{2lsc})  raises three obvious  questions:

$\bullet$ why are these expressions so simple?

$\bullet$ why are these expressions so similar?

$\bullet$ why is the particular function $\xi(\kappa)$ so special?

\noindent The answers lie in the three-way relationship between self-duality, helicity and (quantum mechanical) supersymmetry.

\subsubsection{Self-duality and helicity}
\label{helicity}

Self-dual fields have definite helicity \cite{duffisham1,duffisham2}. Indeed, the self-duality condition (\ref{selfdual}) is just another way of writing the helicity projection:
\ba 
F_{\mu\nu}=\tilde{F}_{\mu\nu} \qquad \Leftrightarrow \qquad \sigma_{\mu\nu}F_{\mu\nu}\left(\frac{{\bf 1}+\gamma_5}{2}\right)=0\ .
\ea
For anti-self-dual fields the other helicity projection vanishes, so that the photon field has the opposite helicity. 

It is well-known that scattering amplitudes for external field lines with like helicities are particularly simple \cite{ttwu,mangano,bernhelicity}. Since the effective action for a self-dual field is the generating function for like-helicity amplitudes, it is consistent that the effective action in a self-dual background should be simple. However, almost all of these helicity amplitude results are for massless particles on internal lines (but see for example \cite{bernmorgan}), while here we see a generalization to massive particles. For example, in massless QCD, the tree level amplitudes with all external helicities alike, or all but one alike vanish identically \cite{mangano}:
\begin{eqnarray}
M[p_1+;p_2+;\dots; p_n+]=0 \ , \quad 
M[p_1-;p_2+;\dots; p_n+]=0\ .
\label{treevanish}
\end{eqnarray}
For massless QED at one loop, amplitudes with all external helicities alike vanish except for the case of four external lines, and these are simple \cite{mahlon}
\begin{eqnarray}
M[p_1+;p_2+;\dots; p_n+]&=&0 \ ,\qquad n\neq 4\nonumber\\[1mm]
M[p_1+;p_2+;p_3+;p_4+]&=&\frac{ie^4}{2\pi^2}\frac{\langle 
12\rangle^*\langle 34\rangle^*}{\langle 12\rangle\langle 
34\rangle}\ .
\label{1lvanish}
\ea
Here the RHS is written in terms of a convenient  spinor-helicity basis \cite{ttwu,mangano,bernhelicity} for the momenta: $\langle ij\rangle\equiv \bar{u}_-(p_i)u_+(p_j)$, with $u_\pm(p)$ being fundamental spinors of definite helicity. Similarly, when all but one external helicity is alike, the amplitudes vanish except for four external lines, where it also has a simple expression
\ba
M[p_1-;p_2+;\dots; p_n+]&=&0  \ ,\qquad n\neq 4\nonumber\\[1mm]
M[p_1-;p_2+;p_3+;p_4+]&=&\frac{ie^4}{2\pi^2}\frac{\langle 
12\rangle\langle 34\rangle^* \langle 24\rangle^*}{\langle 
12\rangle^*\langle 34\rangle\langle 24\rangle}\ .
\label{1lmvanish}
\end{eqnarray}
These massless helicity amplitudes, with $4$ external legs, have been extended to two-loop level \cite{2loophelicity}, where they also exhibit remarkable simplicities.

The results (\ref{2lsp}) and (\ref{2lsc}) provide information about amplitudes with any number of external lines, with all helicities being equal. Massive QED turns out to be rather different from massless QED, but still shows dramatic simplicity when considered in a helicity basis. For example, a new prediction \cite{ds2} is that for massive QED the low energy limit of the one-loop and  two-loop amplitudes with all external helicities $+$ behave as:
\begin{eqnarray}
\Gamma^{(1)}[k_1,\epsilon_1^+;k_2,\epsilon_2^+;
\dots;k_N,\epsilon_N^+]&=&-2\, \frac{(2e)^N}{(4\pi)^2 m^{2N-4}}\,
c^{(1)}_{N/2}\,\, \chi_N\label{1lhel}
\\[2mm]
\Gamma^{(2)}[k_1,\epsilon_1^+;k_2,\epsilon_2^+;
\dots;k_N,\epsilon_N^+]&=&-2 \alpha\pi\,\frac{(2e)^N}{(4\pi)^2 m^{2N-4}}\,
c^{(2)}_{N/2}\,\, \chi_N\ .
\label{2lhel}
\end{eqnarray}
where the momentum dependence of the low energy limit is naturally expressed in terms of symmetrized helicity basis elements $\chi_N$:
\begin{eqnarray}
\chi_N\equiv\frac{(N/2)!}{2^{N/2}} \, \left\{[12]^2[34]^2
\cdots[(N-1)N]^2+{\rm all~perms}\right\}
\label{chin}
\end{eqnarray}
where $[ij]=\langle k_i^+|k_j^-\rangle$ denotes a basic spinor product \cite{ttwu,mangano,bernhelicity}.
The numerical coefficients $c_n^{(1)}$ and $c_n^{(2)}$ appearing in (\ref{1lhel}) and (\ref{2lhel})  come from the weak field expansion of the one-loop and two-loop effective Lagrangians for a self-dual background \cite{ds2}:
\begin{eqnarray}
c_n^{(1)}&=&- \frac{{\mathcal B}_{2n}}{2n(2n-2)}\nonumber\\
c^{(2)}_n &=&
\frac{1}{(2\pi)^2}\left\{
\frac{2n-3}{2n-2}\,{\mathcal B}_{2n-2}
+3 \sum_{k=1}^{n-1}
\frac{{\mathcal B}_{2k}}{2k}
\frac{{\mathcal B}_{2n-2k}}{(2n-2k)} \right\}\,. 
\label{cns}
\end{eqnarray}
In massive QED, the $[+\, +\,+\, \dots +]$ helicity amplitudes for $N\geq 6$ do not vanish at one or two loop, in contrast to the one loop case (\ref{1lvanish}) for massless QED. On the other hand, the amplitudes with all but one helicity alike, the $[-\, +\,+\, \dots +]$ amplitudes, vanish in the low-energy limit in massive QED, a result that is true to {\it all} orders \cite{ds2}. These results have since been generalized  to all helicity combinations (up to 10 external legs) by making an explicit helicity expansion of Ritus's general result (\ref{2lspr}). In fact, in the low energy limit all ``odd -'' amplitudes vanish, to all orders \cite{louise}.

\subsubsection{Self-duality and quantum mechanical supersymmetry}
\label{qmsusy}

Another consequence of the self-duality of the background is that the corresponding Dirac operator has a quantum mechanical supersymmetry. That is, apart from zero modes, the Dirac operator has the same spectrum (but with a multiplicity of 4) as the corresponding scalar Klein-Gordon operator\cite{thooft,jackiwrebbi,dadda}. At one-loop this implies [compare to (\ref{instrelation})]
\ba
{\mathcal L}^{(1)}_{\rm spinor}=-2{\mathcal L}^{(1)}_{\rm scalar}+\frac{1}{2} 
\left(\frac{ef}{2\pi}\right)^2\, \ln\left(\frac{m^2}{\mu^2}\right)
\label{1lsdconnection}
\ea
where $N_0=\left(\frac{ef}{2\pi}\right)^2$ is the zero mode number density. The logarithmic term in (\ref{1lsdconnection}) corresponds to the zero mode contribution, as can be verified directly from the one loop Heisenberg-Euler results restricted to a self-dual background \cite{zmb}. Renormalizing on-shell ({\it i.e.}, $\mu^2=m^2$), we find that the spinor and scalar effective Lagrangians (\ref{1lspsd}) and (\ref{1lscsd}) are proportional to one another for a self-dual background, in such a way that the supersymmetric combination vanishes:
\ba
{\mathcal L}^{(1)}_{\rm spinor}+2{\mathcal L}^{(1)}_{\rm scalar}=0
\label{susyvanish}
\ea
These properties are familiar from one loop instanton physics \cite{thooft,jackiwrebbi,duffisham1}.

Now consider the implications of self-duality of the background at the two loop level. The two-loop the effective action is not simply a log determinant, so the situation is more complicated and we do not expect a SUSY relation like (\ref{susyvanish}) to hold. Nevertheless, the quantum mechanical SUSY of the Dirac operator relates the spinor propagator to the scalar propagator via simple helicity projections \cite{leeleepac}, 
\begin{eqnarray}
{\mathcal S}=- \left({D \hskip -7pt / -m}\right) G
\left(\frac{1+\gamma_5}{2}\right) - G{D \hskip -7pt /
}\left(\frac{1-\gamma_5}{2}\right)+\frac{\left(1+ {D \hskip
-7pt /} G {D \hskip -7pt
/}\right)}{m}\left(\frac{1-\gamma_5}{2}\right)\,.
\nn\\
\label{susyprops}
\end{eqnarray}
This has the consequence that simply doing the Dirac traces in the spinor two loop effective Lagrangian (\ref{2lsplag}), one finds that  it can be written as the sum of two terms involving matrix elements of the scalar propagator, and moreover these are the same two matrix elements of the scalar propagator that appear in the scalar QED effective Lagrangian, but with different numerical coefficients \cite{zmb} :
\begin{eqnarray}
{\mathcal L}^{(2)} &=& \frac{\alpha}{\pi}\int 
 \frac{d^4x^\prime}
{(x-x^\prime)^2}\left[
A\langle x |D_\mu G|x^\prime\rangle \langle x^\prime |D_\mu G|x\rangle 
+ B \langle x | G|x^\prime\rangle \langle x^\prime |D_\mu G D_\mu|x\rangle
\right]\,,\nn\\[2mm]
&&\hskip -1cm {\rm spinor\,\,\, QED}: A=-4\ ,\, B=8\quad ; \quad
{\rm scalar\,\,\, QED}: A=-1\ ,\, B=-1
\label{2lcoeffs}
\end{eqnarray}
 This structure explains why the two loop answers (\ref{2lsp}) and (\ref{2lsc}) for spinor and scalar QED have such a similar form, involving just two terms with different numerical coefficients.

\subsubsection{Propagators in self-dual background}
\label{propagators}

A more prosaic reason for the simplicity of the two-loop expressions (\ref{2lsp}) and (\ref{2lsc}) is that for a self-dual field the square of the matrix $F_{\mu\nu}$ is proportional to the identity matrix:
\ba
F_{\mu\nu}F_{\nu\rho}=-f^2 \delta_{\mu\rho}\ .
\label{sdid}
\ea
This dramatically simplifies the propagators of spinors or scalars in the background field. For example, for a scalar particle, the Lorentz matrix structure in (\ref{focksc}) disappears, leaving (neglecting the unimportant phase factor)
\begin{eqnarray}
&&\hskip -0.6cm G_{\rm scalar}(x,x^\prime)=\left(\frac{ef}{4\pi}\right)^2\int_0^\infty
\frac{dt}{\sinh^2(e f t)}\,e^{ -m^2 t -\frac{ef}{4}(x-x^\prime)^2 
\coth(e f t)}\ ,
\label{xmprop}\\[2mm]
&&\hskip -0.6cm G_{\rm scalar}(p)=\int_0^\infty \frac{dt}{\cosh^2(e f t)}\,e^{ -m^2 t
-\frac{p^2}{ef} \tanh(e f t)}\  .
\label{pmprop}
\end{eqnarray}
Note that these propagators are functions of $(x-x^\prime)^2$ and $p^2$ respectively, which greatly simplifies computations involving the propagators. Also note the duality between the momentum and coordinate space propagators -- the coordinate space form is well suited to a weak-field expansion, while the momentum space form is well suited to a strong-field expansion.
The propa\-gators satisfy simple differential equations
\ba
(p^2+m^2)G(p) &=& 1+\left(\frac{ef}{2}\right)^2\frac{\partial^2 G(p)}{\partial p_\mu^2}\ ,\label{mompropde}\\[1mm]
(-\partial_\mu^2+m^2)G(x,x^\prime) &=& \delta(x-x^\prime)-\left(\frac{ef}{2}\right)^2(x-x^\prime)^2 G(x,x^\prime)\ .\label{spacespropde}
\ea
The simplification of the self-dual propagators is even more dramatic for massless particles, where for scalars
\begin{eqnarray}
G_{\rm scalar}(x,x^\prime)=\frac{e^{
-\frac{ef}{4}(x-x^\prime)^2 }}{4\pi^2(x-x^\prime)^2}\ , \quad G_{\rm scalar}(p)&=&\frac{1-e^{  -\frac{p^2}{ef} }}{p^2}\ .
\label{massless}
\end{eqnarray}

\subsection{Weak field expansions in self-dual case}
\label{sdweak}

The weak field expansions are trivial to derive because we have simple closed-form expressions for the two loop effective Lagrangians in terms of the function $\xi(\kappa)$ in (\ref{xi}), whose analytic properties are completely understood, since it is essentially the digamma function.
To derive the weak field expansion, first recall the definition of $\kappa$ as $\kappa=\frac{m^2}{2ef}$, so that the weak field limit is the large $\kappa$ limit. This weak-field expansion follows directly from (\ref{xi}), together with the asymptotic (large $x$) expansion of the digamma
function \cite{abramowitz}:
\begin{eqnarray}
\psi(x)\sim \ln x-\frac{1}{2x}-\sum_{k=1}^\infty \frac{{\mathcal B}_{2k}}{2k
\,x^{2k}} 
\label{psiasymptotic}
\end{eqnarray}
where ${\mathcal B}_{2k}$ are the Bernoulli numbers. Then 
the weak-field (large $\kappa$) expansion of (\ref{2lsp}) is
\ba
{\mathcal L}^{(2)}_{\rm spinor}(\kappa)\sim -2 \alpha\pi\,
\frac{m^4}{(4\pi)^2}\, 
\sum_{n=2}^{\infty} c^{(2)\,{\rm spinor}}_n \frac{1}{\kappa^{2n}}
\label{2lspweaksd}
\end{eqnarray}
where the two-loop expansion coefficients are (for $n\geq 2$):
\begin{eqnarray}
c^{(2)\,{\rm spinor}}_n =
\frac{1}{(2\pi)^2}\biggl\lbrace
\frac{2n-3}{2n-2}\,{\mathcal B}_{2n-2}
+3\sum_{k=1}^{n-1}
\frac{{\mathcal B}_{2k}}{2k}
\frac{{\mathcal B}_{2n-2k}}{(2n-2k)}
\biggr\rbrace\,.
\label{cn2spinor}
\ea
The leading term in the weak-field expansion (\ref{2lspweaksd}) is 
\ba
{\mathcal L}^{(2)}_{\rm spinor}(\kappa)\sim -\frac{e^6 f^4}{m^4 \pi^4} \frac{5}{384}
\label{2lspweakleading}
\ea
agreeing with Ritus's result (\ref{2lspweak}) with the self-dual replacement $\eta=i\epsilon=f$.

Similarly, for scalar QED the results are almost the same, but with different coefficients:
\ba
{\mathcal L}^{(2)}_{\rm scalar}(\kappa)\sim \alpha\pi\,
\frac{m^4}{(4\pi)^2}\, 
\sum_{n=2}^{\infty} c^{(2)\,{\rm scalar}}_n \frac{1}{\kappa^{2n}}
\label{2lscweaksd}
\end{eqnarray}
where the two-loop expansion coefficients are (for $n\geq 2$):
\begin{eqnarray}
c^{(2)\,{\rm scalar}}_n =
\frac{1}{(2\pi)^2}\biggl\lbrace
\frac{2n-3}{2n-2}\,{\mathcal B}_{2n-2}
+\frac{3}{2}\sum_{k=1}^{n-1}
\frac{{\mathcal B}_{2k}}{ 2k}
\frac{{\mathcal B}_{2n-2k}}{(2n-2k)}
\biggr\rbrace\,.
\label{cn2scalar}
\ea
The leading term in the weak-field expansion (\ref{2lspweaksd}) is 
\ba
{\mathcal L}^{(2)}_{\rm scalar}(\kappa)\sim \frac{e^6 f^4}{m^4 \pi^4} \frac{3}{512}
\label{2lscweakleading}
\ea
agreeing with Ritus's result (\ref{2lscweak}) with the self-dual replacement $\eta=i\epsilon=f$.

Another advantage of having a closed-form expression for the two loop effective Lagrangians is that it becomes simple to study the nonperturbative contributions to the imaginary part. For example, real $\kappa$ is like the magnetic $B$ field case, while imaginary $\kappa$ is like the electric $E$ field case \cite{ds3}. 
When $\kappa$ is imaginary, using the properties of the digamma function we find the exact instanton expansions (here for scalar QED)
\begin{eqnarray}
{\rm Im}{\mathcal L}^{(1)}&=&\frac{m^4}{(4\pi)^3}\frac{1}{\kappa^2} 
\sum_{k=1}^\infty \left(\frac{2\pi\kappa}{k} +\frac{1}{k^2}\right) \, 
e^{-2\pi k \kappa}\ ,\label{sdimag}\\
 \nonumber\\
{\rm Im}{\mathcal L}^{(2)}&=&\alpha\pi\, 
\frac{m^4}{(4\pi)^3}\frac{1}{\kappa^2} \sum_{k=1}^\infty 
\left(2\pi\kappa k -1-3\kappa^2\sum_{l=1}^\infty \frac{(-1)^l {\mathcal 
B}_{2l}}{2l \kappa^{2l}}\right) \, e^{-2\pi k \kappa}\ .\nonumber
\end{eqnarray}
Note that the leading terms in the weak field (large $\kappa$) limit differ by a factor of $\alpha\pi$, just as in the electric field case discussed in Section \ref{2lspinormag}. Also, observe that the two loop imaginary part has a prefactor of the same form as that proposed by Lebedev and Ritus in (\ref{2lsppair}), but in this self-dual case we have a closed-form expression for all terms in this prefactor series. We see moreover that this prefactor series is an asymptotic series. While this self-dual background is unphysical in Minkowski space, these results are suggestive of what to expect for a real electric field background at the two loop level.

\subsection{Strong field limits in self-dual case}
\label{sdstrong}

The strong-field (small $\kappa$) expansion follows directly from (\ref{xi}), together with the small argument expansion of the digamma function \cite{abramowitz}:
\begin{eqnarray}
\psi(x)\sim -\frac{1}{x}-\gamma +\sum_{k=2}^\infty (-1)^k \zeta(k) x^{k-1}\ .
\label{psitaylor}
\end{eqnarray}
This leads to the following strong-field behavior for spinor QED:
\begin{eqnarray}
{\mathcal L}_{\rm spinor}^{(2)}&=& -\frac{2 \alpha m^4}{(4\pi)^3}\,\frac{1}{\kappa^2}
\left[ 3\left(\frac{1}{2}+\gamma \kappa +\kappa \ln \kappa\right)^2 +3\left( \sum_{l=2}^\infty (-1)^l \zeta(l) \kappa^{l-1}\right)^2\right.\nn\\
&&\left. -6\left(\frac{1}{2}+\gamma \kappa +\kappa \ln \kappa\right)\left( \sum_{l=2}^\infty (-1)^l \zeta(l) \kappa^{l-1}\right)\right.\nn\\
&&\left.-\left(1+\gamma+\ln \kappa\right) +\sum_{l=2}^\infty (-1)^l l\, \zeta(l) \kappa^{l-1}\right]\,.
\label{2lspstrongsd}
\end{eqnarray}
The leading strong-field behavior for spinor QED is
\begin{eqnarray}
{\mathcal L}_{\rm spinor}^{(2)} \sim -\frac{e^4 f^2}{32\pi^4} \ln \left(\frac{2e f}{m^2}\right)
\,.
\label{2lspstrongleading}
\end{eqnarray}
Similarly, for scalar QED we find
\begin{eqnarray}
{\mathcal L}_{\rm scalar}^{(2)}&=& \frac{\alpha m^4}{(4\pi)^3}\,\frac{1}{\kappa^2}
\left[ \frac{3}{2}\left(\frac{1}{2}+\gamma \kappa +\kappa \ln \kappa\right)^2 +\frac{3}{2}\left( \sum_{l=2}^\infty (-1)^l \zeta(l) \kappa^{l-1}\right)^2\right.\nn\\
&&\left. -3\left(\frac{1}{2}+\gamma \kappa +\kappa \ln \kappa\right)\left( \sum_{l=2}^\infty (-1)^l \zeta(l) \kappa^{l-1}\right)\right.\nn\\
&&\left.-\left(1+\gamma+\ln \kappa\right) +\sum_{l=2}^\infty (-1)^l l\, \zeta(l) \kappa^{l-1}\right]\,.
\label{2lscstrongsd}
\end{eqnarray}
The leading strong-field behavior for scalar QED is
\begin{eqnarray}
{\mathcal L}_{\rm scalar}^{(2)} \sim \frac{e^4 f^2}{64\pi^4} \ln \left(\frac{2e f}{m^2}\right)
\,.
\label{2lscstrongleading}
\end{eqnarray}

\subsubsection{Strong-field limit and $\beta$-function in self-dual case}
\label{sdbeta}

Since in some sense the self-dual background is the simplest type of background, as it leads to the simplest propagators, this suggests it should be a useful probe for computing $\beta$-functions. From the one-loop Heisenberg-Euler expressions (\ref{1lspsd}) and (\ref{1lscsd}), and the closed-form two-loop expressions (\ref{2lsp}, \ref{2lsc}) we find
\begin{eqnarray}
{\mathcal L}^{(1)}_{\rm scalar}\sim 
\frac{e^2}{48\pi^2}f^2 \ln\left(\frac{ef}{m^2}\right) \, , \qquad
{\mathcal L}^{(1)}_{\rm spinor}&\sim& -\frac{e^2}{24\pi^2}f^2 \ln\left(\frac{ef}{m^2}\right)\,,
\nonumber\\[2mm]
{\mathcal L}^{(2)}_{\rm scalar}\sim 
\frac{e^4}{64\pi^4}f^2 \ln\left(\frac{ef}{m^2}\right)
\, , \qquad
{\mathcal L}^{(2)}_{\rm spinor}&\sim& -\frac{e^4}{32\pi^4}f^2 \ln\left(\frac{ef}{m^2}\right)\,.
\label{2lsdleading}
\end{eqnarray}
Recalling the argument in Section \ref{general} relating the coefficient of the leading strong-field behavior of the effective Lagrangian to the $\beta$-function coefficients, we see that the scalar QED results in (\ref{2lsdleading}) give the correct one and two loop scalar QED $\beta$-function coefficients [compare with (\ref{corr}) and (\ref{scbeta}), and recall that the Maxwell Lagrangian is $f^2$]. However, for spinor QED comparing with (\ref{spbeta}) we see that the correspondence appears to fail. The reason is that in a self-dual background spinor QED has zero modes, and these preclude the massless limit on which is based the argument in Section \ref{general} relating the $\beta$-function and the strong field limit of ${\mathcal L}$. This argument must be modified to account for the zero modes\cite{zmb}.  At one loop the mismatch is due to the infrared divergence of the unrenormalized ${\mathcal L}_{\rm spinor}$, as is familiar from instanton physics\cite{thooft,jackiwrebbi}. But at two loop there is no IR divergence in the bare Lagrangian and the zero modes actually enter via the mass renormalization\cite{zmb}.

\section{Diagrammatic approach to two loop effective Lagrangians}
\label{loops}
\renewcommand{\theequation}{7.\arabic{equation}}
\setcounter{equation}{0}

It is natural to ask if the remarkable simplifications of the two loop results for a self-dual background might extend to even higher loops. To go beyond two loops one should take advantage of the great progress that has been made recently in understanding the structure of higher-loop quantum field theory (without background fields). The general strategy is to manipulate diagrams to reduce the number to a much smaller set of so-called "master diagrams" which need to be computed. This has led, for example, to many new two-loop results for QCD scattering amplitudes \cite{glover}. Kreimer \cite{dirk} and collaborators have discovered an elegant Hopf algebra structure underlying the seemingly messy jumble of ordinary Feynman diagram perturbation theory. I conclude this review with some brief comments and speculations about how these techniques can be extended to incorporate background fields \cite{gvdloops}.

\subsection{Why $\xi$ and $\xi^\prime$ ?}
\label{whyxi}

 The first question to address is why are the simple expressions (\ref{2lsp}) and (\ref{2lsc}) for the two loop Heisenberg-Euler effective Lagrangians in a self-dual background expressed in terms of the function $\xi(\kappa)$ and its derivative, where recall that $\xi$ was defined in (\ref{xi}) as essentially the Euler digamma function. The first hint comes from the following facts that for a self-dual background the scalar propagator loops, evaluated using (\ref{pmprop}), turn out to be simply related to the $\xi(\kappa)$ function:
\ba
\mbox{\olscbf}\quad -\quad\mbox{\olscfr} \quad&\equiv &\int \frac{d^4p}{(2\pi)^4}\left[G(p)-G_0(p)\right]=-\frac{m^2}{(4\pi)^2} \,\frac{ \xi(\kappa)}{\kappa}\ ,
\label{xiint}\\[2mm]
\mbox{\olscbfv}\quad -\quad\mbox{\olscfrv} \quad &\equiv &\int \frac{d^4p}{(2\pi)^4}\left[(G(p))^2-(G_0(p))^2\right]= \frac{\xi^\prime(\kappa)}{(4\pi)^2}\ .
\label{xiprint}
\ea
Here the double lines refer to scalar propagators in the self-dual background and the single line is the free scalar propagator, while the dot on a propagator refers to the propagator being squared. Thus, $\xi$ and $\xi^\prime$ are natural one loop traces for the self-dual background.

Given (\ref{xiint}) and (\ref{xiprint}), we can write the closed-form expressions (\ref{2lsp}) and (\ref{2lsc})  for the two loop effective Lagrangians in diagrammatic form:
\ba
{\rm spinor\,\,QED} &:&\nn\\
\quad\Bigg[~~\mbox{\tlspbf}\quad - \hskip.3cm \mbox{\tlscfr}~~\Bigg]&=&-6\,e^2\, 
\Bigg[~~\mbox{\olscbf}\quad - \hskip.3cm \mbox{\olscfr} ~~\Bigg]^2 +\frac{(ef)^2}{2\pi^2} \Bigg[~~\mbox{\olscbfv}\quad - \hskip.3cm  \mbox{\olscfrv}~~\Bigg]\,,\nn\\
\label{2lspdig}\\
{\rm scalar\,\,QED}&:&\nn\\
 \quad\Bigg[~~\mbox{\tlscbf}\quad - \quad \mbox{\tlscfr}~~\Bigg]&=&\frac{3}{2}\,e^2\,
 \Bigg[~~\mbox{\olscbf}\quad -  \hskip.3cm \mbox{\olscfr}~~\Bigg]^2 -\frac{(ef)^2}{4\pi^2}  \Bigg[~~\mbox{\olscbfv}\quad - \hskip.3cm  \mbox{\olscfrv}~~\Bigg]\,.\nn\\
\label{2lscdig}
\ea
Here the notation is that the triple line loop on the LHS of (\ref{2lspdig}) refers to a spinor propagator in a self-dual background, while the double-line loops [including those on the RHS of  (\ref{2lspdig})] refer to a scalar propagator in the self-dual background. This shows the remarkable result that the two loop fully renormalized answers are expressed naturally in terms of one loop quantities. Qualitatively, we can write:
\ba
\fbox{\begin{minipage}{6.5cm}{$$
{\rm two\,\,loop}=\left({\rm one\,\,loop}\right)^2+\left({\rm one\,\,loop}\right)$$}
\end{minipage}}
\label{2l1l}
\ea
Interestingly, such a relation with two loop quantities being expressed as squares of one loop quantities plus a one loop remainder has been found recently \cite{abdk} in the amplitudes of 4 dimensional super Yang-Mills theory, which is a very different theory from QED. This suggests something deeper is at work here. This is discussed in the next sections.

\subsection{Background field "integration-by-parts" rules}
\label{intbyparts}

Indeed, we can go further than the qualitative statement (\ref{2l1l}) and derive the results (\ref{2lspdig}) and (\ref{2lscdig}) by algebraic means. First, we identify the source of the coefficient factors $-6e^2$ and $\frac{3}{2}e^2$ which appear in front of the $\left({\rm one\,\,loop}\right)^2$ terms in (\ref{2lspdig}) and (\ref{2lscdig}). Note that 
in free QED ({\it i.e.},  with no background field) it is a straightforward exercise to show that in 4 dimensions
\ba
{\rm spinor\,\, QED} : \qquad \mbox{\tlscfr} \quad &=& -6\,e^2\,\bigg[~~\mbox{\olscfr}~~\bigg]^2 \,,
\label{sp2}\\ \nn\\
{\rm scalar\,\, QED} : \qquad\mbox{\tlscfr} \quad &=& ~\frac{3}{2}\,e^2\,\bigg[~~\mbox{\olscfr}~~\bigg]^2\,.
\label{sc2}
\ea
(The loop on the RHS is a scalar loop in each case.)  Thus, the two loop free vacuum diagram can be expressed in terms of a simpler one loop diagram. These results can either be derived by computing each side using dimensional regularization, or a quicker proof follows from an integration-by-parts argument (see below). Notice that the coefficients of the $\left({\rm one\,\,loop}\right)^2$ parts in this free case are exactly the same as the corresponding coefficients in the background field case, for both spinor and scalar QED. This is no accident, as I now illustrate for the case of scalar QED (for spinor QED the argument is similar).

Consider the derivation of (\ref{sc2}) using dimensional regularization and integration-by-parts \cite{bender,vladimirov,chetyrkin}. By purely algebraic manipulations it follows that
\ba 
\tlscfr \hskip.3cm &=&\frac{e^2}{2} \int \frac{d^dp\, d^dq}{(2\pi)^{2d}}\, \frac{(p+q)^2}{(p-q)^2(p^2+m^2)(q^2+m^2)}\nn \\ \nn\\
&=& \frac{e^2}{2} \int \frac{d^dp\, d^dq}{(2\pi)^{2d}}\, \frac{\left[-(p-q)^2+2(p^2+m^2)+2(q^2+m^2)-4 m^2\right]}{(p-q)^2(p^2+m^2)(q^2+m^2)}\nn \\ \nn \\
&=& -\frac{e^2}{2}\bigg[~~\olscfr~~\bigg]^2 -2 e^2 m^2  
\bigg[~~\tlscfrph~~\bigg]
\label{freemanip}
\ea
where the dotted line denotes a massless scalar propagator. The first term has been written as the square of a one loop diagram but the second term is apparently still two loop. However, using simple integration-by-parts manipulations, this two loop diagram can also be written as a square of a one loop diagram:
\ba 
0 &=& \int \frac{d^dp\, d^dq}{(2\pi)^{2d}}\,\frac{\partial}{\partial p_\mu}  \left[\frac{(p-q)_\mu}{(p-q)^2(p^2+m^2)(q^2+m^2)}\right]\nn \\ \nn\\
&=& (d-2) \bigg[~~\tlscfrph~~\bigg] - \int \frac{d^dp\, d^dq}{(2\pi)^{2d}}\, \frac{2p\cdot (p-q)}{(p-q)^2(p^2+m^2)^2(q^2+m^2)}\nn \\ \nn \\
&=& (d-2) \left[\hskip .3cm \tlscfrph \hskip .3cm \right] - \int \frac{d^dp\, d^dq}{(2\pi)^{2d}}\, \frac{[(p-q)^2+(p^2+m^2)-(q^2+m^2)]}{(p-q)^2(p^2+m^2)^2(q^2+m^2)}\nn \\ \nn \\
&=& (d-3) \bigg[~~\tlscfrph~~\bigg] -  \bigg[~~\olscfrv~~\bigg]  \bigg[~~\olscfr~~\bigg]
\nn \\ \nn \\
&=&(d-3) \bigg[~~\tlscfrph~~\bigg] -  \frac{(d-2)}{2m^2}\bigg[~~\olscfr~~\bigg]^2\,. \label{bp}
\ea
Thus, the apparently $2$-loop term on the RHS of (\ref{freemanip}) is a square of a one loop diagram, leading to
\ba 
\tlscfr \hskip .3cm = \frac{e^2}{2}\left(\frac{d-1}{d-3}\right)\bigg[~~\olscfr~~\bigg]^2\,.
\label{free1lsq}
\ea
This reduces to (\ref{sc2}) in $d=4$, and I stress that this result has been derived without doing any integrations, only making simple algebraic manipulations on the integrands.

Now consider the analogous manipulation in a self-dual background \cite{gvdloops}. First, we extend the scalar propagator in a self-dual background to arbitrary dimensions by taking multiple copies of the block diagonal structure of $F_{\mu\nu}$. That is, we take as our definition of "self-dual" fields in $d\neq 4$ the property (\ref{sdid}), since the notion of self-duality in (\ref{selfdual}) is clearly tied to $d=4$. This is equivalent to dimensional regularization in the worldline formalism \cite{csreview}. Then the scalar propagator (\ref{pmprop}) becomes
\begin{eqnarray}
G(p)=\int_0^\infty \frac{dt}{\cosh^{d/2}(eft)} \, e^{-m^2 t-\frac{p^2}{ef}\tanh(eft)}\ .
\label{dsdscprop}
\end{eqnarray}
Then we can repeat the algebraic steps in (\ref{freemanip}) to obtain
\ba 
\tlscbf \hskip.3cm &=& \frac{e^2}{2} \int \frac{d^dp\, d^dq}{(2\pi)^{2d}}\, \frac{1}{(p-q)^2}\left\{ (p+q)^2G(p)G(q)-(ef)^2 \frac{\partial G(p)}{\partial p_\mu}  \frac{\partial G(q)}{\partial q_\mu} \right\}\nn  \\
\nn\\
&=& \frac{e^2}{2}\left(\frac{d-1}{d-3}\right)\bigg[~~\olscbf~~~\bigg]^2+O(f^2)
\label{bfmanip}
\ea
where we have chosen to isolate this particular coefficient of the square of the one loop propagator trace motivated by the free-field result (\ref{free1lsq}). 

\subsection{Algebraic view of mass renormalization}
\label{massren}

The advantage of the manipulation in (\ref{bfmanip}) is that it makes the mass renormalization (which was a very difficult part of previous two loop computations) almost trivial. To see this we subtract the free field two loop diagram from the background field diagram
\ba
 \bigg[~~~\mbox{\tlscbf}\hskip .3cm- \hskip .3cm\mbox{\tlscfr}~~\bigg]=\frac{e^2}{2}\left( \frac{d-1}{d-3}\right) \left\{\bigg[~~~\olscbf~~~\bigg]^2- \bigg[~~\olscfr~~\bigg]^2 \right\}+O(f^2)\nn\\
\label{diff}
\ea
and then simply complete the square in the first terms:
\ba 
\hskip -0.4cm \bigg[~~\olscbf~~\bigg]^2\!\!\!- \bigg[~~\olscfr~~\bigg]^2\!\!\!=
\bigg[~~\olscbf~~ - ~~\olscfr~~\bigg]^2\!\!\!+2\bigg[~~\olscfr~~\bigg]
\bigg[~~\olscbf~~ - ~~\olscfr~~\bigg]\,.
\label{complete}
\ea
The cross-term in (\ref{complete}) is immediately identified with the mass renormalization because 
\ba
\delta m^2=\bigg[\hskip .4cm \olmass \hskip .4cm\bigg]_{p^2=-m^2}=e^2\left( \frac{d-1}{d-3}\right) \bigg[\hskip .3cm\olscfr\hskip .3cm \bigg] \ ,
\label{massshift}
\ea
which can also be derived algebraically. Furthermore, 
\ba 
 \bigg[ \hskip .3cm \olscbf\hskip .3cm -\hskip .3cm \olscfr\hskip .2cm \bigg] = -\frac{\partial {\mathcal L}^{(1)}}{\partial (m^2)}+(f^2\,\,{\rm term})\ .
\label{der1l}
\ea
The $f^2$ term in (\ref{der1l}) contributes to the charge renormalization, and so (\ref{diff}) can be written as
\ba
\bigg[\hskip .3cm \mbox{\tlscbf}\hskip .3cm- \hskip .3cm\mbox{\tlscfr}\hskip .3cm\bigg]=\frac{e^2}{2}\left( \frac{d-1}{d-3}\right) \bigg[\hskip .3cm \mbox{\olscbf}\hskip .3cm- \hskip .3cm \mbox{\olscfr}\hskip .3cm \bigg]^2- \delta m^2\, \frac{\partial {\mathcal L}^{(1)}}{\partial (m^2)}
+O(f^2)\ .\nn\\
\label{massrenorm}
\ea
Observe that the first term is now completely finite, so that we can set $d=4$, and by (\ref{xiint}) we obtain the first term of the final answer (\ref{2lscdig}), the $({\rm one\,\,loop})^2$ piece, without doing any integrals at all! The second term is manifestly the mass renormalization term, and so is absorbed by mass renormalization (\ref{2lmass}). The only remaining divergence can be proportional to the bare Maxwell Lagrangian $f^2$, which is then subtracted by charge renormalization. It is simple to isolate and subtract this piece, leaving an $O(f^4)$ term, whose kernel in the $d\to 4$ limit reduces to a momentum delta function:
\ba
\hskip -1cm O(f^4)&{\buildrel {d\to 4} \over \longrightarrow}&-4\pi^2 e^2 (ef)^2 \int \frac{d^4pd^4q}{(2\pi)^8}\left[G(p) G(q)-G_0(p) G_0(q)\right] \, \delta(p-q)\nonumber\\ \nonumber\\
&=&
-\frac{e^2}{4\pi^2} \, (ef)^2\bigg[ \hskip .4cm \mbox{\olscbfv}\hskip .3cm -\hskip .3cm\mbox{\olscfrv}\hskip .3cm\bigg]
\label{xippiece}
\ea
where in the last step we have used (\ref{xiprint}). Thus, we have derived the diagrammatic form (\ref{2lscdig}) of the fully renormalized two loop scalar QED effective Lagrangian in a self-dual background by essentially algebraic manipulations. It would be interesting to develop these background field ``integration-by-parts" rules into a systematic set of rules that might be applied to even higher loop order.

\section{Conclusions}
\label{conclusions}

To conclude, I reiterate that from seemingly humble beginnings the Heisenberg-Euler result for the one loop effective Lagrangian for spinor QED in a constant background field has led to many applications in a wide range of branches of modern particle physics and quantum field theory. In some sense the Heisenberg-Euler result is analogous to Landau's computation of the partition function for nonrelativistic Landau levels which leads to numerous applications for the magnetic properties of materials. I have only skimmed the surface in this review, but hopefully enough groundwork has been covered to lead the interested reader directly into the more advanced applications.

\section*{Acknowledgments}
I thank Christian Schubert and Holger Gies for collaboration on various aspects of effective Lagrangians, the US DOE for support through grant DE-FG02-92ER40716, and the Rockefeller Foundation for a Bellagio Center residency.



\begin{thebibliography}{200}

\bibitem{he}
W. Heisenberg and H. Euler, ``Consequences of Dirac's Theory of Positrons'', Z. Phys. {\bf 98} (1936) 714.

\bibitem{viki1} V. Weisskopf, ``The electrodynamics of the vacuum based on the quantum theory of the electron'', Kong. Dans. Vid. Selsk.
Math-fys. Medd. XIV No. 6 (1936); English translation in: {\it Early Quantum Electrodynamics: A Source Book}, A. I. Miller, (Cambridge University Press, 1994).

\bibitem{jentschura}
U.~D.~Jentschura, H.~Gies, S.~R.~Valluri, D.~R.~Lamm and E.~J.~Weniger,
``QED effective action revisited,''
Can.\ J.\ Phys.\  {\bf 80}, 267 (2002)
[arXiv:hep-th/0107135].

\bibitem{greiner}
W.~Greiner and J.~Reinhardt,
{\it Quantum Electrodynamics},  (Springer, Berlin, 1992);
W.~Greiner, B.~Muller and J.~Rafelski,
{\it Quantum Electrodynamics Of Strong Fields}, (Springer, Berlin, 1985).

\bibitem{dr-qed}
W.~Dittrich and M.~Reuter,
{\it Effective Lagrangians In Quantum Electrodynamics}, 
Lect.\ Notes Phys.\  {\bf 220}, 1 (Springer, Berlin, 1985).

\bibitem{blau}
S.~K.~Blau, M.~Visser and A.~Wipf,
``Analytical Results For The Effective Action,''
Int.\ J.\ Mod.\ Phys.\ A {\bf 6}, 5409 (1991).

\bibitem{soldati}
R.~Soldati and L.~Sorbo,
``Effective action for Dirac spinors in the presence of general uniform
electromagnetic fields,''
Phys.\ Lett.\ B {\bf 426}, 82 (1998)
[arXiv:hep-th/9802167].

\bibitem{dittrichgies}
W.~Dittrich and H.~Gies,
``Probing the quantum vacuum. Perturbative effective action approach in quantum
electrodynamics and its application,''
Springer Tracts Mod.\ Phys.\  {\bf 166}, 1 (2000).


\bibitem{ek} 
H. Euler and B K\"ockel, ``On the scattering of light from light in the Dirac theory'', Naturwiss. {\bf 23}, 246 (1935).

\bibitem{karplus}
 R. Karplus and M. Neuman, ``The Scattering of Light by Light'', 
Phys. Rev. {\bf 83}, 776 (1951).


\bibitem{sauter}
F. Sauter, Zeit. f. Phys. {\bf 69}, 742 (1931).

\bibitem{schwinger}
J. Schwinger, ``On gauge invariance and vacuum polarization'', Phys. Rev.
{\bf 82} (1951) 664.

\bibitem{schwinger2}
J. Schwinger, ``The Theory of Quantized Fields. V'', 
Phys. Rev. {\bf 93}, 615 (1954),
``The Theory of Quantized Fields. VI'',
Phys. Rev. {\bf 94}, 1362 (1954).

\bibitem{ringwald}
A. Ringwald, ``Pair production from vacuum at the focus of an X-Ray free
electron laser'', Phys. Lett. B {\bf 510} (2001) 107, (hep-ph/0103185);
``Fundamental physics at an X-ray free electron laser,'' in Proceedings of Erice {\it Workshop On Electromagnetic Probes Of Fundamental Physics},  W. Marciano and S. White (Eds.), (River Edge, World Scientific, 2003),
arXiv:hep-ph/0112254.

\bibitem{viki2} 
V. Weisskopf, ``On the self-energy and the electromagnetic field of the electron'', Phys. Rev. {\bf 56} (1939), 72.

\bibitem{gellmannlow}
M. Gell-Mann and F. E. Low, ``Quantum Electrodynamics at Small Distances'', 
Phys. Rev. {\bf 95}, 1300 (1954).

\bibitem{manohar}
A.~V.~Manohar,
``Effective field theories,''
in Proceedings of  Schladming Winter School {\it Perturbative And Nonperturbative Aspects Of Quantum Field Theory}, Eds. H. Latal and W. Schweiger, (Springer-Verlag,  Berlin, 1997);
arXiv:hep-ph/9606222.

\bibitem{donoghue}
J.~F.~Donoghue, E.~Golowich and B.~R.~Holstein,
{\it Dynamics Of The Standard Model}, (Cambridge University Press, 1992).

\bibitem{novikov}
V.~A.~Novikov, M.~A.~Shifman, A.~I.~Vainshtein and V.~I.~Zakharov,
``Calculations In External Fields In Quantum Chromodynamics:. Technical Review
(Abstract Operator Method, Fock-Schwinger Gauge),''
Fortsch.\ Phys.\  {\bf 32}, 585 (1985).

\bibitem{itep}
M.~A.~Shifman,
``ITEP Lectures On Particle Physics And Field Theory.  Vol. 1, 2,''
World Sci.\ Lect.\ Notes Phys.\  {\bf 62}, 1 (1999).

\bibitem{arkadyope}
V.A. Novikov, M.A. Shifman, A.I. Vainshtein and V.I. Zakharov,
``Operator Expansion In Quantum Chromodynamics Beyond Perturbation Theory,''
Nucl.\ Phys.\ B {\bf 174}, 378 (1980);
``Wilson's Operator Expansion: Can It Fail?,''
Nucl.\ Phys.\ B {\bf 249}, 445 (1985)
[Yad.\ Fiz.\  {\bf 41}, 1063 (1985)].

\bibitem{bateman} A. Erd\'elyi (ed.), {\it Higher Transcendental Functions, Vol. I}, (Kreiger, Florida, 1981).

\bibitem{abramowitz} M. Abramowitz and I. Stegun, {\it Handbook of
Mathematical Functions}, (Dover, New York, 1972).

\bibitem{gradshteyn} I.S. Gradshteyn and I.M. Ryzhik, {\it Table of
Integrals, Series and Products}, (Academic Press, New York,
1972).

\bibitem{ww} 
E. Whittaker and G. Watson, {\it Modern
Analysis}, (Cambridge, 1927).

\bibitem{chadha}
S.~Chadha and P.~Olesen,
``On Borel Singularities In Quantum Field Theory,''
Phys.\ Lett.\ B {\bf 72}, 87 (1977).

\bibitem{berndt}  
B. C. Berndt (Ed.), {\it Ramanujan's Notebooks, Vol. II}, (Springer, New York, 1985).

\bibitem{valluri} 
S. R. Valluri, D. R. Lamm and W. J. Mielniczuk, ``Applications of the representation of the Heisenberg-Euler Lagrangian by means of special functions'', Can. J. Phys. {\bf 71} (1993), 389.

\bibitem{nikishov} A.I. Nikishov, ``Pair production by a constant external
field'', Zh. Eksp. Teor. Fiz. {\bf 57} (1969) 1210,
[Sov. Phys. JETP {\bf 30} (1970) 660].

\bibitem{popov}
V. S. Popov, ``Pair production in a variable external field (quasiclassical approximation)'', Sov. Phys. JETP {\bf 34} (1972), 709; ``Pair production in a variable and homogeneous electric field as an oscillator problem'', Sov. Phys. JETP {\bf 35} (1972), 659.


\bibitem{elizaldebook}
E.~Elizalde,
``Ten Physical Applications Of Spectral Zeta Functions,''
Lect.\ Notes Phys.\  {\bf M35}, (Springer-Verlag, Berlin, 1995).

\bibitem{klaus}
K. Kirsten, {\it Spectral Functions in Mathematics and Physics}, (Chapman-Hall, Boca Raton, 2002).

\bibitem{barnes} 
E.W. Barnes, ``The theory of the G-function'', Quart. J. Pure Appl.
Math {\bf XXXI} (1900) 264.

\bibitem{adamchik}
V. Adamchik, ``Symbolic and Numeric computation of the Barnes function'', Conference on Applications of Computer Algebra, Albuquerque, June 2001; ``Contributions to the theory of the Barnes function'', (math.CA/0308086).


\bibitem{salam}
A.~Salam and P.~T.~Matthews,
``Fredholm Theory Of Scattering In A Given Time Dependent Field,''
Phys.\ Rev.\  {\bf 90}, 690 (1953).

\bibitem{salamstrathdee}
A.~Salam and J.~Strathdee,
``Transition Electromagnetic Fields In Particle Physics,''
Nucl.\ Phys.\ B {\bf 90}, 203 (1975).

\bibitem{volkov} 
D. M. Volkov, Z. Phys. {\bf 94}, 25 (1935).

\bibitem{landauqed}  V.B. Berestetskii, E.M. Lifshitz, and L.P. Pitaevskii, {\it Quantum electrodynamics}, (Pergamon Press, New York, 1982). 

\bibitem{brownkibble} 
L. S. Brown and T. W. B. Kibble, ``Interaction of Intense Laser Beams with Electrons'', 
Phys. Rev. 133, A705 (1964).

\bibitem{bunkin}
F. V. Bunkin and I. I. Tugov, ``Possibility of creating electron-positron pairs in a vacuum by the focusing of laser radiation'', Sov. Phys. Dokl. {\bf 14} (1970), 678.

\bibitem{friedavan}
H.~M.~Fried, Y.~Gabellini, B.~H.~J.~McKellar and J.~Avan,
``Pair production via crossed lasers,''
Phys.\ Rev.\ D {\bf 63}, 125001 (2001).

\bibitem{nn} N.B. Narozhnyi and A.I. Nikishov, ``The
simplest processes in a pair-producing field'', Yad. Fiz. {\bf 11} (1970) 1072,
[Sov. J. Nucl. Phys. {\bf 11} (1970) 596].

\bibitem{dhelectric}
G.~V.~Dunne and T.~Hall,
``On the QED effective action in time dependent electric backgrounds,''
Phys.\ Rev.\ D {\bf 58}, 105022 (1998)
[arXiv:hep-th/9807031].

\bibitem{cangemi2+1}
D.~Cangemi, E.~D'Hoker and G.~V.~Dunne,
``Effective energy for QED in (2+1)-dimensions with semilocalized magnetic
fields: A Solvable model,''
Phys.\ Rev.\ D {\bf 52}, 3163 (1995)
[arXiv:hep-th/9506085].

\bibitem{dh3+1}
G.~V.~Dunne and T.~M.~Hall,
``An exact QED(3+1) effective action,''
Phys.\ Lett.\ B {\bf 419}, 322 (1998)
[arXiv:hep-th/9710062].

\bibitem{kimpage}
S.~P.~Kim and D.~N.~Page,
``Schwinger pair production via instantons in a strong electric field,''
Phys.\ Rev.\ D {\bf 65}, 105002 (2002)
[arXiv:hep-th/0005078].

\bibitem{nikishovpreprint}
A.~I.~Nikishov,
``On the theory of scalar pair production by a potential barrier,''
arXiv:hep-th/0111137.

\bibitem{devega}
H.~J.~de Vega and F.~A.~Schaposnik,
``Nonuniform External Fields And Vacuum Properties In A Two-Dimensional Gauge
Theory,''
Phys.\ Rev.\ D {\bf 26}, 2814 (1982).

\bibitem{tomaras}
T.~N.~Tomaras, N.~C.~Tsamis and R.~P.~Woodard,
``Back-reaction in lightcone QED,''
Phys.\ Rev.\ D {\bf 62}, 125005 (2000)
[arXiv:hep-ph/0007166];
``Pair creation and axial anomaly in light-cone QED(2),''
JHEP {\bf 0111}, 008 (2001)
[arXiv:hep-th/0108090].

\bibitem{fried}
H.~M.~Fried and R.~P.~Woodard,
``The one loop effective action of QED for a general class of electric
fields,''
Phys.\ Lett.\ B {\bf 524}, 233 (2002)
[arXiv:hep-th/0110180].

\bibitem{brezin}
E.~Brezin and C.~Itzykson,
``Pair Production In Vacuum By An Alternating Field,''
Phys.\ Rev.\ D {\bf 2}, 1191 (1970);

\bibitem{popovmarinov}
V.~S.~Popov and M.~S.~Marinov,
``E+ E- Pair Production In Variable Electric Field,''
Yad.\ Fiz.\  {\bf 16}, 809 (1972), [Sov. J. Nucl. Phys. {\bf 16} (1973), 449];
``Electron - Positron Pair Creation From Vacuum Induced By Variable Electric Field,''
Fortsch.\ Phys.\  {\bf 25}, 373 (1977).

\bibitem{keldysh}
L. V. Keldysh, ``Ionization in the field of a strong electromagnetic wave'',
Sov. Phys. JETP {\bf 20} (1965) 1307.

\bibitem{perelomov}
A. M. Perelomov, V. S. Popov and M. V. Terent'ev, ``Ionization of atoms in an alternating electric field, I and II'', Sov. Phys JETP {\bf 23}, 924 (1966), {\bf 24}, 207 (1967).

\bibitem{landauqm} 
L.D. Landau and E.M. Lifshitz,   {\it Quantum mechanics : non-relativistic theory}, (Pergamon Press, New York, 1977).


\bibitem{opp}
 J. R. Oppenheimer, ``Three Notes on the Quantum Theory of Aperiodic Effects'', 
Phys. Rev. {\bf 31}, 66 (1928).
 
 \bibitem{carlbook} 
 C.M. Bender and S.A. Orszag, {\it Advanced Mathematical
Methods for Scientists and Engineers} (Mc Graw-Hill, New York, 1978).
 
\bibitem{strassler}
M.~J.~Strassler,
``Field theory without Feynman diagrams: One loop effective actions,''
Nucl.\ Phys.\ B {\bf 385}, 145 (1992)
[arXiv:hep-ph/9205205].

\bibitem{ss1}
M.G. Schmidt, C. Schubert, 
``On the calculation of effective actions by string methods'',
Phys. Lett. {\bf B 318} (1993) 438, hep-th/9309055. 

\bibitem{ss2}
M. G. Schmidt and C. Schubert, ``Worldline Green Functions for
Multiloop Diagrams'', Phys. Lett. {\bf B 331} (1994) 69, hep-th/9403158;
``Multiloop calculations in the
string-inspired formalism: the single spinor-loop in QED'', Phys. Rev.
{\bf D 53} (1996) 2150, hep-th/9410100.

\bibitem{gerry}
D.G.C. McKeon and T.N. Sherry, 
``Radiative effects in a constant magnetic field using the 
quantum mechanical path integral'',
Mod. Phys. Lett. {\bf A9} (1994) 2167.

\bibitem{csreview}
C. Schubert, ``Perturbative quantum field theory in the string-inspired
formalism'', Phys. Rept. {\bf 355} (2001) 73, hep-th/0101036.

\bibitem{cangemider}
D.~Cangemi, E.~D'Hoker and G.~V.~Dunne,
``Derivative expansion of the effective action and vacuum instability for QED
in (2+1)-dimensions,''
Phys.\ Rev.\ D {\bf 51}, 2513 (1995)
[arXiv:hep-th/9409113].

\bibitem{leepacshin}
H.~W.~Lee, P.~Y.~Pac and H.~K.~Shin,
``Derivative Expansions In Quantum Electrodynamics,''
Phys.\ Rev.\ D {\bf 40}, 4202 (1989).


\bibitem{igor}
V.~P.~Gusynin and I.~A.~Shovkovy,
``Derivative expansion for the one-loop effective Lagrangian in QED,''
Can.\ J.\ Phys.\  {\bf 74}, 282 (1996)
[arXiv:hep-ph/9509383];
``Derivative expansion of the effective action for QED in 2+1 and 3+1
dimensions,''
J.\ Math.\ Phys.\  {\bf 40}, 5406 (1999)
[arXiv:hep-th/9804143].

\bibitem{dhborel}
G.~V.~Dunne and T.~M.~Hall,
``Borel summation of the derivative expansion and effective actions,''
Phys.\ Rev.\ D {\bf 60}, 065002 (1999)
[arXiv:hep-th/9902064].

\bibitem{zinnborel} J. Zinn-Justin, {\it Quantum Field Theory and
Critical Phenomena}, (Oxford University Press, 1997).

\bibitem{thooftborel} G. 't Hooft, ``Can we make sense out of `quantum 
chromodynamics'?'', in {\it The Whys of Subnuclear Physics}, ed. A. Zichichi,
(Plenum, New York, 1978); Reprinted in: G. 't Hooft, {\it Under the spell of the
gauge principle}, (World Scientific, Singapore, 1994).

\bibitem{fliegnermass}
D.~Fliegner, P.~Haberl, M.~G.~Schmidt and C.~Schubert,
``Application of the worldline path integral method to the calculation of
inverse mass expansions,''
Nucl.\ Instrum.\ Meth.\ A {\bf 389}, 374 (1997)
[arXiv:hep-th/9702092];
``The higher derivative expansion of the effective action by the string
inspired method. II,''
Annals Phys.\  {\bf 264}, 51 (1998)
[arXiv:hep-th/9707189].

\bibitem{muller}
U.~Muller,
``A basis for invariants in non-Abelian gauge theories,''
arXiv:hep-th/9508031.

\bibitem{form}
J.~A.~M.~Vermaseren,
``New features of FORM,''
arXiv:math-ph/0010025.


\bibitem{vandeven}
A.~E.~M.~van de Ven,
``Index-free heat kernel coefficients,''
Class.\ Quant.\ Grav.\  {\bf 15}, 2311 (1998)
[arXiv:hep-th/9708152].

\bibitem{gieslangfeld}
H.~Gies and K.~Langfeld,
``Quantum diffusion of magnetic fields in a numerical worldline approach,''
Nucl.\ Phys.\ B {\bf 613}, 353 (2001)
[arXiv:hep-ph/0102185];
``Loops and loop clouds: A numerical approach to the worldline formalism  in
QED,''
Int.\ J.\ Mod.\ Phys.\ A {\bf 17}, 966 (2002)
[arXiv:hep-ph/0112198];
K.~Langfeld, L.~Moyaerts and H.~Gies,
``Fermion-induced quantum action of vortex systems,''
Nucl.\ Phys.\ B {\bf 646}, 158 (2002)
[arXiv:hep-th/0205304].

\bibitem{schmidtstamatescu}
M.~G.~Schmidt and I.~O.~Stamatescu,
``Determinant calculations with random walk worldline loops,''
arXiv:hep-lat/0201002;
``Matter Determinants In Background Fields Using Random Walk World Line Loops
On The Lattice,''
Mod.\ Phys.\ Lett.\ A {\bf 18}, 1499 (2003);
``Determinant calculations using random walk worldline loops,''
Nucl.\ Phys.\ Proc.\ Suppl.\  {\bf 119}, 1030 (2003)
[arXiv:hep-lat/0209120].

\bibitem{casimir}
H.~Gies, K.~Langfeld and L.~Moyaerts,
``Casimir effect on the worldline,''
JHEP {\bf 0306}, 018 (2003)
[arXiv:hep-th/0303264].


\bibitem{fry}
M.~P.~Fry,
``QED in inhomogeneous magnetic fields,''
Phys.\ Rev.\ D {\bf 54}, 6444 (1996)
[arXiv:hep-th/9606037];
``Fermion determinants,''
Int.\ J.\ Mod.\ Phys.\ A {\bf 17}, 936 (2002)
[arXiv:hep-th/0111244];
``Fermion determinant for general background gauge fields,''
Phys.\ Rev.\ D {\bf 67}, 065017 (2003)
[arXiv:hep-th/0301097];
``Fermion determinants 2003,'' To appear in the proceedings of 6th Workshop on Quantum Field Theory under the Influence of External Conditions (QFEXT03), Norman, Oklahoma, Sep 2003,
arXiv:hep-th/0311145.

\bibitem{terentev} 
V. S. Vanyashin and M. V. Terentev, ``The vacuum polarization of a charged vector field'',
Sov. Phys. JETP {\bf 21} (1965) 375.

\bibitem{grosswilczek}
D.~J.~Gross and F.~Wilczek,
``Ultraviolet Behavior Of Non-Abelian Gauge Theories,''
Phys.\ Rev.\ Lett.\  {\bf 30}, 1343 (1973);
``Asymptotically Free Gauge Theories. I,''
Phys.\ Rev.\ D {\bf 8}, 3633 (1973);
``Asymptotically Free Gauge Theories. 2,''
Phys.\ Rev.\ D {\bf 9}, 980 (1974).

\bibitem{politzer}
H.~D.~Politzer,
``Reliable Perturbative Results For Strong Interactions?,''
Phys.\ Rev.\ Lett.\  {\bf 30}, 1346 (1973).

\bibitem{skalozub}
V. V. Skalozub, ``Vacuum polarization of a charged vector field in a renormalizable theory'', Sov. J. Nucl. Phys. {\bf 21} (1976), 690.


\bibitem{brownduff}
M.~R.~Brown and M.~J.~Duff,
``Exact Results For Effective Lagrangians,''
Phys.\ Rev.\ D {\bf 11}, 2124 (1975).

\bibitem{bms}
I.~A.~Batalin, S.~G.~Matinyan and G.~K.~Savvidy,
``Vacuum Polarization By A Source - Free Gauge Field,''
Sov.\ J.\ Nucl.\ Phys.\  {\bf 26}, 214 (1977)
[Yad.\ Fiz.\  {\bf 26}, 407 (1977)].

\bibitem{shore}
G.~M.~Shore,
``Symmetry Restoration And The Background Field Method In Gauge Theories,''
Annals Phys.\  {\bf 137}, 262 (1981).

\bibitem{drm}
M.~J.~Duff and M.~Ramon-Medrano,
``On The Effective Lagrangian For The Yang-Mills Field,''
Phys.\ Rev.\ D {\bf 12}, 3357 (1975).

\bibitem{savvidy}
G.~K.~Savvidy,
``Infrared Instability Of The Vacuum State Of Gauge Theories And Asymptotic
Freedom,''
Phys.\ Lett.\ B {\bf 71}, 133 (1977).

\bibitem{ms}
S.~G.~Matinyan and G.~K.~Savvidy,
``Vacuum Polarization Induced By The Intense Gauge Field,''
Nucl.\ Phys.\ B {\bf 134}, 539 (1978).


\bibitem{cw}
S.~R.~Coleman and E.~Weinberg,
``Radiative Corrections As The Origin Of Spontaneous Symmetry Breaking,''
Phys.\ Rev.\ D {\bf 7}, 1888 (1973).

\bibitem{ritusspin}
V. I. Ritus, ``Lagrangian of an intense electromagnetic
field and quantum electrodynamics at short distances'', Zh. Eksp. Teor.
Fiz {\bf 69} (1975) 1517 [Sov. Phys. JETP {\bf 42} (1975) 774].

\bibitem{ritusscal}
V. I. Ritus, ``Connection between strong-field quantum electrodynamics
with short-distance quantum electrodynamics'',
Zh. Eksp. Teor. Fiz {\bf 73} (1977) 807
[Sov. Phys. JETP {\bf 46} (1977) 423].

\bibitem{ginzburg}
V. I. Ritus, ``The Lagrangian Function of an Intense Electromagnetic
Field'', in {\it Proc. Lebedev Phys. Inst.} Vol. {\bf 168}, {\it Issues
in Intense-field Quantum Electrodynamics}, V. I. Ginzburg, ed., (Nova
Science Pub., NY 1987); ``Effective Lagrange function of intense electromagnetic field in QED,''
arXiv:hep-th/9812124.


\bibitem{tsai}
W.~Y.~Tsai and A.~Yildiz,
``Motion Of Charged Particles In A Homogeneous Magnetic Field,''
Phys.\ Rev.\ D {\bf 4}, 3643 (1971);
T.~Goldman, W.~Y.~Tsai and A.~Yildiz,
``Consistency Of Spin-One Theory,''
Phys.\ Rev.\ D {\bf 5}, 1926 (1972).

\bibitem{nielsen}
N.~K.~Nielsen and P.~Olesen,
``An Unstable Yang-Mills Field Mode,''
Nucl.\ Phys.\ B {\bf 144}, 376 (1978);
``Electric Vortex Lines From The Yang-Mills Theory,''
Phys.\ Lett.\ B {\bf 79}, 304 (1978).

\bibitem{olesen}
H.~B.~Nielsen and P.~Olesen,
``A Quantum Liquid Model For The QCD Vacuum: Gauge And Rotational Invariance Of
Domained And Quantized Homogeneous Color Fields,''
Nucl.\ Phys.\ B {\bf 160}, 380 (1979);
J.~Ambjorn and P.~Olesen,
``A Color Magnetic Vortex Condensate In QCD,''
Nucl.\ Phys.\ B {\bf 170}, 265 (1980);
P.~Olesen,
``On The QCD Vacuum,''
Phys.\ Scripta {\bf 23}, 1000 (1981).

\bibitem{yildiz1}
A.~Yildiz and P.~H.~Cox,
``Vacuum Behavior In Quantum Chromodynamics,''
Phys.\ Rev.\ D {\bf 21}, 1095 (1980);
M.~Claudson, A.~Yildiz and P.~H.~Cox,
``Vacuum Behavior In Quantum Chromodynamics. II,''
Phys.\ Rev.\ D {\bf 22}, 2022 (1980).

\bibitem{leutwyler}
H.~Leutwyler,
``Vacuum Fluctuations Surrounding Soft Gluon Fields,''
Phys.\ Lett.\ B {\bf 96}, 154 (1980);
``Constant Gauge Fields And Their Quantum Fluctuations,''
Nucl.\ Phys.\ B {\bf 179}, 129 (1981).

\bibitem{duffisham1}
M.~J.~Duff and C.~J.~Isham,
``Selfduality, Helicity, And Supersymmetry: The Scattering Of Light By Light,''
Phys.\ Lett.\ B {\bf 86}, 157 (1979).

\bibitem{duffisham2}
M.~J.~Duff and C.~J.~Isham,
``Selfduality, Helicity, And Coherent States In Nonabelian Gauge Theories,''
Nucl.\ Phys.\ B {\bf 162}, 271 (1980).

\bibitem{grisaru}
M.~T.~Grisaru and H.~N.~Pendleton,
``Some Properties Of Scattering Amplitudes In Supersymmetric Theories,''
Nucl.\ Phys.\ B {\bf 124}, 81 (1977);
M.~T.~Grisaru, H.~N.~Pendleton and P.~van Nieuwenhuizen,
``Supergravity And The S Matrix,''
Phys.\ Rev.\ D {\bf 15}, 996 (1977);
S.~M.~Christensen, S.~Deser, M.~J.~Duff and M.~T.~Grisaru,
``Chirality, Selfduality, And Supergravity Counterterms,''
Phys.\ Lett.\ B {\bf 84}, 411 (1979).

\bibitem{kallosh}
R.~E.~Kallosh,
``Superselfduality,''
JETP Lett.\  {\bf 29}, 172 (1979)
[Pisma Zh.\ Eksp.\ Teor.\ Fiz.\  {\bf 29}, 192 (1979)];
``Structure Of Divergences In Supergravitation. (In Russian),''
JETP Lett.\  {\bf 29}, 449 (1979)
[Pisma Zh.\ Eksp.\ Teor.\ Fiz.\  {\bf 29}, 493 (1979)].

\bibitem{drzeta}
W.~Dittrich and M.~Reuter,
``Effective QCD Lagrangian With Zeta Function Regularization,''
Phys.\ Lett.\ B {\bf 128}, 321 (1983).

\bibitem{elizalde}
E.~Elizalde,
``Effective Lagrangian For Ordinary Quarks In A Background Field,''
Nucl.\ Phys.\ B {\bf 243}, 398 (1984);
E.~Elizalde and J.~Soto,
``Zeta Regularized Lagrangians For Massive Quarks In Constant Background Mean Fields,''
Annals Phys.\  {\bf 162}, 192 (1985);
E.~Elizalde and J.~Soto,
``Exact Effective Actions For Quarks In Pure And Selfdual Mean Fields,''
Nucl.\ Phys.\ B {\bf 260}, 136 (1985).

\bibitem{mcarthur}
T.~D.~Gargett and I.~N.~McArthur,
``Derivative Expansion Of One-Loop Effective Actions For Yang-Mills  Fields,''
J.\ Math.\ Phys.\  {\bf 39}, 4430 (1998).

\bibitem{mcarthur2}
S.~M.~Kuzenko and I.~N.~McArthur,
``On the background field method beyond one loop: A manifestly covariant
derivative expansion in super Yang-Mills theories,''
JHEP {\bf 0305}, 015 (2003)
[arXiv:hep-th/0302205].

\bibitem{rss}
M.~Reuter, M.~G.~Schmidt and C.~Schubert,
``Constant external fields in gauge theory and the spin 0, 1/2, 1 path
integrals,''
Annals Phys.\  {\bf 259}, 313 (1997)
[arXiv:hep-th/9610191].

\bibitem{buchbook}
I.~L.~Buchbinder and S.~M.~Kuzenko,
{\it Ideas And Methods Of Supersymmetry And Supergravity: Or A Walk Through
Superspace}, (IOP Press, Bristol, 1998).

\bibitem{ftsusy}
E.S. Fradkin, A.A. Tseytlin, 
``Quantum properties of higher dimensional and dimensionally
reduced supersymmetric theories'', Nucl. Phys. {\bf B 227} (1983) 252.

\bibitem{bkt}
I.~L.~Buchbinder, S.~M.~Kuzenko and A.~A.~Tseytlin,
``On low-energy effective actions in N = 2,4 superconformal theories in  four
dimensions,''
Phys.\ Rev.\ D {\bf 62}, 045001 (2000)
[arXiv:hep-th/9911221].


\bibitem{thooft}
G.~'t Hooft,
``Computation Of The Quantum Effects Due To A Four-Dimensional 
Pseudoparticle,'' Phys.\ Rev.\ D {\bf 14}, 3432 (1976)
[Erratum-ibid.\ D {\bf 18}, 2199 (1978)].

\bibitem{jackiwrebbi}
R.~Jackiw and C.~Rebbi,
``Degrees Of Freedom In Pseudoparticle Systems,''
Phys.\ Lett.\ B {\bf 67}, 189 (1977);
``Spinor Analysis Of Yang-Mills Theory,''
Phys.\ Rev.\ D {\bf 16}, 1052 (1977).

\bibitem{schwarz}
A.~S.~Schwarz,
``On Regular Solutions Of Euclidean Yang-Mills Equations,''
Phys.\ Lett.\ B {\bf 67}, 172 (1977);

\bibitem{dadda}
A.~D'Adda and P.~Di Vecchia,
``Supersymmetry And Instantons,''
Phys.\ Lett.\ B {\bf 73}, 162 (1978).

\bibitem{cc}
R.~D.~Carlitz and D.~B.~Creamer,
``Light Quarks And Instantons,''
Annals Phys.\  {\bf 118}, 429 (1979).

\bibitem{kwon} 
O.~K.~Kwon, C.~Lee and H.~Min,
``Massive field contributions to the QCD vacuum tunneling amplitude,''
Phys.\ Rev.\ D {\bf 62}, 114022 (2000)
[arXiv:hep-ph/0008028].

\bibitem{gpy}
D.~J.~Gross, R.~D.~Pisarski and L.~G.~Yaffe,
``QCD And Instantons At Finite Temperature,''
Rev.\ Mod.\ Phys.\  {\bf 53}, 43 (1981).

\bibitem{diakonov}
D.~Diakonov, N.~Gromov, V.~Petrov and S.~Slizovskiy,
``Quantum weights of dyons and of instantons with non-trivial holonomy,''
arXiv:hep-th/0404042.

\bibitem{chiral}
G.~V.~Dunne, A.~W.~Thomas and S.~V.~Wright,
``Chiral extrapolation: An analogy with effective field theory,''
Phys.\ Lett.\ B {\bf 531}, 77 (2002)
[arXiv:hep-th/0110155].


\bibitem{fock}
V.~Fock,
``Proper Time In Classical And Quantum Mechanics,''
Phys.\ Z.\ Sow. {\bf 12}, 404 (1937);

\bibitem{nambu}
Y.~Nambu,
``The Use Of The Proper Time In Quantum Electrodynamics,''
Prog.\ Theor.\ Phys.\  {\bf 5}, 82 (1950).


\bibitem{itzyksonzuber}
C. Itzykson, J. Zuber, {\it Quantum Field Theory},
McGraw-Hill 1985.


 \bibitem{dunsch}
G.~V.~Dunne and C.~Schubert,
``Two-loop Euler-Heisenberg QED pair-production rate,''
Nucl.\ Phys.\ B {\bf 564}, 591 (2000)
[arXiv:hep-th/9907190].

\bibitem{lebedev}
S.~L.~Lebedev and V.~I.~Ritus,
``Virial Representation Of The Imaginary Part Of The Lagrange Function Of The
Electromagnetic Field,''
Sov.\ Phys.\ JETP {\bf 59}, 237 (1984)
[Zh.\ Eksp.\ Teor.\ Fiz.\  {\bf 86}, 408 (1984)].

\bibitem{ritusmass} 
V. I. Ritus,
``Method Of Eigenfunctions And Mass Operator In Quantum Electrodynamics Of A
Constant Field,''
Sov.\ Phys.\ JETP {\bf 48}, 788 (1978)
[Zh.\ Eksp.\ Teor.\ Fiz.\  {\bf 75}, 1560 (1978)].

\bibitem{affleck} 
I.~K.~Affleck, O.~Alvarez and N.~S.~Manton,
``Pair Production At Strong Coupling In Weak External Fields,''
Nucl.\ Phys.\ B {\bf 197}, 509 (1982).

\bibitem{dittrichjpa}
W.~Dittrich,
``Weisskopf-Schwinger Lagrangian With Radiative Corrections,''
J.\ Phys.\ A {\bf 10}, 833 (1977).

\bibitem{fliegner3}
D. Fliegner, M. Reuter, M. G. Schmidt, C. Schubert, 
``The two-loop
Euler-Heisenberg Lagrangian in dimensional renormalization'', 
Teor. Mat. Fiz. {\bf 113} (1997) 289
[Theor. Math. Phys. {\bf 113} (1997) 1442], hep-th/9704194.

\bibitem{kors}
B. K{\"o}rs, M.G. Schmidt, 
``The effective two loop Euler-Heisenberg action for scalar and
spinor QED in a general constant background field'', 
Eur. Phys. J. {\bf C 6} (1999) 175, 
hep-th/9803144.

\bibitem{sato}
H.~T.~Sato and M.~G.~Schmidt,
``World-line approach to the Bern-Kosower formalism in two-loop Yang-Mills theory,''
Nucl.\ Phys.\ B {\bf 560}, 551 (1999)
[arXiv:hep-th/9812229];
H.~T.~Sato, M.~G.~Schmidt and C.~Zahlten,
``Two-loop Yang-Mills theory in the world-line formalism and an  Euler-Heisenberg type action,''
Nucl.\ Phys.\ B {\bf 579}, 492 (2000)
[arXiv:hep-th/0003070];
B.~Kors and M.~G.~Schmidt,
``Two-loop Feynman diagrams in Yang-Mills theory from bosonic string  amplitudes,''
arXiv:hep-th/0003171.

\bibitem{km}
S.~M.~Kuzenko and I.~N.~McArthur,
``Low-energy dynamics in N = 2 super QED: Two-loop approximation,''
JHEP {\bf 0310}, 029 (2003)
[arXiv:hep-th/0308136];
``Relaxed super self-duality and effective action,''
arXiv:hep-th/0403082;
``Relaxed super self-duality and N = 4 SYM at two loops,''
arXiv:hep-th/0403240.

\bibitem{crewther}
R.~J.~Crewther,
``Nonperturbative Evaluation Of The Anomalies In Low-Energy Theorems,''
Phys.\ Rev.\ Lett.\  {\bf 28}, 1421 (1972);
M.~S.~Chanowitz and J.~R.~Ellis,
``Canonical Anomalies And Broken Scale Invariance,''
Phys.\ Lett.\ B {\bf 40}, 397 (1972).

\bibitem{anomaly}
S.~L.~Adler, J.~C.~Collins and A.~Duncan,
``Energy-Momentum-Tensor Trace Anomaly In Spin 1/2 Quantum
Electrodynamics,'' Phys.\ Rev.\ D {\bf 15}, 1712 (1977);
J.~C.~Collins, A.~Duncan and S.~D.~Joglekar,
``Trace And Dilatation Anomalies In Gauge Theories,''
Phys.\ Rev.\ D {\bf 16}, 438 (1977).

\bibitem{pagels} H.~Pagels and E.~Tomboulis,
``Vacuum Of The Quantum Yang-Mills Theory And Magnetostatics,''
Nucl.\ Phys.\ B {\bf 143}, 485 (1978).

\bibitem{fujikawa} 
K.~Fujikawa,
``A Nondiagramatic Calculation Of One Loop Beta Function In QCD,''
Phys.\ Rev.\ D {\bf 48}, 3922 (1993).

\bibitem{hansson} 
J.~Grundberg and T.~H.~Hansson,
``The QCD trace anomaly as a vacuum effect (The vacuum is a medium is
the message!),'' Annals Phys.\  {\bf 242}, 413 (1995)
[arXiv:hep-th/9407139].

\bibitem{shifman}
M.~A.~Shifman and A.~I.~Vainshtein,
``Operator Product Expansion And Calculation Of The Two Loop
Gell-Mann-Low Function,'' Sov.\ J.\ Nucl.\ Phys.\  {\bf 44}, 321 (1986)
[Yad.\ Fiz.\  {\bf 44}, 498 (1986)].

\bibitem{bornsen}
J.~P.~Bornsen and A.~E.~van de Ven,
``Three-loop Yang-Mills beta-function via the covariant background field  method,''
Nucl.\ Phys.\ B {\bf 657}, 257 (2003)
[arXiv:hep-th/0211246].

\bibitem{peskin}
M.~E.~Peskin and D.~V.~Schroeder,
{\it An Introduction To Quantum Field Theory},  (Addison-Wesley, Reading, 1995).

\bibitem{gorishnii}
S.~G.~Gorishnii, A.~L.~Kataev, S.~A.~Larin and L.~R.~Surguladze,
``The Analytical Four Loop Corrections To The QED Beta Function In The Ms Scheme And To The QED Psi Function: Total Reevaluation,''
Phys.\ Lett.\ B {\bf 256}, 81 (1991).

\bibitem{jost} 
R. Jost and J. M. Luttinger, Helv. Phys. Acta {\bf 23} (1950), 201.

\bibitem{rosner}
J. Rosner, ``Sixth-order contribution to $Z_3$ in finite quantum electrodynamics'', Phys. Rev. Lett. {\bf 17} (1966), 1190; ``Higher-order contributions to the divergent part of $Z_3$ in a model of quantum electrodynamics'', Ann. Phys. {\bf 44} (1967), 11; E. de Rafael and J. Rosner, ``Short-distance behavior of quantum electrodynamics and the Callan-Symanzik equation for the photon propagator'', Ann. Phys. {\bf 82} 91974), 369.

\bibitem{larin87}
S.~G.~Gorishnii, A.~L.~Kataev and S.~A.~Larin,
``Analytical Four Loop Result For Beta Function In QED In Ms And Mom Schemes,''
Phys.\ Lett.\ B {\bf 194}, 429 (1987).

\bibitem{4loop}
T.~van Ritbergen, J.~A.~M.~Vermaseren and S.~A.~Larin,
``The four-loop beta function in quantum chromodynamics,''
Phys.\ Lett.\ B {\bf 400}, 379 (1997)
[arXiv:hep-ph/9701390];
J.~A.~M.~Vermaseren, S.~A.~Larin and T.~van Ritbergen,
``The 4-loop quark mass anomalous dimension and the invariant quark  mass,''
Phys.\ Lett.\ B {\bf 405}, 327 (1997)
[arXiv:hep-ph/9703284].

\bibitem{bender}
C.~M.~Bender, R.~W.~Keener and R.~E.~Zippel,
 ``New Approach To The Calculation Of F(1) (Alpha) In Massless Quantum
Electrodynamics,''
Phys.\ Rev.\ D {\bf 15}, 1572 (1977).

\bibitem{broadhurst}
D.~J.~Broadhurst, R.~Delbourgo and D.~Kreimer,
``Unknotting the polarized vacuum of quenched QED,''
Phys.\ Lett.\ B {\bf 366}, 421 (1996)
[arXiv:hep-ph/9509296];
D.~J.~Broadhurst,
``Four-loop Dyson-Schwinger-Johnson anatomy,''
Phys.\ Lett.\ B {\bf 466}, 319 (1999)
[arXiv:hep-ph/9909336],
``Dimensionally continued multi-loop gauge theory,''
arXiv:hep-th/9909185.

\bibitem{zmb}
G.~V.~Dunne, H.~Gies and C.~Schubert,
``Zero modes, beta functions and IR/UV interplay in higher-loop QED,''
JHEP {\bf 0211}, 032 (2002)
[arXiv:hep-th/0210240].

\bibitem{ds1}
G.~V.~Dunne and C.~Schubert,
``Closed-form two-loop Euler-Heisenberg Lagrangian in a self-dual 
background,'' Phys.\ Lett.\ B {\bf 526}, 55 (2002)
[arXiv:hep-th/0111134].

\bibitem{ds2}
G.~V.~Dunne and C.~Schubert,
``Two-loop self-dual Euler-Heisenberg Lagrangians. I: Real part and 
helicity amplitudes,'' JHEP {\bf 0208}, 053 (2002)
[arXiv:hep-th/0205004].

\bibitem{ds3}
G.~V.~Dunne and C.~Schubert,
``Two-loop self-dual Euler-Heisenberg Lagrangians. II: Imaginary part 
and Borel analysis,'' JHEP {\bf 0206}, 042 (2002)
[arXiv:hep-th/0205005].

\bibitem{ttwu} 
R. Gastmans and T. T. Wu, 
{\it The ubiquitous photon: helicity method for QED and QCD}, (Oxford Univerity Press,
New York, 1990).

\bibitem{mangano} 
M.~L.~Mangano and S.~J.~Parke,
``Multiparton Amplitudes In Gauge Theories,''
Phys.\ Rept.\  {\bf 200}, 301 (1991);

\bibitem{bernhelicity}
Z.~Bern, G.~Chalmers, L.~J.~Dixon and D.~A.~Kosower,
``One loop N gluon amplitudes with maximal helicity violation via collinear
limits,''
Phys.\ Rev.\ Lett.\  {\bf 72}, 2134 (1994)
[arXiv:hep-ph/9312333].

\bibitem{bernmorgan}
Z.~Bern and A.~G.~Morgan,
``Massive Loop Amplitudes from Unitarity,''
Nucl.\ Phys.\ B {\bf 467}, 479 (1996)
[arXiv:hep-ph/9511336].

\bibitem{mahlon} 
G.~Mahlon,
``One loop multi - photon helicity amplitudes,''
Phys.\ Rev.\ D {\bf 49}, 2197 (1994)
[arXiv:hep-ph/9311213].

\bibitem{2loophelicity}
Z.~Bern, A.~De Freitas and L.~Dixon,
``Two-loop helicity amplitudes for gluon gluon scattering in QCD and  
supersymmetric Yang-Mills theory,''
JHEP {\bf 0203}, 018 (2002)
[arXiv:hep-ph/0201161];
T.~Binoth, E.~W.~N.~Glover, P.~Marquard and J.~J.~van der Bij,
``Two-loop corrections to light-by-light scattering in supersymmetric QED,''
JHEP {\bf 0205}, 060 (2002)
[arXiv:hep-ph/0202266].

\bibitem{louise}
L.~C.~Martin, C.~Schubert and V.~M.~Villanueva Sandoval,
``On the low-energy limit of the QED N - photon amplitudes,''
Nucl.\ Phys.\ B {\bf 668}, 335 (2003)
[arXiv:hep-th/0301022].


\bibitem{leeleepac}
C.~Lee, H.~W.~Lee and P.~Y.~Pac,
``Calculation Of One Loop Instanton Determinants Using Propagators With
Space-Time Dependent Mass,'' Nucl.\ Phys.\ B {\bf 201}, 429 (1982).

\bibitem{glover}
E.~W.~Glover,
``Progress in NNLO calculations for scattering processes,''
Nucl.\ Phys.\ Proc.\ Suppl.\  {\bf 116}, 3 (2003)
[arXiv:hep-ph/0211412];
Z.~Bern,
``Recent progress in perturbative quantum field theory,''
Nucl.\ Phys.\ Proc.\ Suppl.\  {\bf 117}, 260 (2003)
[arXiv:hep-ph/0212406].

\bibitem{dirk}
D.~Kreimer,
``On the Hopf algebra structure of perturbative quantum field theories,''
Adv.\ Theor.\ Math.\ Phys.\  {\bf 2}, 303 (1998)
[arXiv:q-alg/9707029];
A.~Connes and D.~Kreimer,
``Renormalization in quantum field theory and the Riemann-Hilbert  problem. I: The Hopf algebra structure of graphs and the main theorem,''
Commun.\ Math.\ Phys.\  {\bf 210}, 249 (2000)
[arXiv:hep-th/9912092];
D. Kreimer, {\it Knots and Feynman Diagrams}, (Cambridge Univ Press, 2000).

\bibitem{gvdloops}
G.~V.~Dunne,
``Two-loop diagrammatics in a self-dual background,''
JHEP {\bf 0402}, 013 (2004)
[arXiv:hep-th/0311167].

\bibitem{abdk}
C.~Anastasiou, Z.~Bern, L.~Dixon and D.~A.~Kosower,
``Planar amplitudes in maximally supersymmetric Yang-Mills theory,''
Phys.\ Rev.\ Lett.\  {\bf 91}, 251602 (2003)
[arXiv:hep-th/0309040].


\bibitem{vladimirov}
A.~A.~Vladimirov,
``Method For Computing Renormalization Group Functions In Dimensional Renormalization Scheme,''
Theor.\ Math.\ Phys.\  {\bf 43}, 417 (1980).

\bibitem{chetyrkin}
K.~G.~Chetyrkin and F.~V.~Tkachov,
``Integration By Parts: The Algorithm To Calculate Beta Functions In 4 Loops,''
Nucl.\ Phys.\ B {\bf 192}, 159 (1981).

\end{thebibliography}
\end{document}